\documentclass[aps,twocolumn,prd,superscriptaddress]{revtex4-2}

\usepackage[utf8]{inputenc}

\usepackage{amsmath}
\usepackage{bbm}
\usepackage{mathtools}
\usepackage{amsfonts}
\usepackage{mathrsfs}
\usepackage{bm}
\usepackage{slashed}
\usepackage{tensor}
\usepackage{bbold}
\usepackage{MnSymbol}
\usepackage{textcomp}
\usepackage[export]{adjustbox}

\usepackage{graphicx}
\usepackage{float}
\usepackage{color}
\usepackage{array}
\usepackage[abs]{overpic}

\usepackage{placeins}

\usepackage{makecell}
\usepackage{subcaption}

\usepackage{xspace}
\usepackage{siunitx}
\usepackage{xfrac}
\usepackage{hyperref}
\usepackage[nameinlink]{cleveref}
\usepackage{appendix}

\usepackage{tikz}
\usetikzlibrary{shapes, arrows}
\tikzstyle{roundbox} = [rectangle, draw, text centered, rounded corners, 
minimum height=2em, minimum width=2em, draw=black, fill=blue!10]
\tikzstyle{process} = [rectangle, draw, minimum height=1em, 
minimum width=3em, text centered, draw=black, fill=green!10]
\tikzstyle{integration} = [ellipse, draw, text centered, minimum height=1em, 
minimum width=3em, draw=black, fill=red!10]

\newcommand{\LEGO}{LEGO\textsuperscript{\textregistered}}

\usepackage{xifthen}
\usepackage{xcolor}
\hypersetup{
	colorlinks,
	linkcolor={red!75!black},
	citecolor={blue!75!black},
	urlcolor={blue!75!black}
}

\usepackage{booktabs}
\usepackage{multirow}

\newcolumntype{C}{>{$}c<{$}}
\AtBeginDocument{
	\heavyrulewidth=.08em
	\lightrulewidth=.05em
	\cmidrulewidth=.03em
	\belowrulesep=.65ex
	\belowbottomsep=0pt
	\aboverulesep=.4ex
	\abovetopsep=0pt
	\cmidrulesep=\doublerulesep
	\cmidrulekern=.5em
	\defaultaddspace=.5em
}

\makeatletter
\def\l@subsubsection#1#2{}
\makeatother

\graphicspath{{./figures/}{./figures/equations/}{./figures/results/}}

\newcommand{\gettitle}{Towards quantitative precision in functional QCD I}

\newcommand{\getHeidelbergAffiliation}{\affiliation{Institut f{\"u}r Theoretische Physik, Universit{\"a}t Heidelberg, Philosophenweg 16, 69120 Heidelberg, Germany}}
\newcommand{\getZuerichAffiliation}{\affiliation{Institut f{\"u}r Theoretische Physik, ETH Z{\"u}rich, Wolfgang-Pauli-Str. 27, 8093 Z{\"u}rich, Switzerland}}
\newcommand{\getEMMIAffiliation}{\affiliation{ExtreMe Matter Institute EMMI, GSI, Planckstr. 1, 64291 Darmstadt, Germany}}
\newcommand{\getDarmstadtAffiliation}{\affiliation{Institut f\"ur Kernphysik (Theoriezentrum), Technische Universit\"at Darmstadt, 64289 Darmstadt, Germany}}

\hypersetup{
	pdftitle={\gettitle},
	pdfauthor={Sattler},
	pdfkeywords={discontinuous galerkin}
	{functional renormalization group} {effective potential}
	{phase transition} {numerical methods} {phase structure},
	bookmarksopen=true,
	bookmarksopenlevel=2,
	bookmarksnumbered=true
}

\begin{document}

\title{\gettitle}

\author{Friederike Ihssen}
\getHeidelbergAffiliation
\getZuerichAffiliation
\author{Jan M. Pawlowski}
\getHeidelbergAffiliation
\getEMMIAffiliation
\author{Franz R. Sattler}
\getHeidelbergAffiliation
\author{Nicolas Wink}
\getDarmstadtAffiliation

\begin{abstract}

Functional approaches are the only first principle QCD setup that allow for direct computations at finite density. Predictive power and quantitative reliability of the respective results can only be obtained within a systematic expansion scheme with controlled systematic error estimates. Here we set up such a scheme within the functional renormalisation group (fRG) approach to QCD, aiming for full apparent convergence. In the current work we test this setup, using correlation functions and observables in 2+1 flavour vacuum QCD as a natural benchmark case. While the current work includes many evolutionary improvements collected over the past two decades, we also report on three novel important developments: (i) A comprehensive systematic error analysis based on the modular nature of the fRG approach. (ii) The introduction of a fully automated computational framework, allowing for unprecedented access and improvement of the fRG approach to QCD. (iii) The inclusion of the full effective potential of the chiral order parameter. This also gives access to all-order scattering events of pions and to the full momentum dependence of correlation functions, which is a first application of the automated computational framework (ii). The results compare very well to other state-of-the-art results both from functional approaches and lattice simulations, and provide data on general multi-scattering events of pions and the sigma mode for the first time. 

\end{abstract}

\maketitle

\tableofcontents

\newpage

\section{Introduction}
\label{sec:introduction}

Functional QCD encompasses versatile diagrammatic approaches such as the functional renormalisation group (fRG) and Dyson-Schwinger equations (DSEs) for the first principle description of quantum chromodynamics (QCD). For recent reviews see~\cite{Dupuis:2020fhh, Fu:2022gou, Fischer:2018sdj}. At finite density or baryon chemical potential $\mu_B$, and in particular for $\mu_B/T\gtrsim 3$, it is the only first principle QCD approach that allows for direct computations, since lattice simulations are obstructed by the sign problem in this regime. Still, results from lattice simulations in the vacuum and and at finite temperature and vanishing density provide crucial benchmark results for the functional QCD approach. In particular, a necessary condition for quantitative reliability of functional QCD predictions on the phase structure of QCD is the fulfilment of all available lattice and functional QCD benchmarks at vanishing density, for a recent discussion see \cite{Fu:2023lcm}. A significant recent milestone for the prediction of the QCD phase structure with functional QCD was achieved in \cite{Fu:2019hdw}: the results there meet the lattice benchmarks at vanishing chemical potential, and in particular that of the curvature of the phase boundary, \cite{Fu:2023lcm, Bernhardt:2023ezo}. Its quantitative reliability regime covers the regime $\mu_B / T \lesssim 4$, and this fRG study was corroborated by subsequent Dyson-Schwinger studies, \cite{Gao:2020qsj, Gao:2020fbl, Gunkel:2021oya}. The setup in \cite{Fu:2019hdw, Gao:2020qsj, Gao:2020fbl, Gunkel:2021oya} also allowed for an estimate of the location of the critical end point at about $(\mu, T)_\textrm{CEP} \approx  (\qty{600}{\MeV}, \qty{115}{\MeV}) -(\qty{650}{\MeV}, \qty{105}{\MeV})$. While this agreement across different functional approaches with different resummation patterns and systematic approximation schemes is highly non-trivial and promising, a full quantitative prediction of the phase structure for $\mu_B/T \gtrsim 4$ requires a systematic improvement of the current approximation level. 

The present work is dedicated to the first step in this endeavour by improving on the approximation scheme and the truncation level, incorporating more automation, and providing a conceptual and practical analysis of systematic error estimates. These advancements pave the way to significantly push the boundaries of reliable predictions for the phase structure at higher densities. 

The advances presented in the following are deeply rooted in the diagrammatic, modular structure of functional approaches. This structure is also at the heart of their universal applicability, as well as making them tailor-made for applications to QCD at finite temperature and density as well as realtime processes as already eluded to above: they do not suffer from the sign problem which obstructs respective progress on the lattice. 

These advantages are paid for with intricacies within the systematic error control, which are inherent to all diagrammatic approaches for strongly correlated systems and originates in the apparent absence of a small expansion parameter. However, their highly modular structure allows to easily combine rapidly converging systematic expansion schemes of physical subsystems and across different functional approaches. This rectifies the above mentioned intricacies: in the vertex expansion, QCD scattering processes are expanded in the number of single scatterings, including the full momentum-dependent distribution functions of the respective scattering process. In the derivative expansion part, all scattering orders are taken into account, but their momentum distribution is expanded in powers of momenta. In the fRG approach this scheme is sustained by the fact, that the theory is solved by integrating out scattering events momentum-shell by momentum-shell, and momenta can be measured in the respective infrared cutoff $k$, the expansion being one in $p^2/k^2$. Both expansion schemes come with their own systematics, which have been tested in a plethora of different models with intricate dynamics, encompassing that present in QCD. Moreover, the combination allows for a thorough evaluation of the convergence behaviour. In summary, it allows to not only rely on results and error estimates from the fRG approach to QCD, \cite{Gies:2002hq, Pawlowski:2003hq, Fischer:2004uk, Braun:2008pi, Braun:2009gm, Mitter:2014wpa, Braun:2014ata, Rennecke:2015eba, Cyrol:2016tym, Cyrol:2017ewj, Cyrol:2017qkl, Corell:2018yil, Fu:2019hdw, Braun:2020ada}; but also from the plethora of results and analyses within low energy effective theories in QCD, see~\cite{Dupuis:2020fhh}. 

For example, we utilise the vertex expansion in the glue subsystem of QCD, where it has been proven to be highly efficient and rapidly converging~\cite{Cyrol:2016tym}. In the quark-meson system at hadronic energy scales we use a field gradient expansion of interactions, while in general retaining the full propagators. The gradient expansion in both, vertices and propagators, called the derivative expansion, has been proven to be highly efficient and rapidly converging in low-energy effective theories, see e.g.,~\cite{Dupuis:2020fhh, Fu:2022gou} and references therein. Its rapid convergence also allows for the quantitative determination of critical exponents whose accuracy rivals that of the conformal bootstrap, see \cite{Balog:2019rrg, Dupuis:2020fhh}. The present expansion scheme in the quark-meson sector with the full propagators comprises an improved derivative expansion scheme with even faster convergence. We coin this general modular combination of systematic expansions of subsystems the \LEGO-expansion. A detailed explanation is given in \Cref{sec:ApparentConvergence}, and particularly \Cref{sec:Lego}.

The present work makes progress towards the completion of the low-energy part of this modular scheme, while a further work, \cite{FHPT2024}, will concentrate on full momentum dependences and Fierz completeness, building upon \cite{Mitter:2014wpa, Cyrol:2016tym, Cyrol:2017ewj, Fu:2022uow, Fu:2024ysj}. 
We take the full meson field-dependence of all considered vertices into account using finite element methods (FE methods). In contrast, the momentum dependence is limited to that required for semi-quantitative accuracy. This was tested and proven in a comparison of functional QCD with full momentum dependences, \cite{Pawlowski:2003hq, Fischer:2004uk, Mitter:2014wpa, Cyrol:2016tym, Cyrol:2017ewj, Cyrol:2017qkl, Corell:2018yil}, and functional QCD with an emphasis on the dynamics of mesonic composites \cite{Gies:2002hq, Braun:2008pi, Braun:2009gm, Braun:2014ata, Rennecke:2015eba, Fu:2019hdw, Braun:2020ada}. Furthermore, our fully automated and numerical setup allows to easily scan different regulators, providing a systematic estimate of the associated error. 

In summary, we aim at a fully systematic setup for functional QCD with the functional renormalisation group, including a thorough discussion of systematic error estimates and the extension of the expansion scheme beyond the level discussed explicitly in the present work. 
The combination of these tasks is extensive, and in the present work we concentrate on vacuum QCD. Further works will be dedicated to QCD at finite temperature and density. 
In \Cref{sec:QCDtrunc} we introduce the effective action used in this combined scheme and discuss the relevant approximations and quantities. 
We proceed by discussing the generalised flow equation in a setup with fundamental and composite fields in \Cref{sec:FunFlowsQCD}.
The systematic treatment of QCD within the fRG is detailed in \Cref{sec:ApparentConvergence}, considering the convergence of the given expansion, as well as optimisation procedures and extensions of the scheme. 
We describe the full numerical setup in \Cref{sec:SystematicsQCD}, which is an automated process beginning with the generation of flow equations from a given truncation up to the computation of final results.
The results are presented in \Cref{sec:Results}. Here we combine a discussion of specific benchmark observables such as the pion decay constant and the mass of the scalar $\sigma$-mode with one of momentum-dependent (gauge-fixed) quark, gluon and mesonic correlation functions and the effective potential of the chiral order parameter in comparison with results in the literature. Finally we close in \Cref{sec:outlookQCD} with a brief summary and a discussion of the next steps towards a fully quantitative resolution of the QCD phase structure. Many details on the notation, the derivations as well as results in two-flavour QCD and further ones in 2+1 flavour QCD are deferred to Appendices.

\section{Functional QCD}
\label{sec:QCDtrunc}

In this Section, we discuss the effective action of QCD and detail our approximation and expansion scheme. In \Cref{sec:SQCD+notation}, we provide the classical action with gauge-fixing and ghost terms and establish the general notation. \Cref{sec:EmergentCompositesAction} discusses the inclusion of composite fields for resonant interaction channels. Finally, in \Cref{sec:QCDEffAct}, we provide explicit expressions of our approximation to the effective action for all subsystems. 
Furthermore, we discuss the underlying expansion scheme, which is based on a systematic vertex expansion. An especially important part of the expansion scheme is its use of renormalisation group invariant building blocks, which is detailed in \Cref{sec:RG-invariant expansion}. This provides the theoretical foundation for quantitative reliability and convergence of functional QCD in combination with \Cref{sec:SystematicsQCD}, which delves into systematic error control for an fRG setup with emergent composites and a systematic \LEGO-expansion.

\subsection{Classical action and notation}
\label{sec:SQCD+notation}

With only a few exceptions, functional approaches to QCD are usually formulated in terms of the gauge-fixed theory: the gauge fixing is mandatory for the existence of the gluon propagator, a key building block in the diagrammatic functional approaches. Here, we will make a standard choice and use the Landau gauge. The classical gauge-fixed action reads
\begin{align}
	S_\textrm{QCD}= S_A[A] +S_\textrm{gf}[A] +S_\textrm{gh}[A,c, \bar c] +S_q[A,q,\bar q] \,.
	\label{eq:SQCD}
\end{align}
The glue dynamics is carried by the classical Yang-Mills action $S_A$, while the Dirac action $S_q$ contains the quark dynamics. They are given by
\begin{align}\nonumber
	S_A[A] =            & \, \frac14 \int_x F_{\mu\nu}^a F_{\mu\nu}^a\,,                           \\[1ex]
	S_q [A,q,\bar q]  = & \, \int_x \,\bar q \left( \gamma_\mu D_\mu+m_q-\gamma_0 \mu_q \right)\,q \,. 
	\label{eq:SA+Sq}
\end{align}
The quark field $q$ comprises all quarks used in the computation, in the most general case all three families, $q=u,d,s,c,b,t$.
In the present work, we concentrate on the quantitative computation of the off-shell quantum dynamics in the strongly correlated infrared regime of QCD below approximately \qty{1}{\GeV}. Thus, we consider the 2+1 flavour case $q=(u,d,s)$ with only one heavy quark, as well as the two-flavour case $q=(u,d)$ with only the light flavours. We note in passing, that while the heavier quarks are subleading for the off-shell dynamics, the inclusion of $c,b$ is interesting for phenomenological purposes, e.g.~the computation of heavy quark transport in heavy-ion collisions. As the $c,b$ off-shell dynamics only gives rise to subleading effects in the glue, $u,d,s$ and mesonic off-shell dynamics, their inclusion can be done in a subsequent (modular) step. 

The quark mass term is discussed in detail in \Cref{sec:EmergentCompositesAction}. Here, we only note that we use a flavour diagonal mass matrix, given in general by %
\begin{align} 
	m_q=~\textrm{diag}(m_u,m_d,m_s,m_c,m_b,m_t)\,.
\label{eq:mqall}
\end{align} 
Using the isospin-symmetric approximation, this reduces to
\begin{align} \nonumber 
N_f=2:\ m_q=&\, (m_l,m_l)\,,\\[1ex] 
N_f=2+1:\ m_q=&\,(m_l,m_l,m_s)\,. 
\end{align} 
We have also introduced the quark chemical potential in \labelcref{eq:SA+Sq}, which is given by 
\begin{align}
\mu_q=\textrm{diag}(\mu_l,\mu_l)\,,\qquad \mu_q=\textrm{diag}(\mu_l,\mu_l,\mu_s)\,, 
\label{eq:muq}
\end{align}
in the isospin-symmetric case for two flavours and 2+1 flavours respectively. In the more general case, each heavier quark has its own chemical potential, simply extending the flavour-diagonal matrix $\mu_q$ analogously to the mass matrix. The covariant derivative is given by
\begin{align}
	D_\mu = \partial_\mu - {\textrm{i}}\,g_s\,A_\mu\,,
	\label{eq:CovDer}
\end{align}
and the quarks live in the fundamental representation of the gauge group. 
The components of the field-strength tensor in the Yang-Mills action read
\begin{align}
	F_{\mu\nu}=F^a_{\mu\nu}t^a \,,\quad 	F_{\mu\nu}^a =\partial_\mu A^a_\nu -\partial_\nu A_\mu^a +g_s f^{abc} A_\mu^b A^c_\nu\,,
	\label{eq:Fstrength}
\end{align}
with the su($N_c$) Lie algebra
\begin{align}
	[t^a\,,\,t^b] = {\textrm{i}} f^{abc} t^c\,, 
	\label{eq:LieAlg}
\end{align}
where $t^a $ are the generators of su($N_c$) in a given representation, and $f^{ abc}$ are its structure constants. Ghosts and gluons carry the adjoint representation of the gauge group. We will use the notation 
\begin{align}
	(t_\textrm{ad}^a)^{bc} = ( t^a)^{bc} \,,\qquad ( t_\textrm{f})^{BC} = (T^a)^{BC} \,,
\end{align}
where lowercase letters $b,c$ label indices in the adjoint representation and capital letters $B,C$ label indices in the fundamental representation. 
Furthermore, the generators in the fundamental and adjoint representations are normalised with 
\begin{align}
	{\textrm{tr}}\,t^a t^b = \frac12  \delta^{ab}\,,\qquad \qquad  {\textrm{tr}} \,T^a T^b = \frac{1}{2N_c}  \delta^{ab}\,,
	\label{eq:LieNorm}
\end{align}
where the traces are taken in the respective representation. 
Hence, in component notation, the matter part of the QCD action reads 
\begin{align} 
	S_q [A,q,\bar q]  = & \, \int_x \,\bar q^B  \left( \gamma_\mu D^{BC}_\mu+m_q-\gamma_0 \mu_q\delta^{BC}  \right)\,q^C \,,  
	\label{eq:SqComponents} 
\end{align}
with 
\begin{align}
D_\mu^{BC} = \delta^{BC} \partial_\mu - {\textrm{i}} g A^{BC}_\mu \,.
\end{align}
Finally, the gauge-fixing part of the classical action for a covariant gauge consists of the gauge-fixing term $S_\textrm{gf}$ and the corresponding ghost action $S_\textrm{gh}$,
\begin{align}
	S_\textrm{gf}= \frac{1}{2\xi} \int_x\, (\partial_\mu A_\mu^a)^2 \,,\qquad  S_\textrm{gh}=	\int_x \, \bar c^{\,a}\left(-\partial_\mu D_\mu^{ab}\right) c^b \,,
	\label{eq:Sgauge}
\end{align}
where we take the Landau gauge, i.e. $\xi\to 0$, in explicit computations. On the technical side, this choice comes with a significant reduction of the size of the flows and the number of dressings that have to be computed. The latter originates in the fact that the purely transverse system is closed in the Landau gauge, while for $\xi\neq 0$ the longitudinal system feeds into the transverse one. For a detailed discussion see \cite{Fischer:2008uz, Cyrol:2016tym, Dupuis:2020fhh, Pawlowski:2022oyq}, for functional results (DSE) with $\xi\neq 0$ see \cite{Aguilar:2015nqa, Huber:2015ria, Napetschnig:2021ria}. Therefore, this choice is more than conceptual: it reduces approximation artefacts in the system and therefore also minimises the systematic error. Finally, from now on we shall only consider the physics case $N_c=3$ with $a,b,c,...=1,...,8$ and $A,B,C,...=1,2,3$.

\subsection{Emergent composites} 
\label{sec:EmergentCompositesAction}

At large energy scales QCD, is well-described by the (perturbative) dynamics of quarks and gluons, the fundamental degrees of freedom in QCD. At low energy scales, the gluonic dynamics decouple. In this regime, the QCD dynamics is dominated by that of the lightest hadrons. They emerge from resonant channels of four-quark scatterings (mesons) or higher-order scatterings of quarks (baryons and higher-order resonances). In the present functional renormalisation group approach with emergent composites (or dynamical hadronisation), this emergence of low-energy degrees of freedom is captured in a natural way, which we explain in more detail in \Cref{sec:EmergentMesons}.

In the vacuum, the most dominant channel is the pseudoscalar one which carries the dynamics of the pions $\boldsymbol{\pi}$. It is followed by the scalar channel, which carries the dynamics of the $\sigma$-mode. In \cite{Mitter:2014wpa, Cyrol:2017ewj}, it has been shown that the additional fluctuations induced by the other tensor channels are negligible and can safely be dropped.

Furthermore, this relevance ordering also persists at finite temperatures and chemical potentials with $\mu_B/T \lesssim 6$. Beyond this regime, the dominant channel switches rapidly to the diquark channel, see \cite{Braun:2019aow}. 
A respective conservative estimate of the relevance ordering at finite temperature and density leads us to $\mu_B/T \lesssim 4$ as a bound for quantitative reliability within the approximation discussed in the present work.
This quantitative reliability bound is corroborated by further considerations concerning non-trivial pion and $\sigma$ dispersions in \cite{Fu:2019hdw} which hint at a possible moat regime \cite{Pisarski:2021qof} in QCD. 

We conclude that for $\mu_B/T \gtrsim 4$ the present approximation has to be systematically improved. We emphasise that this poses merely a technical challenge, in contradistinction to the sign problem for lattice simulations at large chemical potential. The current work sets the stage for the systematic resolution of this task, as we outline and set up an easily extendable framework for functional QCD.

In the fRG approach with emergent composites the effective action of QCD explicitly contains the fields arising from the resonant channels. In the present case these are the pions $\boldsymbol{\pi}$ and the $\sigma$-mode. For $N_f$ flavours, the scalar mesonic matrix fields in the singlet-octet basis are given by 
\begin{align}
	\Sigma=\left(\sigma^i +i \,\pi^i\right) T_{f}^i \,,\quad \textrm{with}	\quad  \left(T^i _{f}\right) = ( T_f^0\,,\, T^a_f )  \,, 
	\label{eq:Sigma}
\end{align}
where $T^a_f$ with $a=1,...,N_f^2-1$ are the generators of SU($N_f$) with the Lie algebra \labelcref{eq:LieAlg} and the singlet matrix is given by 
\begin{align}
T^0_f =\frac{1}{\sqrt{2 N_f} }\mathbbm{1}\,.
\end{align}
The mesonic fields couple to the quarks in a scalar-pseudoscalar combination. In the presence of the full SU($N_f$) flavour symmetry this leads to 
\begin{align}
\int_x \bar q \,\Sigma_5 \, q\,,\qquad \textrm{with}\qquad  	\Sigma_5 = \left(\sigma^i +i \gamma_5 \pi^i\right) T_{f}^i\,,   
\end{align}
with the scalar modes $\sigma^i$ and the pseudoscalar Goldstone modes $\pi^i$, for more details see e.g.~\cite{Jungnickel:1996aa, Schaefer:2008hk, Mitter:2013fxa, Herbst:2013ufa, Rennecke:2016tkm, Wen:2018nkn, Fu:2018qsk, Fu:2019hdw}. 

In the following we only consider the part of the multiplet, that contributes sizably to the off-shell dynamics of QCD: this choice is informed by the mesonic mass hierarchy in physical QCD, compared to the chiral symmetry breaking scale $k_\chi$, below which the mesonic fluctuations contribute to the off-shell dynamics. This scale is given by $k_\chi\approx \qty{400}{\MeV}$, see \cite{Cyrol:2017ewj, Fu:2019hdw}, and we find the same scale in the present work, see \labelcref{eq:ChiralSymmbScale} in \Cref{sec:InitialConditions} and the discussion there. We conclude that only the three physical pions with $m_\pi \approx \qty{140}{\MeV}$ contribute sizably to the off-shell dynamics. All further mesonic excitations carry masses $m_\textrm{mes} \gtrsim \qty{500}{\MeV}$, the lightest one being lightest scalar resonance, the light quark $\sigma$ mode, and the kaons. Their off-shell dynamics is already suppressed at $k_\chi$ and their contributions to the off-shell dynamics is subleading. Moreover, for this off-shell analysis one also has to consider coloured composites such as diquarks: while they are not related to asymptotic states, they may contribute to the off-shell dynamics. However, their effective mass scales are higher than that of the $\sigma$-mode and we can safely drop them for the computations in the present work. At sufficiently large chemical potential, though, they have to be considered. Finally, the off-shell irrelevance argument applies in particular to quark-composites that contain the heavier quarks, whose larger current quark mass leads to comparably large masses of the respective hadronic and diquark resonances in the loops. This structural argument has been confirmed in an explicit computation two-flavour QCD within Fierz-complete computations, where all mesonic channels are taken into account, \cite{Mitter:2014wpa, Cyrol:2017ewj}: dropping all channels but the scalar-pseudoscalar channels in the off-shell dynamics carried by the diagrams had a minimal impact on the results. While being of subleading importance for the off-shell dynamics modes such as the kaons, general scalar and vector modes or diquarks, may be relevant for on-shell scattering processes, thermodynamic observables such as the pressure or more generally gain importance at finite chemical potential. 

Indeed, the light and strange chiral condensates are carried by the diagonal part of $\sigma^i T^i$, that couples to the quark mass operator. They have to be considered at least as background fields. This leaves us with $(\sigma,\vec \pi)$ in two-flavour QCD and $(\sigma^0,\sigma^8,\vec \pi)$ in 2+1 flavour QCD. 

We illustrate this first for $N_f=2$, where this combination comprises the scalar $\sigma$-mode $\sigma_l$, that couples to $\bar l l$, and the pseudoscalar pions, 
\begin{subequations} 
	\label{eq:phiNf2}
\begin{align}
\phi_l = \phi^i_l \tau^i_2 \,,\qquad \tau_2=\frac12 (\mathbb{1}, i \gamma_5 \bm \sigma)\,,
	\label{eq:phiNf21}
\end{align}
and $\bm \sigma$ are the Pauli matrices.
The component fields $\sigma^0$ and $\pi^a$ are given by 
\begin{align}
	(\phi^{i}_l) = \left(\sigma^{\ }_l, \boldsymbol{\pi}\right)\,, \quad  \textrm{with} \quad \boldsymbol{\pi} = (\pi^1, \pi^2, \pi^3) \,. 
	\label{eq:phiNf22}
\end{align}
\end{subequations} 
The poles masses of these fields are about $\qty{140}{\MeV}$ (pions) and $\qty{500}{\MeV}$ ($\sigma$-mode). As already discussed above, these masses have to be compared with the chiral symmetry breaking (cutoff) scale $k_\chi \approx \qty{400}{\MeV}$, below which these degrees of freedom get dynamical. Accordingly, the pions by far dominate the mesonic infrared off-shell dynamics of physical QCD also for $N_f>2$. Note also that the choice \labelcref{eq:phiNf2} implies a maximal breaking of the axial $U_A(1)$ symmetry, which will be discussed further in \Cref{sec:Matter}. While this is a subleading effect for the off-shell dynamics considered here, it is relevant for an access to the full hadronic mass spectrum, for the intricacies of axial $U_A(1)$ restoration in the chiral phase transition at finite temperature, as well as further observables.  

We now proceed with the 2+1 flavour case. Taking into account the above arguments, the two-flavour field $\phi_l$ carries the most important part of the off-shell dynamics and we simply embed it into the three-flavour case. The only additional degree of freedom considered is the scalar $\sigma_s$ that couples to the strange mass operator $\bar s s$. The two scalar fields $\sigma_l,\sigma_s$ can be constructed from the singlet-octet basis \labelcref{eq:Sigma} in three-flavour QCD. For details on the relation between $(\sigma_l, \sigma_s)$ and the singlet-octet basis see e.g.~\cite{Jungnickel:1996aa, Schaefer:2008hk, Mitter:2013fxa, Herbst:2013ufa, Rennecke:2016tkm, Wen:2018nkn, Fu:2018qsk, Fu:2019hdw}. Here, we simply quote the relation 
\begin{align}
	\begin{pmatrix} \sigma^{\ }_l \\[.5ex] \sigma^{\ }_s \end{pmatrix}  =
	\frac{1}{\sqrt{3}} \begin{pmatrix} \sqrt{2} &  1
		\\[.5ex]  1 & -\sqrt{2} \end{pmatrix}\,
	\begin{pmatrix} \sigma^{\ }_0 \\[.5ex]  \sigma^{\ }_8 \end{pmatrix} \,, 
	\label{eq:SigmaSigmas}
\end{align}
for the two scalar components of the 2+1-flavour mesonic composite, that couple to the mass operators $\bar l l, \bar s s$. As mentioned before, $\sigma_s$ is not added for the relevance of its off-shell contributions but only because its expectation value drives the strange constituent quark mass. In summary, we arrive at the 2+1 flavour mesonic composite $\phi$ with 
\begin{subequations}
	\label{eq:phiNf2+1}
\begin{align}
\phi= \phi^i \tau^i\,,\qquad 	(\phi^i)=\left((\phi^i_l)\,,\,\sigma_s\right)\,. 
	\label{eq:phiNf2+11}
\end{align}
In \labelcref{eq:phiNf2+11} we have simply embedded the two-flavour mesonic composite into the three-flavour matrix with
\begin{align}
\tau =\tau^{\ }_{2+1} =\left( \Biggl(  \begin{matrix} \tau_2 & 0 \\[1ex] 0 & 0 \end{matrix}\Biggr) \,,\,   \left( \begin{matrix} 0 & 0 \\[1ex] 0 & \frac{1}{\sqrt{2}} \end{matrix}\right)\right) \,,
\label{eq:tau2+1}
\end{align}
\end{subequations}
where the vector $\tau_{2}$ of the two-flavour generators is defined in \labelcref{eq:phiNf21}. It is already worth emphasising here that the emergent composites are only introduced as an efficient book-keeping device of (resonant) four-quark channels and multi-scatterings in these channels. The latter are potentially important close to the resonance: in the present fRG approach to QCD, the occurrence of channel fields in the action does not signal an effective field theory setup, but a reparametrisation of first-principles QCD. 
The emergent composite $\phi$ is only introduced by means of the emergent composite approach in QCD (also called dynamical hadronisation) and simply absorbs emerging QCD dynamics from the corresponding quark channel. This is discussed in \Cref{sec:EmergentMesons}, for further details on the emergent composite approach in QCD, see \cite{Braun:2014ata, Mitter:2014wpa, Rennecke:2015eba, Cyrol:2017ewj, Fu:2019hdw, Fukushima:2021ctq}, which are based on \cite{Gies:2001nw, Gies:2002hq, Pawlowski:2005xe, Floerchinger:2009uf}, for reviews see \cite{Dupuis:2020fhh, Fu:2022gou}.

\subsection{QCD effective action and expansion scheme}
\label{sec:QCDEffAct}
\begin{figure*}%
	\centering%
	\includegraphics[width=.65\linewidth]{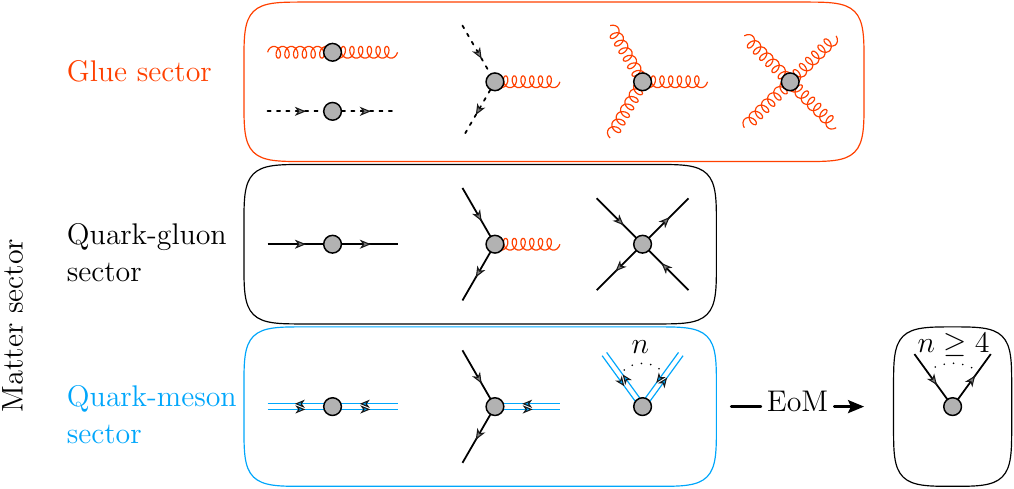}%
	\caption{Diagrammatic depiction of the truncation used in the present work. The arrows indicate the flow of the fermion charge, not the momentum. The vertices and propagators in the pure glue sector are contained in the orange frame, and the matter sector is split into the quark-gluon sector, contained in the black frame and the quark-meson sector, contained in the blue frame. All propagators and vertices involving $\Phi_f$, are evaluated on the EoM of $\phi$, and the purely mesonic correlation functions are computed for all constant fields $\phi$. The diagrammatic rules are provided in \Cref{fig:diag_notation} in \Cref{app:flows}. 
	Note that the mesonic composites $\phi$ are shown as double lines i.e. two-quark composites with opposite arrows. On the equations of motion of $\phi$ they are related to a series in powers of quark-bilinears. Hence, two- and higher point functions of the composites $\phi$ translate to four- and higher-order multi-quark scatterings, as indicated in the quark-meson sector depiction.
			\hspace*{\fill}}%
	\label{fig:system_overview}%
\end{figure*}

In this Section we discuss the approximation of the effective action used in the present work. As already mentioned previously, this work sets the stage for a systematic improvement of the fRG approach to QCD at finite $T$ and $\mu_B$ building on \cite{Braun:2014ata, Mitter:2014wpa, Rennecke:2015eba, Cyrol:2017ewj, Cyrol:2017qkl, Fu:2019hdw}, in order to finally allow for quantitative access to QCD for $\mu_B/T\gtrsim 4$.

To begin with, quantitative access to the dynamics in the vicinity of a potential critical end point -- or rather the onset regime of new phases -- requires the inclusion of a full effective potential of the critical mode. Note, that this is not relevant for the quantitative prediction of the location of this regime which is a direct consequence of smallness of the critical regime. Indeed, the non-universal aspects of the phase structure including the location of the onset regime of new phases are governed by the dynamics of the soft modes in QCD, see \cite{Braun:2023qak}.

The present approach is readily extended to fully momentum-dependent correlation functions, either included directly in the flow, or within the iteration procedure as set up in \cite{Helmboldt:2014iya} and used in \cite{IKPS2023}.
This step is already prepared in this work, as all threshold functions are computed numerically. Importantly, this also allows to arbitrarily switch regulator functions, which enables us to study the regulator dependence of the results to obtain a part of the systematic error estimate.

We defer the full inclusion of momentum-dependent correlation functions in QCD to future work and use cutoff-scale-dependent quantities with the exception of the gluon and ghost two-point functions. For these we use data from \cite{Cyrol:2017ewj} obtained in two-flavour QCD as external input, following the procedure put forward in \cite{Braun:2014ata}. Still, already here we make use of the full momentum-dependent setup and read out the full momentum structure of the quark mass and wave function, as well as the full meson propagator. This gives direct access to the pion decay constant $f_\pi$ and the pion pole mass, and latter allows us to directly adjust the physical current quark masses in our system. 

The full scale-dependent effective action is split into a pure glue part and a matter part, 
\begin{align}
	\Gamma_k[\Phi] = \Gamma_{\textrm{glue},k}[A,c,\bar c] + \Gamma_{\textrm{mat},k}[\Phi]\,, 
	\label{eq:effAct}
\end{align}
where the explicit expressions for the pure glue and matter part are provided in \labelcref{eq:Gammaglue,eq:Gammamatter}. 
The superfield $\Phi$ is defined as the combination of the superfield of the fundamental fields $\Phi_f$ and the emergent composites $\phi$ in \labelcref{eq:Sigma}, 
\begin{align}
	\Phi = \left(\Phi_f,\phi \right) \,, \quad \Phi_f = \left(A, c, \bar c, q, \bar q\right)\,. 
	\label{eq:Superfield}
\end{align}
The effective action is expanded in $n$-point scattering vertices, 
\begin{subequations} 
	\label{eq:GammaVertexExpansion}	
\begin{align}
\Gamma[\Phi]= \sum_{\boldsymbol{n}}  \int_{\boldsymbol{p}}  \frac{1}{\boldsymbol{n}!} \Gamma_{\Phi_{i_1} \cdots\Phi_{i_n} }^{(n)}(\boldsymbol{p}) \Phi_{i_n}(p_n)\cdots  \Phi_{i_1}(p_1)\,,
\end{align}
with 
\begin{align} 
	\boldsymbol{p} =(p_1,...,p_n)\,,\qquad \boldsymbol{n}! = n_{A}! \cdots n_{\phi}! \,,
	\label{eq:boldpn}
\end{align} 
with $n_{\Phi_i}$ being the number of derivatives with respect to the field $\Phi_i$ in the $n$-point correlation function. 
For an explanation of the notation and rules for field derivatives, see \Cref{app:Diagrammatics}. The $n$-point scattering vertices carry the one-particle irreducible part of the interaction with the full momentum dependence, including all tensor channels. 

The $S$-matrix of QCD can be directly constructed from tree-level diagrams of these correlation functions. In general they read, 
\begin{align}
	\hspace{-.1cm} \Gamma_{\Phi^n}^{(n)}(	\boldsymbol{p}) = \sum_{i=1}^{N_{\Phi^n}}\!  \lambda_{\Phi^n}^{(i)}(	\boldsymbol{p}) {\cal T}^{(i)}_{\Phi^n}(	\boldsymbol{p})\,,
	\label{eq:CompleteBasisGn}
\end{align}
with $	\boldsymbol{p}$ defined in \labelcref{eq:boldpn}. The subscript $\Phi^n$ denotes some combination of $\Phi_{i_1} \cdots\Phi_{i_n}$. 
The scalar functions $ \lambda_{\Phi^n}^{(m)}$ are the dressings of a complete tensor basis $\{{\cal T}^{(i)}_{\Phi^n}\}$ of $\Gamma_{\Phi^n}$, where $m=1,...,N_{\Phi^n}$. 
\end{subequations}
While \labelcref{eq:GammaVertexExpansion} represents $\Gamma[\Phi]$ in terms of a vertex expansion about $\Phi=0$, we generally consider vertex expansions about a solution of the equations of motion (EoM) for $\Phi$. Note that $\Phi=0$ is such a solution but it may not minimise the effective action. 

The modular form of functional approaches also suggests to combine the vertex expansion in the fundamental QCD degrees of freedom, $\Phi_f$ in \labelcref{eq:Superfield}, with a momentum (derivative) expansion in the emergent dynamical low-energy degrees of freedom. In vacuum, this only concerns the scalar-pseudoscalar modes in $\phi$. Then, the full $\phi$-dependence of all $n$-point correlation functions of $\Phi_f$ is taken into account and expanded about constant backgrounds in powers of $p^2/\Lambda_\textrm{QCD}^2$, i.e. in powers of derivatives $\partial^n \phi(x)$. 
The systematics of such an expansion and the respective systematic error estimates are discussed in \Cref{sec:ApparentConvergence}. 

The following three Sections discuss general properties of the present expansion scheme. We introduce the approximation used for the pure glue part in \Cref{sec:PureGlue}, the matter part in \Cref{sec:Matter} and its underlying full momentum dependence in \Cref{sec:DerMatter}. The details of the renormalisation group (RG) invariant expansion scheme in are discussed in \Cref{sec:RG-invariant expansion}.

For a diagrammatic overview we depict the $n$-point functions, propagators and vertices considered in the present work in \Cref{fig:system_overview}: The vertices and propagators in the pure glue sector are contained in the orange frame, whereas the matter sector is split into the quark-gluon sector, in the black frame, and the quark-meson sector, in the blue frame. If evaluated on the EoM for $\phi$, the latter sector is converted into multi-scattering vertices of quarks and anti-quarks, as also indicated in \Cref{fig:system_overview}. This is the diagrammatic depiction of the property that the QCD effective action with emergent composite fields reduces to the standard QCD
effective action on the $\Phi_f$-dependent solution of the EoM for $\phi$, 
\begin{align}
	\Gamma_\textrm{QCD}[ \Phi_f]= \Gamma\bigl[ \Phi_f, \phi_0\left[\Phi_f\right]\bigr]\,, 
\label{eq:GComp=GQCD}
\end{align}
for a more detailed discussion see \cite{Fu:2019hdw}. 
Here, we only note that \labelcref{eq:GComp=GQCD} entails that the correlation functions $\Gamma^{(n)}_{\textrm{QCD},\Phi_f^n}$ are sums of correlation functions of $\Gamma^{(m)}_{\Phi^n}[ \Phi]$ with $m\leq n$, the prefactors being $\Phi_f$-derivatives of $ \phi_0[\Phi_f]$. Explicit examples are discussed in \cite{Fu:2019hdw}. 

All propagators and vertices, which involve the fundamental QCD fields $A,c,\bar c,q,\bar q$, are computed on the solution 
\begin{align}
\phi_0=(\sigma_{l,0},\boldsymbol{\pi}=0, \sigma_{s,0})\,,
 \label{eq:phi0sigma0}
 \end{align}
 of the equation of motion for $\phi$. 
This solution carries the information about chiral symmetry breaking and is directly proportional to the light quark condensate. In turn, the propagators and vertices of the mesonic emergent field $\phi$ are evaluated for all constant mesonic backgrounds. This incorporates all order mesonic scatterings which is especially important in the chiral limit and the scaling regime of phase transitions.

\subsubsection{Pure glue part} 
\label{sec:PureGlue}

For the discussion of our approximation of the full momentum dependence, we first retain the full momentum dependence of the dressings of the tensor structures taken into account. We restrict ourselves to an approximation which only includes the leading (classical) tensor structures of all vertices and allow for general dressings or form factors for these tensor structures. Then, the approximation of the pure glue action contains the kinetic term of the gluon as well as three- and four-gluon vertices with running vertex factors for the classical tensor structure. Furthermore, we have the kinetic term of the ghost and the classical ghost-gluon vertex.
For more details, see the full setup in terms of diagrammatic rules in \Cref{app:Diagrammatics}.

All computations are performed in momentum space, and hence we parametrise the interacting part in momentum space as, 
\begin{widetext} 
\begin{align}\nonumber
	\Gamma_{\textrm{glue},k}[A,c,\bar c]  =&\,
	\frac12 \sumint\limits_p \, A^a_\mu(p) \, p^2 \left[  Z_A(p)\,\Pi^\perp_{\mu\nu}(p) 
		+\frac{ 1}{\xi}Z^\parallel_A(p)\,\Pi^\parallel_{\mu\nu}(p) \right] \, A^a_\nu(-p)  \\[2ex]\nonumber
		&\hspace{-2.3cm}  + \frac{1}{3!} \sumint\limits_{p_1,p_2} \lambda_{A^3}(p_1,p_2)  \left[ {\cal T}_{A^3}^{(1)}(p_1,p_2)\right]^{a_1 a_2 a_3}_{\mu_1\mu_2\mu_3}
	\prod_{i=1}^3 A^{a_i}_{\mu_i}(p_i)  
 +  \frac{1}{4!}\sumint\limits_{p_1,p_2,p_3} \hspace{-.3cm}\lambda_{A^4}(p_1,p_2,p_3)   \left[ {\cal T}_{A^4}^{(1)}(p_1,p_2,p_3)\right]^{a_1 a_2 a_3 a_4}_{\mu_1\mu_2\mu_3\mu_4}
	\prod_{i=1}^4 A^{a_i}_{\mu_i}(p_i)    \\[2ex]
	&\hspace{-2.3cm}	+ 	\sumint\limits_p   Z_c(p)  \bar c^{\,a}(p) p^2 \delta^{ab}c^b(-p) + \sumint\limits_{p_1,p_2} \lambda_{c \bar c A}(p_1,p_2)  \left[ {\cal T}_{c\bar c A}^{(1)}(p_1,p_2)\right]^{a_1 a_2 a_3}_{\mu}
	\bar c^{a_2} (p_2) c^{a_1}(p_1) A^{a_3}_\mu(-(p_1+p_2))      \,, 
	\label{eq:Gammaglue}
\end{align}
\end{widetext} 
with the transversal and longitudinal projection operators $\Pi^\bot(p),\Pi^\parallel(p)$, defined in \labelcref{eq:ProjectionOps}, as well as the respective dressings $ Z_A(p)$ of the transverse gluon two point function and $Z^\parallel_A(p)$ of the longitudinal one. 

At finite temperature $T$, the temporal integration is running from $0$ to $1/T$. In momentum space this leads to sums over Matsubara frequencies, 
\begin{align}
	\int_x = \int_0^{1/T} dx_0 \int d^3x\,, \qquad \sumint\limits_p = T \sum_n \int \frac{d^3 p}{(2 \pi)^3}\,,
	\label{eq:finiteTx}
\end{align}
with $p_0 = 2 \pi T \,n$ for gauge fields and ghosts.
However, in this work we consider QCD only in the vacuum and thus all Matsubara sums reduce to integrals of $p_0$ from $-\infty$~to~$\infty$.

The tensor structures ${\cal T}^{(1)}_{A^n}$ in \labelcref{eq:Gammaglue} are the classical tensor structures of the Yang-Mills action in \label{eq:SA-Sq}
\begin{align}
	{\cal T}_{A^n}^{(1)} = 	\left[ S_\textrm{A}+S_\textrm{gf}\right]^{(n)}_{A^n}\,,\qquad {\cal T}_{c\bar c A^n}^{(1)} = 	\left[ S_\textrm{gh}\right]^{(n+2)}_{c\bar c A^n}\,, 
\end{align}
evaluated at $g=1$ and $A,c,\bar c=0$. 

\Cref{eq:Gammaglue} is a rather sophisticated approximation to the glue sector of QCD. However, while it carries the full momentum dependences of dressings of the classical, primitively divergent tensor structures, it misses the non-classical ones: A basis of tensor structures for the three-gluon vertex contains eight transverse tensors, for the four-gluon vertex a basis contains 41 transverse tensors, see e.g.~\cite{Eichmann:2015nra}.

Finally, the ghost-gluon vertex accommodates a further longitudinal tensor structure. Note that longitudinal tensor structures do not contribute to the closed dynamical system of transverse propagators and vertices in the Landau gauge, but are potentially relevant for the dynamical generation of the mass gap. This has been discussed thoroughly in the literature, see in particular \cite{Aguilar:2011xe, Binosi:2012sj, Aguilar:2021uwa, Papavassiliou:2022wrb, Ferreira:2023fva, Cyrol:2016tym, Eichmann:2021zuv, Horak:2022aqx}. For a detailed discussion of the momentum dependences and computations in pure glue and $N_f=2$ flavour QCD, including also further tensor structures, see \cite{Cyrol:2016tym, Cyrol:2017ewj}. 

The dressings $Z$, $\lambda$ in \labelcref{eq:Gammaglue} are related by Slavnov-Taylor identities (STIs) and allow also for the definition of different avatars of the momentum-dependent running coupling. To that end, we evaluate the dressings at the symmetric point and divide by the appropriate powers of the wave functions, 
\begin{subequations}
\label{eq:StrongCouplings}
\begin{align}
g_{A^3}(\bar p)= \frac{ \lambda_{A^3}(\bar p) }{ Z_A^{3/2} (\bar p)}\,, \qquad  g_{A^4}(\bar p)= \frac{\lambda_{A^4}(\bar p)}{Z_A^{2} (\bar p)}\,,
\label{eq:StrongCouplingsGlue}
\end{align}
where $\bar p$ is a symmetric point configuration for the three-point vertex and the four-point vertex (for a definition, see \labelcref{eq:symmetricPoint}). 
The avatar of the strong running coupling from the ghost-gluon vertex follows as 
\begin{align}
g_{c\bar c A}(\bar p)= \frac{ \lambda_{c\bar c A}(\bar p) }{ Z_A^{1/2} (\bar p) Z_c(\bar p)}\,. 
	\label{eq:StrongCouplingsGhostGlue}
\end{align}
\end{subequations}
The running couplings in \labelcref{eq:StrongCouplings} are renormalisation group invariant, while the vertex dressings are not. 
Both dressings and couplings are $k$-dependent and the $k$-dependence of the running couplings is related, but not identical, to their $\bar p$-dependence. The conversion of the RG-variant dressings and wave functions into RG-invariant running couplings is an example of a more general structure, which allows to define RG-invariant vertex dressings. This will be discussed and utilised for a manifestly RG-invariant expansion scheme in \Cref{sec:RG-invariant expansion}. 

Here, we proceed with a discussion of the asymptotic behaviour of the couplings. For asymptotically large momenta and a vanishing cutoff scale, the pure glue couplings agree,
\begin{align}
g_i(p)	= g_s(p) +O_i( g_s^3)\,,\quad \textrm{for} \quad \frac{p^2}{\Lambda^2_{\textrm{QCD}}}\gg 1\,,
	\label{eq:AgreeAvatars}
\end{align}
with $i=A^3,A^4, c\bar c A$. Here, $O_i$ captures higher order terms that vanish with $g_i\to 0$ and $g_s(p)$ is the unique renormalised coupling at large $p$. \Cref{eq:AgreeAvatars} follows from quantum gauge invariance, as comprised by the Slavnov-Taylor identities. It also reflects the two-loop universality of the running coupling.
Note that the latter only applies to the RG-running of the coupling in mass-independent RG-schemes and does not extend to the full momentum dependence. An example for the latter statement is the Taylor coupling, which can be constructed from the dressings of ghost and gluon propagators in the Landau gauge, 
\begin{align} 
	g^{\ }_T(p) = \frac{g_s}{ Z^{1/2}_A(p) Z_c(p)} \,,
\end{align}
where $g_s$ is the renormalised coupling at some large, perturbative momentum scale, see e.g.~\cite{Hauck:1998hv, vonSmekal:2009ae}. It has the same perturbative RG-running as the other avatars $g_i$ in \labelcref{eq:AgreeAvatars} and differs from the ghost-gluon coupling only by the dressing of the ghost-gluon vertex, 
\begin{align}
	g_{c\bar c A}(p) =  g^{\ }_T(p)\, \lambda_{c\bar c A}(p)\,. 
	\label{eq:Taylor-GhostGluon}
\end{align}
The identical RG-running of the left and right hand side originates in the non-renormalisation theorem for the ghost-gluon vertex in the Landau gauge, $\mu\partial_\mu \lambda_{c\bar c A}(p)=0$. Evidently, the momentum dependence differs by that of the vertex dressing, $\lambda_{c\bar c A}(p)$. It is also worth noting that the RG-running of all avatars of the strong coupling in \labelcref{eq:AgreeAvatars} agrees in all renormalisation group (RG) schemes at one loop, while it only persists at two-loop for mass-independent RG-schemes. 
With the infrared momentum cutoffs used in the present fRG approach, the underlying RG-scheme is mass-dependent and thus leads to modified Slavnov-Taylor identities at finite $k$. 

In turn, for small momenta, $p/\Lambda_{\textrm{QCD}}\lesssim 1$, the couplings do not agree anymore. Indeed, they differ significantly: the respective Slavnov-Taylor identities depend on scattering kernels which are trivial in the perturbative regime, but grow large in the infrared. For an overview of the resulting vertex expansion, see e.g.~\Cref{app:Diagrammatics}, for related literature in functional approaches to Yang-Mills theory and QCD (fRG and DSE) see \cite{Fischer:2008uz, Aguilar:2009nf, Mitter:2014wpa, Cyrol:2016tym, Cyrol:2017ewj, Binosi:2016nme, Huber:2018ned, Huber:2020keu}.

In the approximation used in the current work, we shall use the relation between the $k$-running and the $\bar p$-running to approximate the couplings \labelcref{eq:StrongCouplings} by their counterparts at $\bar p=0$ in the loop diagrams: this approximation is based on the fact that the loop momenta $q$ in the fRG approach are restricted to $ q^2 \lesssim k^2$ and the $\bar p$ dependence of the couplings for these loop momenta is subleading. 
This approximation is well-tested in comparison to full results, see in particular \cite{Mitter:2014wpa, Cyrol:2016tym,Cyrol:2017ewj}. Hence, we use the approximation
\begin{align}
g_{i,k}(p^2\lesssim k^2) \approx g_{i,k}(0)\,, \quad \textrm{with }\quad i=A^3\,,\,A^4\,,\,c\bar cA\,.
\label{eq:Localg}
\end{align}
This concludes our discussion of the approximation in the pure glue sector.

\subsubsection{Matter part} 
\label{sec:Matter}

The discussion of the pure glue part in the last Section already illustrated the occurrence of RG-invariant building blocks in the effective action. 
We use this pattern again for the expansion of the matter part of the effective action. As in pure glue part, we restrict ourselves to couplings in the limit of a vanishing symmetric point configuration $\bar p\to 0$ as used for the couplings in \labelcref{eq:Localg}. This leads us to 
\begin{align}\nonumber
	\Gamma_{\textrm{mat},k}[\Phi] = \int_x \,\Biggl\{&  \,\bar q \, Z_q \left( \gamma_\mu D_\mu-\gamma_0 \mu_q \right)q \\[1ex]\nonumber
	                 & - \lambda_q(\rho) \, \left[ \left(\bar q  \,Z_q\,\tau^0 q\right)^2 + \left(\bar q \,Z_q\,\bm \tau q \right)^2 \right]          \\[1ex]\nonumber
	                 & + \bar q\, \left[  Z_q Z_\phi^{1/2} \,h_\phi(\rho)\,\phi^i \tau^i\right]\,q                                  \\[1ex]\nonumber 
	                 & +\frac{1}{2} \left( Z_\phi^{1/2} \partial_\mu \phi \right)^2 + V(\rho)\\[1ex]
	                 & -\frac14 \textrm{tr}\, c_\phi \cdot \left(Z_\phi^{1/2} \phi\right)	\Biggr\}\,,
	\label{eq:Gammamatter}
\end{align}
with the superfield \labelcref{eq:Superfield}, the mesonic composites \labelcref{eq:phiNf2} ($N_f=2$) or 	\labelcref{eq:phiNf2+1} ($N_f=2+1$), the quark chemical potential $\mu_q$. In our present approximation, $\sigma_s$ is the only mesonic composite with strangeness, and $\rho$ contains the two chiral invariants 
\begin{align}
\rho=\left(\rho_l,\rho_s\right) \,,\quad \textrm{with}\quad \rho_l = \frac{\sigma_{l}^2+\boldsymbol{\pi}^2}{2}\,,\qquad   \rho_s = \frac{\sigma_{s}^2}{2}\,.
\label{eq:rho-rhos}
\end{align}
The four-quark terms in the second line of \labelcref{eq:Gammamatter} include the two-flavour scalar and pseudoscalar channels as well as the scalar strange channel, see \labelcref{eq:tau2+1}, and we use a uniform coupling $\lambda_q$ for $u,d,s$. The quark flavours are split into the two light flavours $l$ and the heavy ones, 
\begin{align} 
	q= ( l,s,c,b,t)\,,\quad \textrm{with} \quad l=(u,d)\,, 
\end{align}
using a isospin-symmetric approximation. Furthermore, we restrict ourselves to a flavour diagonal matrix of Yukawa couplings $h_\phi(\rho)$, 
\begin{align}
	h_\phi =\textrm{diag}\left( h_{q_i}\right)\,, \qquad  h_u=h_d=h_l\,,
	\label{eq:hphi}
\end{align}
with isospin symmetry. The matrices of the quark and meson wave functions $Z_q$ and $Z_\phi$ are diagonal, 
\begin{align}
	Z_q =\textrm{diag}\left( Z_{q_i}\right)\,, \qquad Z_\phi= \textrm{diag}\left( Z_{\phi_l},  Z_{\phi_l}, Z_{\phi_s}\right)\,,
\end{align}
with the isospin-symmetric choice $ Z_u=Z_d=Z_l$ and a separate $Z_s$ for the strange quark. From now on we restrict ourselves to the 2+1 flavour case, dropping the heavy quarks $(c,b,t)$. We have also computed results in two-flavour QCD, and the respective results \Cref{app:TwoFlavourQCD}. In our current approximation to 2+1 flavour QCD, the Yukawa term in the third line of \labelcref{eq:Gammamatter} reads 
\begin{align}
 \left\{ Z_l \, h_l \,\bar l Z_{\phi_l}^{1/2}  \left[\tau^0_2 \sigma_l + \bm{\tau} \cdot \bm \pi\right] l +  \frac{1}{\sqrt{2}} Z_s\, h_s\,\bar s  \,Z_{\phi_s}^{1/2} \sigma_s \,s\right\} \,.  
		\label{eq:ExplicitYukTerm}
\end{align}
The different normalisations of the light and strange quark term arise from the rotation \labelcref{eq:SigmaSigmas} between the octet basis and the $\sigma_l,\sigma_s$ basis, and the wave functions and Yukawa couplings depend on $\rho$. The constituent quark masses follow from \labelcref{eq:ExplicitYukTerm} in terms of the light and strange quark condensates, 
\begin{align}
	m_l= \frac{1}{2} h_l \,\left( Z_{\phi_l}^{1/2} \sigma_l\right) \, , \quad\quad m_s =\frac{1}{\sqrt{2}} h_s \, \left( Z_{\phi_s}^{1/2}\sigma_s\right) 
	\, .
	\label{eq:constituent_masses}
\end{align}
As mentioned above, in the present work we do not include the full infrared dynamics of the strange sector: the infrared dynamics of QCD is dominated by the pions, and strange composites are too heavy to contribute below the chiral symmetry breaking cutoff scale $k_\chi \approx \qty{400}{\MeV}$, see \labelcref{eq:ChiralSymmbScale} and the discussion in \Cref{sec:ResonantChannels}. 

The expectation values of $\sigma_l,\sigma_s$ are computed from the cutoff-dependent full mesonic effective potential $V(\rho)$ in \labelcref{eq:Gammamatter}, which depends on the two chiral invariants $\rho_l,\rho_s$, see \labelcref{eq:phi0sigma0}. The respective equations of motion are given by 
\begin{align}
	\frac{\partial V(\rho)}{\partial \sigma_l} = Z_{\phi_l}^{1/2} c_{\sigma_l}\,,\qquad 	\frac{\partial V(\rho)}{\partial \sigma_s} = Z_{\phi_s}^{1/2} c_{\sigma_s}\,,
\end{align}
where we have used that the explicit breaking term in the last line of \labelcref{eq:Gammamatter} reads explicitly 
\begin{align}
\frac14 \textrm{tr}\, c_\phi \cdot \left(Z_\phi^{1/2} \phi\right) = c^{\ }_{\sigma_l}  Z_{\phi^{\ }_l}^{1/2}\sigma^{\ }_l + \frac{1}{\sqrt{2}}  c^{\ }_{\sigma_s}Z^{1/2}_{\phi^{\ }_s}  \sigma^{\ }_s\,, 
\end{align}
where the trace $\textrm{tr}$ is the flavour and Dirac trace and the prefactor factor $1/4$ normalises the Dirac part of the trace. The matrix $c_\phi$ in the breaking term is given by 
\begin{align}
	\quad c_\phi=\textrm{diag}(c_{\sigma_l},c_{\sigma_l},c_{\sigma_s})\,. 
\end{align}
The invariants $\rho_l, \rho_s$ can be constructed from the natural singlet-octet basis in three-flavour QCD discussed in \Cref{sec:EmergentCompositesAction}, see also~\cite{Schaefer:2008hk, Mitter:2013fxa, Herbst:2013ufa, Rennecke:2016tkm, Wen:2018nkn, Fu:2018qsk, Fu:2019hdw}. \Cref{eq:SigmaSigmas} and full flavour symmetry in the UV imply a symmetric mass term and identical wave functions $Z_{\phi_l} =Z_{\phi_s}=Z_\phi$. Hence, in the ultraviolet we obtain for the mesonic mass term contained in $V(\rho)$, 
\begin{align}
\frac12 Z_\phi\, m_\phi^2\,  (\sigma_0^2 + \sigma_8^2) = Z_\phi\, m_\phi^2\, \left(\rho_l + \rho_s\right) \,. 
\label{eq:massphirhos}
\end{align}
with the RG-invariant mass function $m_\phi^2$. We emphasise that the full mesonic effective potential $V(\phi)$ depends on the three SU(3) flavour invariants 
\begin{align} 
\left( \rho_l + \rho_s\right)  \,,\qquad \frac{1}{24} \left( \rho_l - 2\rho_s \right)\,, \qquad  \rho_l \sqrt{\rho_s}  \,, 
\label{eq:ChiralINvariants}
\end{align}
represented in \labelcref{eq:ChiralINvariants} as functions of the scalar condensate $\rho_l$ and the strange condensate $\rho_s$, for a thorough discussion see \cite{Jungnickel:1996aa}. However, the by far dominant ultraviolet contributions to the full effective potential $V(\rho)$ stem from the quark loop. With the Yukawa term \labelcref{eq:ExplicitYukTerm}, the quark loop contribution is a sum of contributions from the different quark flavours and the effective potential reduces to a sum of a potential for the light quark condensate and the same potential for the strange quark condensate. Moreover, the quantum corrections to the Yukawa couplings $h_l,h_s$ are only functions of the light condensate $\rho_l$ and the strange condensate $\rho_s$ respectively, which sustains the sum property of the potential. 

The lowest order term in the effective potential is given by \labelcref{eq:massphirhos}. In turn, in the infrared regime with chiral symmetry breaking the strange dynamics is decoupled due to the large strange current quark mass: the strange part of the effective potential stops running and the light meson part is only fed by its own fluctuations. In conclusion, for the chiral dynamics in the scalar-pseudoscalar sector the potential $V( \rho)$ is well approximated by a sum of potentials for $\rho_l$ and $\rho_s$ respectively, 
\begin{align} 
	V(\rho) \approx  V_l( \rho_l) + V_s( \rho_s)\,. 
	\label{eq:Vs-V-UV}
\end{align}
The strange part $V_s$ simply lacks the non-trivial chiral dynamics of the scalar-pseudoscalar sector in the infrared, and only drives the strange condensate from its small value in the ultraviolet to its infrared value. This suffices for the quantitative access to the scalar-pseudoscalar dynamics, but does not allow us to evaluate the dynamics of anomalous axial symmetry breaking. The present approximation simply implements maximal $U_A(1)$ breaking and neglects the topological instanton-induced contributions to the flow, see \cite{Pawlowski:1996ch}. The inclusion of the latter is important for a description of the restoration of chiral symmetry within the thermal phase transition, see \cite{Braun:2020mhk}, and will be studied in the forthcoming work. 

In practice, the above discussion entails that we only need the relation between the physical expectation values (at $k=0$) of the sigma mode $\sigma_{l,0}$ and the strange scalar $\sigma_{s,0}(\sigma_{l,0})$. This relation involves the physical Yukawa coupling at vanishing momentum and the difference of the light and strange constituent quark masses, 
\begin{align} 
\Delta m_{sl} = m_s-m_l\,. 
\label{eq:Deltamsl}
\end{align}
Using the definition \labelcref{eq:constituent_masses}, we obtain the relation
\begin{align}
Z_{\phi_s}^{1/2}	\sigma_{s,{0}} =&\,  \frac{1}{\sqrt{2} }
	Z_\phi^{1/2} \sigma_{l,{0}} +
	\frac{\sqrt{2}}{h_\phi} \Delta m_{sl}\,.
	\label{eq:sigmas-sigma2}
\end{align}
In the infrared, the mass difference $\Delta m_{sl}$ is determined by the ratio of kaon and pion decay constants and can be used to determine $c_{\sigma_s}$. This is tantamount to fixing the strange current quark mass. This will be explained in \Cref{sec:strangeQuark}, see in particular \labelcref{eq:fK-fpi-sigmas}. In the ultraviolet the mass difference is significantly reduced. However, its running is irrelevant due to the large (chiral) cutoff mass and we will simply keep the infrared value for all cutoff scales. We have checked its irrelevance explicitly, see \Cref{sec:strangeQuark}. 

Finally, we note that at finite temperature, the integral in \labelcref{eq:Gammamatter} is given in \labelcref{eq:finiteTx}, and the Matsubara frequencies for the different species of fields are given by 
\begin{align}
	A,c,\bar c,\phi:\  p_0= 2 \pi T \,n\,,\quad q,\bar q:\  p_0= 2 \pi T \,\left( n+\frac12\right)\,.
	\label{eq:AllMatsubaras}
\end{align}
In the present work we concentrate on the vacuum case with $T=0$ and thus the Matsubara sum reduces to a $p_0$-integral from $-\infty$ to $\infty$.

\subsubsection{Momentum dependences in the matter part} 
\label{sec:DerMatter}

We proceed by sketching the derivation of \labelcref{eq:Gammamatter} from fully momentum-dependent expressions, similarly to \Cref{sec:PureGlue} on the pure glue part of the effective action. We do this at the example of the quark-gluon dressing $\lambda^{(1)}_{q\bar q A}$ of the classical tensor structure 
\begin{align}
{\cal T}^a_\mu = {\textrm{i}} \gamma_\mu T^a\,,
\label{eq:Tau1}
\end{align}
which originates in the Dirac term in \labelcref{eq:Gammamatter}. If we consider its full momentum dependence, the quark-gluon term reads 
\begin{align}
	- \sumint\limits_{p_1,p_2} \lambda^{(1)}_{q\bar q A}(p_1,p_2) \left[\bar q(p_1) {\cal T}^a_\mu \,q(p_2)\right] A^a_\mu(-p_1-p_2)\,. 
	\label{eq:QuarkGluonTerm}
\end{align}
Then, the coupling is defined at the symmetric point analogously to the other avatars of the strong couplings \labelcref{eq:StrongCouplingsGlue,eq:StrongCouplingsGhostGlue},

\begin{align}
	g_{q\bar q A}(\bar p) = \frac{\lambda^{(1)}_{q\bar q A}(\bar p)}{Z^{1/2}_q(\bar p)Z_q(\bar p)}\,. 
	\label{eq:StrongCouplingQuarkGluon}
	\end{align}
We now extend \labelcref{eq:Localg} to $i=q\bar q A$, resulting in a momentum-independent, but cutoff-dependent 
quark-gluon coupling. The same procedure is applied to the dressings of the four-quark term and the quark-meson term, 
and we are left with the cutoff-dependent couplings 
\begin{align}
g_{q\bar q A}=(g_{l\bar l A}\,,\,g_{s\bar s A})\,,\qquad \lambda_q(\rho)\,,\qquad h_\phi(\rho)\,. 
	\label{eq:MatterCouplings} 
\end{align}
Within the present scheme, we may also consider the $\rho$-dependence of $g_{q\bar q A}$ as well as that of the other avatars of the strong coupling. The computational setup described in \Cref{sec:SystematicsQCD} allows for this in a straightforward way, however it increases the computational costs and overall complexity. In the present work we refrain from adding this layer of complexity, and even reduce the $\rho$-dependence further. Instead of accommodating the full $\rho$-dependence we evaluate the couplings on the solution of the equation of motion $\rho_0$. Accordingly, we consider the couplings 
\begin{align}
g_i(\rho_0)\,,\quad \lambda_q(\rho_0)\,,\quad h_\phi(\rho_0)\,, 
\label{eq:AllCompulings} 
\end{align}
with $\rho_0 = \phi_0^2/2$ and $i=c\bar cA, A^3, A^4,l\bar l A, s\bar s A$. Moreover, the wave functions will also be evaluated on the background $\rho_0$, whereas the masses are extracted from the full potential $V(\rho)$ at $\rho_0$. Note however, that the wave functions in \labelcref{eq:Gammamatter} still carry the momentum dependence of the attached fields. We elucidate this at the example of the bilinear term of the quarks. A parametrisation of the full bilinear is given by 
\begin{align} 
 \sumint\limits_p Z_q(p) \, \bar q(-p)\left[{\textrm{i}}   \slashed{p}+ M_q(p)\right] q(p)\,,  
 \label{eq:Quarkbilinear} 
\end{align} 
with the quark wave function $Z_q(p)$, the dressing of the Dirac term, and the quark mass function $M_q(p)$, The latter is the renormalisation group invariant part of the full dressing $Z_q(p) M_q(p)$ of the scalar term. Both scalar dressing functions also carry field-dependences, in particular of the mesonic composites $\phi$. In the present approximation we will drop the momentum dependence of the quark mass function within loops and use 
\begin{align} \nonumber 
	 M_l(p) \approx &\, M_l(0) =\frac12 h_l \, \left( Z_\phi^{1/2} \sigma_{l,0}\right) \,, \\[1ex]
	 M_s(p) \approx &\, M_s(0) =\frac{1}{\sqrt{2}}  h_s \,  \left( Z_{\phi_s}^{1/2} \sigma_{s,0}\right)  \,, 
\label{eq:ApproxMq}
\end{align} 
where we suppressed the $\rho$-dependence of the wave functions and the Yukawa couplings. \Cref{eq:ApproxMq} entails that we identify $M_q(p)$ with the cutoff-dependent and field-dependent mass functions \labelcref{eq:constituent_masses}. Furthermore, we evaluate \labelcref{eq:constituent_masses} on the solutions of the cutoff-dependent equations of motion for $\sigma_l$ and $\sigma_s$ with the approximation 
\begin{align}
	h_\phi(\rho) \approx h_\phi(\rho_0)\,. 
	\label{eq:YukApprox}
\end{align}
This approximation has also been used in \cite{Fu:2019hdw} and its accuracy is corroborated with a comparison to the $\rho$-dependent results in \cite{Mitter:2014wpa, Braun:2014ata, Rennecke:2015eba, Cyrol:2017ewj}. The cutoff dependence of $  \sigma_{l,0}\,, \sigma_{s,0}$ captures qualitatively the momentum dependence of the mass function. This is sufficient in order to obtain quantitative results for the couplings and observables considered here, for a more detailed discussion see \Cref{sec:SystematicsQCD}. We provide further support for these considerations by also computing the full momentum dependence of $M_q(p)$ on the solution of the truncated system. We shall see that it compares well with the full result, see \Cref{fig:QuarkMomdep}. 

While the momentum-dependent mass function is approximated with \labelcref{eq:ApproxMq}, the Dirac term still carries the full momentum dependence, 
\begin{align}
	\int_x \,Z_q \,\bar q  \slashed{\partial}\, q =  \sumint\limits_p Z_q(p) \, \bar q(-p) \,{\textrm{i}}   \slashed{p}\, q(p)\,. 
	\label{eq:FullKineticQuark}
\end{align}
Similarly, the scalar part of the four-quark term is given by 
\begin{align}\nonumber 
	 \int_x  \lambda_q(\rho)  Z_q^2\,  \left(\bar q \tau^0 q\right)^2 = &  \hspace{-.3cm}\sumint\limits_{p_1,p_2,p_3} \hspace{-.3cm}\lambda_q(\rho) \\[1ex]
	&\hspace{-2.5cm}   \times \left[\prod_{i=1}^4 Z^{1/2}_q(p_i)\right]\ \bar q(p_1) \tau^0 q(p_2)\,  \bar q(p_3) \tau^0 q(p_4)\,,
\end{align}
with $p_4 = -(p_1+p_2+p_3)$. 

In contrast to the functional QCD computations done so far, we take into account the full mesonic effective potential $V(\rho)$, which comprises all orders of mesonic self-scatterings. This is a significant extension of the approximations used so far, and the full effective action, and hence also the full effective potential, carries chiral symmetry as the breaking is solely encoded in the linear breaking term. 

The resulting generalised flow equation, introduced in \labelcref{eq:GenFlow}, is manifestly chirally invariant due to the shift in the zero mode first considered in \cite{Fu:2019hdw}, and the effective mesonic potential can only depend on the chiral invariant $\rho$ defined in \labelcref{eq:rho-rhos}. 

Lastly, we consider the fully momentum-dependent wave functions of quarks and mesons $Z_q$ and $Z_\phi$, which are evaluated for constant mesonic fields $\rho$, but not fully fed back into the flow, see \Cref{sec:Wavefuncts}. In our approximation we assume a very weak $\rho$-dependence of the mesonic wave function, $\partial_\rho Z_\phi(\rho) \approx 0$. This is supported by results in the first order of the derivative expansion with $Z_\phi(\rho)$ and we use the wave functions on the physical point $\rho_0$, 
\begin{align}
Z_\phi(\rho) \approx Z_\phi(\rho_0) \,.
\end{align}
In this approximation, we have in particular $Z_{\pi} = Z_{\sigma_l}=Z_{\phi_l}$. In the ultraviolet, we also have $Z_{\phi_s} =Z_{\phi_l}$, but they differ in the infrared. The higher derivatives of the wave function would account for momentum-dependent mesonic self-scatterings. We drop them in this calculation, since these momentum-dependent scatterings are heavily suppressed at low energy scales and higher-order mesonic dynamics with and without momentum dependence are even more suppressed at high energies.

\subsection{RG-invariant approximation scheme}
\label{sec:RG-invariant expansion}

The systematic expansion scheme of the present setup relies on a vertex expansion in terms of the $n$-point correlation functions $\Gamma_k(p_1,...,p_n)$, where we also partially use an average momentum approximation ($p^2 \approx k^2$). 
The scheme is completed by the inclusion of multi-scatterings of resonant interaction channels, represented by emergent composites: these multi-scatterings event are included in terms of an effective potential of the composite, i.e. they are rewritten as self-interactions of the composite. We consider the mesonic composite $\phi$ with the effective potential $V(\rho)$ and $\rho$-dependent couplings and wave functions. Such an expansion scheme, and in particular the average momentum approximation, is optimised by parametrising the effective action in terms of renormalisation group (RG) invariant $n$-point functions, 
\begin{align}
	\bar\Gamma^{(n)}_{k,\Phi_{a_1} \cdots \Phi_{a_n}}(p_1,...,p_n)= 
	\frac{\Gamma_{k,\Phi_{a_1} \cdots \Phi_{a_n}}^{(n)}(p_1,...,p_n)}{\prod_{i=1}^n Z^{1/2}_{\Phi_{a_i}}(p_i)}\,.
	\label{eq:RGinvGn}
\end{align}
The vertex dressings of the RG-invariant vertices $\bar\Gamma^{(n)}$ follow directly from the dressings of $\Gamma^{(n)}$ in \labelcref{eq:CompleteBasisGn}. We readily obtain 
\begin{align}
	\hspace{-.1cm}\bar \Gamma_{\Phi^n}^{(n)}(p_1,...,p_n) = \sum_{i=1}^{N_{\Phi^n}}\!  \bar \lambda_{\Phi^n}^{(i)}(p_1,...,p_n) {\cal T}^{(i)}_{\Phi^n}(p_1,...,p_n)\,,
	\label{eq:CompleteBasisGnRGinv}
\end{align}
with 
\begin{align} 
	\bar \lambda_{\Phi^n}^{(i)}(p_1,...,p_n) = \frac{ \lambda_{\Phi^n}^{(i)}(p_1,...,p_n)} {\prod_{i=1}^n Z^{1/2}_{\Phi_{a_i}}(p_i)}\,.
\label{eq:barlambda}
\end{align}
The vertices $\bar\Gamma^{(n)}$ defined in \labelcref{eq:RGinvGn}, are RG-invariant but not $k$-independent: We start with the underlying homogeneous renormalisation group equation for the scale-dependent effective action $\Gamma_k$ with the RG-scale $\mu$
\begin{align} 
\mu\frac{d \Gamma_k}{d \mu}=0\  \longrightarrow\  \left( \mu\frac{d}{d \mu} + \frac12 \sum_i \eta_{\Phi_{a_i}}^{(\mu)}\right)   \Gamma^{(n)}_{k,\Phi_{a_1} \cdots \Phi_{a_n}} =0\,, 
\label{eq:RGmu}
\end{align}
with 
\begin{align} 
\eta_{\Phi_a}^{(\mu)}= - \mu\frac{d \log Z_{\Phi_a}(p)}{d \mu} \,, 
	\label{eq:etamu}
\end{align}
as discussed in \cite{Pawlowski:2005xe}. It follows readily from \labelcref{eq:RGmu,eq:etamu} that the vertices $\bar\Gamma^{(n)}$ and the dressings $\bar\lambda_{\Phi^n}^{(i)}$ are RG-invariant, 
\begin{align} 
 \mu\frac{d}{d \mu}   \bar \Gamma^{(n)}_{k,\Phi_{a_1} \cdots \Phi_{a_n}} =0=  \mu\frac{d}{d \mu}   \bar \lambda^{(i)}_{\Phi_{a_1} \cdots \Phi_{a_n}} \,.
	\label{eq:RGmuInv}
\end{align}
In summary, a reparametrisation of the effective action in terms of RG-invariant vertices and field constitutes a manifestly RG-invariant expansion scheme that captures the scaling-properties of the theory at each approximation order. Moreover, the momentum-dependences in the denominator take care of exceptional momentum dependences and the $\bar\Gamma^{(n)}$ typically have a milder angular and momentum dependence. This has been confirmed both in QCD, \cite{Mitter:2014wpa, Cyrol:2016zqb, Cyrol:2017ewj} as well as in quantum gravity \cite{Denz:2016qks}. Accordingly, expansion schemes with a reduced momentum dependence typically show a better convergence in terms of a parametrisation in $\bar\Gamma^{(n)}$. 

Therefore, we use the reparametrisation of the effective action \labelcref{eq:effAct} in terms of the RG-invariant vertices, using the transformation 
\begin{align}
	\Phi _{a}(p) \to  Z^{1/2}_{\Phi_a}(p) \Phi_{a}(p)\,, 
	\label{eq:barPhi}
\end{align}
where the transformed superfield $\Phi$ has absorbed the wave functions of all component fields. After this transformation, \labelcref{eq:Gammamatter} reads 
\begin{align}\nonumber
	\Gamma_k[\Phi] = & \,\Gamma_{\textrm{glue},k} [A,\bar c,c] + 	\int_x \Bigl\{\bar q  \left( \gamma_\mu D_\mu-\gamma_0 \mu_q \right)q \\[1ex]\nonumber
	& -  \lambda_q(\rho) \left[ \left(\bar q \tau^0 q\right)^2 + \left(\bar q \bm \tau q \right)^2 \right]          \\[1ex]\nonumber
& + \bar q\,\left[ h_\phi(\rho)\,\phi^i \tau^i\right] \,q         \\[1ex]    
	& +\frac{1}{2}  \left(\partial_\mu \phi \right)^2 +  V( \rho) -c_{\sigma_l}  \sigma_l- \frac{1}{\sqrt{2}} c_{\sigma_s} \sigma_s
	\Bigr\}\,,
	\label{eq:BareffAct}
\end{align}
with 
\begin{align} 
	\bar q\,\left[ h_\phi(\rho)\,\phi^i \tau^i\right] \,q   =  h_l \,\bar l  \left[\tau^0_2 \sigma_l + \bm{\tau} \cdot \bm \pi\right] l +  \frac{1}{\sqrt{2}}h_s\,\bar s  \,\sigma_s \,s\,, 
\end{align}
with $h_l=h_l(\rho)$ and $h_s=h_s(\rho)$. We will discuss the consequences of this reparametrisation for the flow equation in the next section. The new component fields $\Phi_a$ are RG-invariant but cutoff-dependent, 
\begin{align}
	\mu\frac{d \Phi_a}{d \mu}=0\,,\qquad \partial_t \Phi_a(p)= -\frac12 \eta^{\ }_{\Phi_a}(p) \,\Phi_a(p)\,,
	\label{eq:RG+FRGscalingbarPhi}
\end{align}
with 
\begin{align}
\quad \eta^{\ }_{\Phi_a}(p) = - \partial_t \log Z_{\Phi_a}(p)\,, 
\label{eq:etaPhi}
\end{align}
not to be confused with the $\eta^{(\mu)}$'s in \labelcref{eq:etamu} that are the anomalous dimensions counting the anomalous $\mu$-scaling. 

For the sake of a concise representation we provide explicit expressions for all propagators and vertices in the approximation \labelcref{eq:Gammamatter} with the RG-variant fields in \Cref{app:Diagrammatics}. These expression have the wave functions already removed in the sense of our RG-invariant scheme, but are easily restored as multiplicative factors, if necessary.

\section{Functional flows in QCD} 
\label{sec:FunFlowsQCD}

In the present Section we discuss the functional renormalisation group approach to QCD based on the general functional flow equation \labelcref{eq:GenFlow}, discussed in \Cref{sec:GenflowsQCD}. As already prepared in the previous Sections, the QCD effective action is expanded in propagators and scattering vertices of gluons, ghosts and quarks, but also includes propagators and scattering vertices of emergent dynamical degrees of freedom for pions and $\sigma$-mode $\phi_l= (\sigma_l,\boldsymbol{\pi})$ and well as the strange condensate $\sigma_s$, see \Cref{sec:EmergentCompositesAction}. These composite fields are introduced with the emergent composite fRG approach, which is also called dynamical hadronisation in QCD. This is detailed in \Cref{sec:EmergentMesons}. 

In the following we derive and discuss the flows of vertices and propagators in all sub-sectors of QCD, the projection procedures for obtaining these flow equations from the functional flow \labelcref{eq:GenFlow} of the effective action are summarised in full detail in \Cref{app:Projections}.

\subsection{Generalised flows in QCD}
\label{sec:GenflowsQCD}

In the fRG approach to QCD, the path integral of gauge-fixed QCD is augmented with an infrared cutoff term that suppresses quantum fluctuations with momentum scales $p^2\lesssim k^2$ for a given infrared cutoff scale $k$. This infrared suppression is implemented with an additional bilinear term $\Delta S_k[\Phi]$ in the classical action in the path integral with 
\begin{align}
\Delta S_k[\Phi]= \frac12 \sumint\limits_p \Phi_i(-p)\,R^{ij}_k(p) \Phi_j(p)\,,  
\label{eq:CutoffTerm}
\end{align}
and the block-diagonal regulator matrix $R_k$ is specified in \Cref{app:regs}. 

The flow equation entails, how the effective action evolves, if the the cutoff scale is lowered successively. The input is the QCD effective action $\Gamma_{k=\Lambda_\textrm{UV}}[\Phi]$ at an initial high energy scale $\Lambda_\textrm{UV} \gg \Lambda_\textrm{QCD}$ in the perturbative, asymptotically free regime with weakly interacting quarks and gluons, where all the propagators and vertex dressings tend towards their trivial UV forms: only the UV-relevant dressings show a perturbative scale- and momentum-dependence. A typical initial cutoff scale $\Lambda_\textrm{UV}$ is in the range $\Lambda_\textrm{UV} \in (10, 10^2)\unit{\GeV}$. For scales $k\lesssim \qty{5}{\GeV}$, QCD gets increasingly non-perturbative and enters the strongly correlated infrared regime with confinement and chiral symmetry breaking. In this regime, the degrees of freedom decouple successively, starting with the gluons at $ k\approx \qty{1}{\GeV}$, the light and strange quarks and the $\sigma$-mode at $ k\approx 0.3-\qty{0.5}{\GeV}$ and finally the pions at $ k\approx \qty{0.1}{\GeV}$. 

This has the important consequence that low-energy effective theories (LEFTs) of QCD can be systematically embedded in full QCD (QCD-assisted LEFTs), as the ultraviolet classical action with the physical cutoff scale $\Lambda_\textrm{LEFT}$ of typically $\Lambda_\textrm{LEFT}\approx 0.5 - \qty{1}{\GeV}$ is directly related to the effective action of QCD at this scale. A prominent example is chiral perturbation theory: it emerges as the final LEFT in the deep infrared and is ruled by the off-shell pion dynamics. 

\begin{figure}[t]
	\centering
	\includegraphics[width=\linewidth]{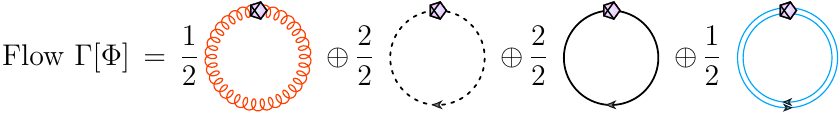}
	\caption{Flow equation \labelcref{eq:GenFlow} of the effective action~\labelcref{eq:effAct} with $\dot \phi=0$. A detailed account of the diagrammatic notation is given in \Cref{fig:diag_notation}, the truncation used in the present work is depicted in \Cref{fig:system_overview}. \hspace*{\fill}}
	\label{fig:FlowGamma}
\end{figure}

In the presence of sources for the emergent composites, the functional flow equation for the effective action takes the form, \cite{Pawlowski:2005xe, Fu:2019hdw, Ihssen:2023nqd},
\begin{subequations} \label{eq:GenFlow}
\begin{align}
	\partial_t \Gamma_k [\Phi]  + \int \dot{\Phi}_a \left(\frac{\delta \Gamma_k[\Phi]}{\delta \Phi_a} + c_{\sigma_i} \delta_{a\mathbf{\sigma}_i} \right) =\textrm{Flow}[\Phi] \,,
	\label{eq:GenFlowEq}
\end{align}
with $i=l,s$ and $(\sigma_i) = (\sigma_l,\sigma_s)$. The diagrammatic part $\textrm{Flow}[\Phi]$ in \labelcref{eq:GenFlowEq} is given by 
\begin{align}
	\textrm{Flow}[\Phi]= \frac{1}{2} G_{ac} [ \Phi]\,\left(\partial_t\delta^c_b + \frac{\delta \dot{\Phi}_b}{\delta  \Phi_c}\right) R^{ab} \,,
	\label{eq:GenFlowDiagrams}
\end{align}
see \Cref{fig:FlowGamma}. In \labelcref{eq:GenFlowEq}, the parameter $t= \ln (k/\Lambda_\textrm{UV})$ is the RG-time and the propagator is given by
\begin{align}
	G_k [\Phi] = \frac{1}{\Gamma_k^{(2)}[\Phi] + R_k } \,, \qquad G_{ab} = (G_k [\Phi])^{\ }_{\Phi_a \Phi_b} \,.
	\label{eq:PropPhi}
\end{align}
\end{subequations}
Note that the $t$-derivative of the effective action is taken at fixed $\Phi$, $\partial_t\Phi=0$, for both the RG-variant and the RG-invariant fields discussed in \Cref{sec:RG-invariant expansion} and \labelcref{eq:GenFlow} holds true for both effective actions. The $t$-scaling of the RG-invariant fields is encoded in the second term on the left hand side of \labelcref{eq:GenFlowEq}: for the RG-invariant field we find 
\begin{align}
	\dot{ \Phi} =  \dot \phi -\frac12 \eta^{\ }_{\Phi}  \Phi \,,
\label{eq:dotPhi}
\end{align}
while the $\eta$-part is missing for the RG-variant fields. Moreover, the $-\eta$ term from $\delta \dot{\Phi}/\delta\Phi$ in the last and $\partial_t  R$ in the penultimate term in \labelcref{eq:GenFlowDiagrams} combine to the common $(\partial_t -\eta) \bar R$ term obtained from a rescaling 
\begin{align}
	R_{\Phi^a} \to \frac{1}{Z} R_{\Phi^a}\,. 
	\label{eq:rescalingR}
\end{align}
This connects the regulators of the RG-variant and RG-invariant fields. In summary, the generalised flow \labelcref{eq:GenFlow} is manifestly invariant under rescalings. 

This argument can also be turned around and we simply choose $\eta^{\ }_\Phi$ such that
\begin{align} 
	\partial_t \bar Z_\Phi(p)\equiv 0\,. 
\label{eq:KinCan}
\end{align}
\Cref{eq:KinCan} implies that $\eta^{\ }_\Phi(p)$ accommodates for the non-trivial momentum and cutoff scale dependence of the kinetic term. Within such a setup, the correlation functions $\Gamma^{(n)}$ of the effective action are straight away the RG-invariant ones in \labelcref{eq:RGinvGn}, as the wave functions are trivial. 

For the numerical solution of the flow equation \labelcref{eq:GenFlow}, the effective action is expanded in $n$-point correlation functions 
of the fundamental fields $A,c,\bar c,q\bar q$, complemented with a derivative expansion in $p^2/\Lambda^2_\textrm{QCD}$ of the dynamics of the emergent mesonic field $\phi=(\sigma,\boldsymbol{\pi})$. In summary, flows for $n$-point correlation functions take the structural form 
\begin{align} \nonumber 
	\left( \partial_t  - \frac12 \sum_{i=1}^n \eta^{\ }_{\Phi_{a_i}}(p_i)\right)   \Gamma^{(n)}_{k,\Phi_{a_1} \cdots \Phi_{a_n}}(p_1,...,p_n) & \\[1ex]
	 - \frac12 \eta^{\ }_{\Phi_{a_0}}(p_0) \Phi_{a_0}(p_0) \Gamma^{(n+1)}_{k,\Phi_{a_0} \cdots,\cdots, \Phi_{a_n}}(p_0, ...,&p_{n}) \notag\\[1ex]
	&\hspace{-3cm}= \textrm{Flow}^{(n)}_{\Phi_{a_1} \cdots \Phi_{a_n}}(p_1,...,p_n) \,,
	\label{eq:FlowsRGinvGn}
\end{align}
where the right hand side constitutes the $n$th $\Phi$-derivative of \labelcref{eq:GenFlowDiagrams}, depicted in \Cref{fig:FlowGamma}. The notation for $\Phi$-derivatives is defined in \labelcref{eq:NotationDerivatives} in \Cref{app:Diagrammatics}. In \labelcref{eq:FlowsRGinvGn} we have dropped the dependence of the $n$-point functions $\Gamma^{(n)}$ on the constant mesonic background $\phi$. We shall use \labelcref{eq:FlowsRGinvGn} in order to project out the flows of the RG-invariant vertex dressings within our approximation of the effective action, \labelcref{eq:BareffAct} with \labelcref{eq:Gammaglue}. We also note that \labelcref{eq:GenFlow} is completely $c_\sigma$-independent, which is ensured by the subtraction in the second term on the left-hand side, discussed for the first time in \cite{Fu:2019hdw}. In \Cref{fig:FlowGamma} we depict the diagrammatic representation of the diagrammatic part of the flow, \labelcref{eq:GenFlowDiagrams}, with the diagrammatic notation explained in \Cref{fig:diag_notation}.

\subsection{Matter sector with emergent composites}
\label{sec:EmergentMesons}

Emergent composites are sourced in \labelcref{eq:GenFlow} by the general RG-transformation or differential reparametrisation $\dot \phi$. This reparametrisation is linked to the expectation value of the flow of the scale-dependent composite operator $\hat \phi_k$ via $ \dot \phi = \langle \partial_t \hat \phi_k \rangle
$, which we have not specified yet. In fact, it is not necessary to specify $\hat \phi_k$ explicitly, since $\dot \phi$ can be defined implicitly in terms of the fundamental (mean) fields. Its local and global existence is subject to several constraints, most notably the locality of $\dot\phi$ in field and momentum space, for a detailed discussion see the forthcoming work \cite{IP2024}. In the present Ansatz we parametrise $\dot \phi$ in terms of the bilinear $\bar q \tau q$
\begin{align}
	\dot \phi =\dot A_\phi\, \bar q \tau q\,, 
\label{eq:Genhadronrel}
\end{align}
with the dynamical hadronisation matrix $\dot A_\phi$. We not in passing that \labelcref{eq:Genhadronrel} is both, momentum-local and local in field space, which guarantees its existence. In the present work we will use the flavour-symmetric approximation 
\begin{align}
\dot A_\phi=\textrm{diag}(\dot A_l,\dot A_s)\,,\qquad \textrm{with}\qquad \dot A_l=\dot A_s=\dot A_k\,. 
	\label{eq:dotAphi}
\end{align}
With \labelcref{eq:dotAphi} the relation \labelcref{eq:Genhadronrel} reduces to 
\begin{align}
	\dot \phi =\dot A_k\, \bar q \tau q\,, 
	\label{eq:hadronrel}
\end{align}
with a global dynamical hadronisation function $\dot A_k$. The hadronisation function $\dot A_k$ in \labelcref{eq:hadronrel} is chosen the same for the light composite $\phi_l$ and the strange one, $\sigma_s$, see \labelcref{eq:dotAphi}: in the ultraviolet this identification comes with full flavour symmetry. In turn, in the infrared this is an approximation which is used together with the flavour-symmetric one in the Yukawa coupling, 
\begin{align}
	h_l=h_s=h_\phi\,.
	\label{eq:hpgi=hl=hs}
\end{align}
Both, \labelcref{eq:hadronrel} and 	\labelcref{eq:hpgi=hl=hs} are supported by the mild variation of $h_\phi$ at all scales, see also \Cref{fig:hPhiNf2Nf2p1} in \Cref{app:TwoFlavourQCD}, where the $k$-dependence of $h_{\phi}$ is shown for both two-flavour and 2+1 flavour QCD. The hadronisation function is used in \Cref{sec:Yukawa} to implement dynamical hadronisation, analogously to \cite{Fu:2019hdw, Rennecke:2015}. There, $\dot A_k$ has been chosen to transfer the t-channel of the scalar-pseudoscalar four-quark vertex to interactions of the effective field $\phi$ and we perform the same procedure here. 

\Cref{eq:hadronrel} entails with this diagonal choice for $\dot A_\phi$, that $\phi$ has the same quantum numbers as the scalar-pseudoscalar mesons, and it also supports the interpretation of $\phi$ as a scalar-pseudoscalar quark-composite. Note however that while the flow \labelcref{eq:hadronrel} is linear in 
$\bar q \tau q$, the full emergent composite is not. Indeed, solving the equations of motion for $\phi$ leads to a complicated composite in terms of $\bar q \tau q$, other quark composites as well as gluons. In any case, the scalar-pseudoscalar mesonic resonances leave their trace in the spectrum of $\phi$, and it is for this reason that in particular $\phi_l=(\sigma_l,\boldsymbol{\pi})$ is commonly called $\sigma$-mode and pions. 

We have already discussed below \labelcref{eq:MatterCouplings}, that all flows are evaluated on the solution of the equations of motion of the mesonic fields, 
\begin{subequations}
\label{eq:eomQCD}
\begin{align}
		\left.\sqrt{2 \rho_i} \,\partial_{ \rho_i} V(\rho)\right \vert_{\rho=\rho_{0,k}}  =   c_{\sigma_i,k} \,, \qquad \textrm{with}\qquad i=l,s\,,
\label{eq:EoMrho}
\end{align}
and we use these equations to compute correlation functions and observables at the physical point. The scale dependence of the RG-invariant coefficients $c_\sigma=(c_{\sigma_l}, c_{\sigma_s})$ follows readily from \labelcref{eq:GenFlow}, 
\begin{align}
	\partial_t c_{\sigma,k} =\frac{ \eta_{\phi_l,k}}{2} c_{\sigma_l,k}\qquad 	\partial_t c_{\sigma_s,k} =\frac{ \eta_{\phi_s,k}}{2} c_{\sigma_s,k}\,,
\label{eq:dtc}
\end{align}
where we included the flow of $c_{\sigma_s}$ only for the sake of completeness, and concentrate on $c_{\sigma_l}$ from now on. \Cref{eq:dtc} simply reflects the rescaling of all quantities with appropriate powers of the wave functions, underlying the RG-invariant scheme. In particular, there are no loop contributions in the flow of $c_{\sigma,k}$. We emphasise that the expansion around $\rho_0$ leads to a $c_\sigma$-dependence to the couplings which is not present in the full effective action. 
\end{subequations}
We also consider flows on the solution EoM $\rho^{(\chi)}_0=\rho^{(\chi)}_0(c_{\sigma_l}=0)$ in the chiral limit of 2+1 flavour QCD with a fixed strange quark mass, $c_{\sigma_s}\neq 0$. The respective extension of the results to either the chiral limit or the physical point serves as a stability check of the present approximation. 

In the following, we briefly discuss some adjustments to this general setup, which are necessary for the quantitative evaluation of pole masses \Cref{sec:Wavefuncts} as well as the dynamical hadronisation procedure \Cref{sec:Yukawa}.

Explicit projections for the flows of all discussed parameters are collected in \Cref{app:Projections}. As the expressions quickly turn very large, the generation of flow equations is automated and uses software discussed in \Cref{sec:automatedGen}. Therefore, we also refrain from showing explicit expressions for the flow equations.

\subsubsection{Emergent mesons}
\label{sec:Wavefuncts}

The current setup includes the full field dependence of the mesonic potential $ V( \rho)$, which accommodates all orders of momentum-independent mesonic self-scatterings. Specific $n$-point vertices are obtained as the respective $n$th order $\phi $-derivatives of the potential. An important example are the mesonic curvature masses, which are given by
\begin{align}\nonumber 
	m_\pi^2  = &\partial_{\rho_l} V_l( \rho_l) \,,  \\[1ex]
	m_\sigma^2 = &\partial_{ \rho_l} V_l(\rho_l) + 2 \rho_l\,\partial_{ \rho_l}^2 V(\rho_l) \,.
\label{eq:rawMass}
\end{align}
In the present RG-invariant scheme the full meson two-point function takes the form 
\begin{align}
\Gamma_{\phi_a \phi_b}^{(2)}= p^2 \delta^{ab} + \partial_{\phi_a}\partial_{\phi_b} V(\rho) \,, 
\label{eq:G2phi} 
\end{align}
and the flow of the meson two-point function follows from \labelcref{eq:GenFlow} with two $\phi$-derivatives. For example, the pion flow reads 
\begin{align}
	\partial_t m_\pi^2 -\eta_{\phi_l}(p) \,\left( p^2 + m_\pi^2\right) =  \frac{\textrm{Flow}^{(2)}_{\pi_i \pi_i}(p)}{N_f^2-1}  \,, 
	\label{eq:FlowG2pi}
\end{align}
where the right hand side constitutes the contribution from the pion derivatives of the loop terms on the right hand side in 
\labelcref{eq:GenFlow} including a sum over all pions $i=1,...,3$. 

\Cref{eq:FlowG2pi} naturally singles out the pion mass parameter $m_\pi^2$ as the physical pole mass of the pion: assuming that $\eta_{\phi_l}$ has no singularities at $p^2=-m_\pi^2$, the evaluation of \labelcref{eq:FlowG2pi} on this momentum leads to 
\begin{align}
	\partial_t m_\pi^2  = \frac{\textrm{Flow}^{(2)}_{\pi_i \pi_i}(p^2=-m_\pi^2)}{N_f^2-1} \,. 
	\label{eq:PoleMass}
\end{align}
We emphasise that \labelcref{eq:PoleMass}, while being natural, is a choice. In the present work we work entirely in the Euclidean domain and refrain from using \labelcref{eq:PoleMass}. Instead we define $m_\pi^2$ as the Euclidean curvature mass at vanishing momentum with 
\begin{align}
	\left( \partial_t -\eta_{\phi_l}\right) m_\pi^2  = \frac{\textrm{Flow}^{(2)}_{\pi_i \pi_i}(0)}{N_f^2-1} \,,\qquad  \eta_{\phi_l} = \eta_{\phi_l}(p=0) \,,
	\label{eq:CurvatureMass}
\end{align}
accompanied with a relation for the anomalous dimension, 
\begin{align}
	\eta_{\phi_l}(p) = \frac{\textrm{Flow}^{(2)}_{\pi_i \pi_i}(p) - \textrm{Flow}^{(2)}_{\pi_i \pi_i}(0)}{(N_f^2-1) p^2 } -m_\pi^2 \frac{\eta_{\phi_l}(p)-  \eta_{\phi_l}}{p^2}  \,,
	\label{eq:AnomalousDim}
\end{align}
or equivalently
\begin{align}
	\eta_{\phi_l}(p) = \frac{\textrm{Flow}^{(2)}_{\pi_i \pi_i}(p) - \textrm{Flow}^{(2)}_{\pi_i \pi_i}(0)}{(N_f^2-1) (p^2 + m_\pi^2) } +\frac{m_\pi^2 \eta_{\phi_l}}{p^2 + m_\pi^2}  \,,
	\label{eq:AnomalousDimEasier}
\end{align}
where we use the latter expression explicitly for $p > 0$. For $p=0$, \labelcref{eq:AnomalousDim} reduces to an equation for $\eta_{\phi_l}=\eta_{\phi_l}(p=0)$ with 
\begin{align}
	\eta_{\phi_l}= \left[ \frac{\partial_{p^2}\textrm{Flow}^{(2)}_{\pi_i \pi_i}(p)}{(N_f^2-1)} -m_\pi^2 \partial_{p^2}\eta_{\phi_l}(p)\right](p=0) \,. 
	\label{eq:AnomalousDim0}
\end{align}
We use only cutoff-dependent anomalous dimensions within the flows but extract the fully momentum-dependent anomalous dimensions and wave functions on the solutions. Since we compute the flow of vertices at $p=0$, it is tempting to only use $\eta_{\phi_l}$ in the flow diagrams. Within the current RG-invariant expansion scheme the wave function is fully encoded in the anomalous dimension and the diagrams only contain terms proportional to 
\begin{align}
\eta_{\phi}(q) R_\phi(q)= \Bigl( \eta_{\phi_l}(q) R_{\phi_l}(q)\,,\, \eta_{\phi_s}(q) R_{\phi_s}(q)\Bigr) \,,
\label{eq:etaR}
\end{align}
with the loop momentum $q^2\lesssim k^2$. In \labelcref{eq:etaR} we have only considered $\eta_{\phi_s} R_{\phi_s}$ for the sake of completeness, as we do not consider $\phi_s$-loops in the current work due to their irrelevance. To optimise the convergence of the expansion scheme, we evaluate the anomalous dimension at the peak of the integrands at $q^2\approx k^2$, hence using 
\begin{align} \label{eq:loopapprox}
	\eta_{\phi}(q) R_\phi(q) \approx \eta_{\phi}(q=k) R_{\phi}(q)\,,
\end{align}
in the loops. We have checked that alternatively one may also use $\eta_{\phi}(p=0)$, only leading to minor changes. 

In a final step we compute the momentum-dependent anomalous dimensions on the solution with \labelcref{eq:AnomalousDimEasier}.
The $t$-integral provides us with the wave function at $k=0$, 
\begin{align}
	Z_\phi(p) = \exp\left\{- \int_{\Lambda_\textrm{UV}}^0 \frac{d k'}{k'} \eta_{\phi,k'}(p)\right\}\,,
	\label{eq:Zphip}
\end{align} 
which includes both the light mesonic composites and the strange scalar composite with $Z_\phi=(Z_{\phi_l},Z_{\phi_s})$. We also have used $Z_{\phi,\Lambda_\textrm{UV}}(p) =1$. This allows us to determine the pole mass of the pion via an analytic continuation of the propagator including the wave function. Moreover, the comparison with the pion curvature mass provides an error estimate for the commonly used approximation, where these are identified. 

Finally, we note that one can also use the cutoff-dependent $\eta_{\phi}$ for estimating the full momentum dependence. This is of benefit for applications where \labelcref{eq:Zphip} is not computed. Then we may assume 
\begin{align} 
	Z_{\phi}(p=\gamma_1 k) \approx Z_{\phi,k}(0) \,,\quad \textrm{or}\quad Z_{\phi}(p=\gamma_2 k) \approx Z_{\phi,k}(k)\,,
\label{eq:Zk-p} 
\end{align}
with slightly different $\gamma_{1,2} \approx 1$. Similar relations can also be used for the other fields, and we will specifically consider them for the quarks in the next Section, \Cref{sec:QuarkZ}. Moreover, we put them to work for the gauge field, see \Cref{sec:GluonGhost} and the respective results for the 2+1 flavour gluon propagator discussed in \Cref{sec:GlueResults}. 

The first relation in \labelcref{eq:Zk-p} is readily checked with 
\begin{align}\label{eq:Zatkandp0}
Z_{\phi,k}(0) = \exp\left\{- \int_{\Lambda_\textrm{UV}}^k \frac{d k'}{k'} \eta_{\phi,k'}\right\}\,, 
\end{align}
comparing the outcome with \labelcref{eq:Zphip}. The second relation requires $Z_{\phi,k}(k)$, which is obtained with 
\begin{align}\label{eq:etaBarPhik}
&Z_{\phi,k}(k)= \notag\\[1ex]
&\hspace{3mm} \exp\left\{- \int_{\Lambda_\textrm{UV}}^k \frac{d k'}{k'} \left( \eta_{\phi,k}(k) - k \,(\partial_p Z_{\phi,k}(p))\Big\vert_{p=k}\right)\right\}\,.
\end{align}
The combination of the two terms in the exponent is $-\partial_t \log Z_{\phi,k}(k)$, the second term taking into account the derivative of the momentum argument $p=k$.

\subsubsection{Yukawa interaction}
\label{sec:Yukawa}

The constituent quark mass is directly linked to the scalar mesonic fields $\sigma=(\sigma_l,\sigma_s)$, which couples to the quarks through the Yukawa couplings, see~\labelcref{eq:constituent_masses,eq:ApproxMq}. Consequently, the flows of the Yukawa couplings are derived from the quark two-point function, for a detailed derivation and the full expressions for the flows see \cite{Fu:2019hdw}. Here, we only quote the essential steps in the derivation within the current approximation. 

The generalised flow \labelcref{eq:GenFlow} is projected onto the scalar part $\bar q q$ at vanishing momentum. This contains the flow of the quark mass $\partial_t m_q(\phi)= (\partial_t \log h_\phi) m_q$. Concentrating on the light quarks and dividing by the factor $\sigma_l$ leads to  
\begin{align}
	&\partial_t h_\phi \, - \left(\frac12 \eta_{\phi_l,k} + \eta_{l,k} \right) h_\phi+ m_\pi^2 \dot{ A}_{l,k} \notag\\
	& \hspace{3cm}= - \frac{1}{4 N_c N_f } \frac{1}{\sigma_l} \mathrm{tr} \left[\textrm{Flow}_{\bar ll}^{(2)}\right]\,, 
\label{eq:yukFlow}
\end{align}
where we have used $h_l=h_s=h_\phi$, see \labelcref{eq:hpgi=hl=hs}. Importantly, n the present flavour-symmetric approximation the flow contains a term proportional to the universal hadronisation function $\dot{A}_{l,k}=\dot A_k$. The flow on the right-hand side of \labelcref{eq:yukFlow} stems from a $q\bar q$-derivative of the right hand side of \labelcref{eq:GenFlow} and comprises loops with both mesonic and gluonic interactions. 

Similarly, we deduce the flow for the four-quark couplings by projecting onto the quark-bilinear $(\bar q \tau q)(\bar q \tau q)$. We use the flavour symmetric approximation discussed in the introduction of \Cref{sec:EmergentMesons}, which also fits to the choice of a uniform dressing function $\lambda_q$ of the four-quark term in \labelcref{eq:Gammamatter}. Its flow is determined from the relevant light quark sector and we arrive at 
\begin{align}
	\partial_t \lambda_q (\rho) - 2 \eta_q \lambda_q (\rho) - \dot{A}_{k} \, h_l = {\textrm{Flow}}_{(\bar l\tau_2 l)(\bar l \tau_2 l)}^{(4)}\,. 
\end{align}
Now we choose $\dot A_k$ such that $\lambda_q \equiv 0$ for all cutoff scales, 
\begin{align}
	\dot{A}_l & =-\frac{1}{{h}_l}\,{
		\textrm{Flow}}^{(4)}_{(\bar q\tau_2 q)(\bar q\tau_2 q)} \,. 
	\label{eq:dotAChoice}
\end{align}
\Cref{eq:dotAChoice} effectively absorbs the flow of the scalar-pseudoscalar four-quark scattering vertex $(\bar q \tau q)(\bar q \tau q)$ as well as higher order scatterings in this channel into momentum-dependent quark-meson scatterings 
\begin{align}
h_\phi G_\phi(p) h_\phi\,,
\end{align} 
and higher order diagrams generated by the effective potential $V(\rho)$ and $\rho$-dependences of the couplings.

\subsubsection{Quark dressings and gauge consistency}
\label{sec:QuarkZ}	

Due to the reparametrisation we applied to our fields, we have absorbed the wave function of the quark into the quark and antiquark fields. Therefore, the flows only depend on the anomalous dimension of the quark. 

In contradistinction to the anomalous dimension of the meson it is uniquely defined by the prefactor of the Dirac term \labelcref{eq:FullKineticQuark} in the effective action. Hence, its flow is obtained by projecting the functional flow \labelcref{eq:GenFlow} for the effective action onto the Dirac part of the quark two-point function, see \labelcref{eq:etas} in \Cref{app:Projections}. This leads us to 
\begin{align} 
	\eta_q(p) =  \frac{1}{4 N_c N_f} \mathrm{tr} \,\left[\frac{{\textrm{i}} \slashed{p}}{p^2}\,\textrm{Flow}_{\bar qq}^{(2)}(p)\right]\,.
\end{align}	
We close this discussion with a comment on the gauge consistency of the quark wave function: Perturbatively, the quark self-energy diagram in QCD at high $k$ only receives a logarithmic contribution from the gluon, that is proportional to the gauge-fixing parameter $\xi$,
\begin{align}
	\includegraphics[valign=c,width=0.15\textwidth]{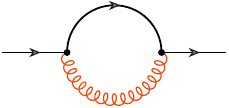} \sim \xi \,\left[p^2\,\log \frac{p^2}{\mu^2}\right] \,.
\end{align}
In the Landau gauge this contribution vanishes and the Dirac term does not require (one-loop) renormalisation. Indeed, the diagram vanishes identically as no momentum-dependence can be generated in the absence of a reference scale such as the RG-scale $\mu$. In the presence of an infrared cutoff scale, the latter argument is not valid anymore and we expect a deviation from this relation for large cutoff scales and $p\neq 0$. At $p=0$ it should still hold as one-loop universality holds true trivially in the fRG approach. Indeed, we see that $\eta_q(p = 0)$ obtains no contributions at perturbative scales $k \gg \Lambda_{\text{QCD}}$. As expected, we also observe that this is not true for $\eta_q(p\neq 0)$. We emphasise that this is not a truncation artefact and holds true in fully momentum-dependent approximations, see also \cite{Mitter:2014wpa, Cyrol:2017ewj}. 

Therefore, in a fully momentum-dependent setup, the initial condition for $Z_{q,\Lambda_\textrm{UV}}(p)$ and other parameters such as the avatars of the couplings have to be chosen such that the resulting $Z_q(p\gg\Lambda_\text{QCD}) = 1$ at $k=0$. Before resolving the above intricacy of gauge consistency for the quark wave function we remark that the quark wave function is absent from the flows in the present RG-invariant scheme, and during the flow it only occurs implicitly in terms of its anomalous dimension in the flow of the couplings such as the quark-gluon avatar of the strong coupling. However, it implicitly also impacts the initial condition thereof. The respective gauge-consistent initial condition and that of the other avatars is discussed later in \Cref{sec:GaugeConsistency}. 

\begin{figure}
	\centering
	\includegraphics[width=0.5\textwidth]{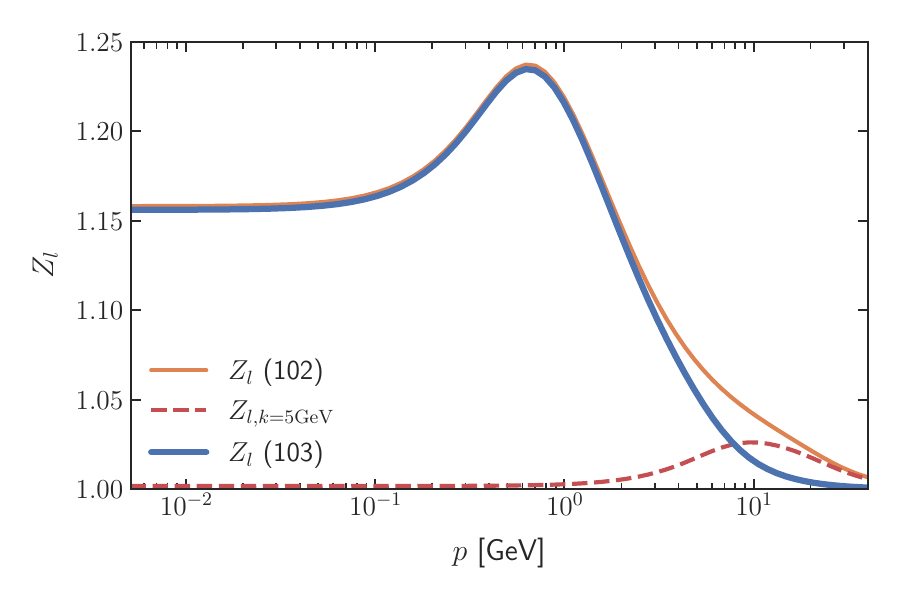}
	\caption{Light quark wave function $Z_l(p)$, obtained from the rough approximation \labelcref{eq:Zqp} to gauge consistency.
		\hspace*{\fill}}
	\label{fig:Zq-Procedure}
\end{figure}

In the quark wave function, we choose an effective procedure to resolve the intricacy of gauge consistence directly. We start this discussion with the integrated flow representation of the wave function, 
\begin{align}
	Z_q(p) = Z_{q,\Lambda_\textrm{UV}}(p) \exp\left\{- \int_{\Lambda_\textrm{UV}}^0 \frac{d k'}{k'} \eta_{q,k'}(p)\right\}\,,
	\label{eq:ZqpFull}
\end{align} 
where $Z_{q,\Lambda_\textrm{UV}}(p)$ is determined by gauge consistency and is subject to fine tuning. 
In the present work we resort to an effective implementation: we start at a large cutoff scale $\Lambda \gtrsim \qty{20}{GeV}$, where we set  $Z_{q,\Lambda_\textrm{UV}}(p)=1$ as a first approximation. 
Then, \labelcref{eq:ZqpFull} is a simple integral over the anomalous dimension $\eta_q(p)$ which is not informed by $Z_{q,\Lambda_\textrm{UV}}(p)$. 
This integral exhibits a logarithmic running at large $k$ and $p$, triggered by the additional cutoff scale $k$. In the large-$k$-regime, which is dominated by one-loop effects, this is a pure cutoff artefact which should be partially compensated by the initial condition, and is partially subtracted by the infrared flow.
We opt to implement this in the simplest way and subtract the flow in the one-loop regime $k\gtrsim \Lambda_\textrm{pert}$ with $\Lambda_\textrm{pert} \approx \qty{5}{\GeV}$, see \Cref{fig:pertAlpha} in \Cref{app:plots}.
This wave function is simply obtained by 
\begin{align}
	Z_q(p) = \exp\left\{- \int_{\Lambda_\textrm{pert}}^0 \frac{d k'}{k'} \eta_{q,k'}(p)\right\}\,.
	\label{eq:Zqp}
\end{align} 
The resulting wave function \labelcref{eq:Zqp} for the light quarks is depicted in \Cref{fig:Zq-Procedure} together with the full integral \labelcref{eq:ZqpFull} and the effective correction $Z_{q,k=\qty{5}{\GeV}}$. Of course, for the strange sector we use the same procedure.

Similar to the mesonic wave functions, it is also instructive to compare the full, momentum-dependent wave function $Z_q(p)$ with that obtained by the identification in \labelcref{eq:Zk-p}, applied to the quark wave function. 
The respective $Z_q$'s are obtained from the integration of the flow and we find 
\begin{align}\label{eq:Zqp0}
	Z_{q,k}(0) = \exp\left\{- \int_{\Lambda_{\textrm{UV}}}^k \frac{d k'}{k'} \eta_{q,k'}\right\}\,, 
\end{align}
and 
\begin{align}\label{eq:Zqpk}
	&Z_{q,k}(k)= \notag \\[1ex]
	&\hspace{3mm}\exp\left\{-\int_{\Lambda_\textrm{UV}}^k \frac{d k'}{k'} \left( \eta_{q,k}(k) - k \,(\partial_p Z_{q,k}(p))\Big\vert_{p=k}\right)\right\}\,.
\end{align}
The constituent quark masses at vanishing momentum, $M_{l,s}(0)$, are the product of the $\sigma$ expectation value and the Yukawa coupling, see \labelcref{eq:ApproxMq}. 
The latter flow is extracted from that of the quark mass function at vanishing momentum, as discussed in \Cref{sec:Yukawa}. 
The same flow equation can be evaluated at general momenta to compute the fully momentum-dependent and $\sigma$-dependent quark-mass functions. We note for this purpose that the mass function is proportional to the cutoff-dependent (and strictly speaking also momentum-dependent) expectation value $\sigma_{l,0}$ of the $\sigma$-mode and $\sigma_{s,0}$ of the scalar strange mode, which is evident from the approximation used in the flows, \labelcref{eq:ApproxMq}. Accordingly, the flow of the mass function $M_q(p)$, evaluated on $\phi_0=(\sigma_{l,0},\sigma_{s,0})$, is only a partial $t$-derivative. The difference to the total derivative may be big as $\sigma_{l,0}$ rises from $\sigma_{l,0}\approx 0$ to $\sigma_{l,0} \approx \qty{60}{\MeV}$. This argument also applies to $\sigma_{s,0}$. Ignoring the additional $\sigma_0$-dependence of $h_\phi(\rho_0)$ we arrive at 
\begin{align}
	\partial_t h_\phi(p) & = - \frac{1}{8 N_c \sigma_l} \mathrm{tr}\,\left[\textrm{Flow}_{\bar ll}^{(2)}\right](p)\,,
	\label{eq:Flowhphip}
\end{align}
where $h_l=h_s=h_\phi$, see \labelcref{eq:hpgi=hl=hs}. The two mass functions are obtained by multiplying the result with $\sigma_{l,0}/2 \,, \,\sigma_{s,0}/\sqrt{2}$ at $k=0$. In the present work we compute the mass functions using \labelcref{eq:Flowhphip}, and use the regulator dependence as a systematic error estimate of the result.  

%

\subsubsection{Inclusion of strange quark}
\label{sec:strangeQuark}

In this Section, we finalise the setup of the strange quark sector, for details on the embedding into our setup, see \Cref{sec:Matter}. The strange current quark mass is encoded in the parameter $c_{\sigma_s}$ in the explicit breaking term in the last line in \labelcref{eq:Gammamatter}. A respective observable, that can be used to fix the current quark mass, is the ratio of pion and kaon decay constants. The latter decay constants are given by
\begin{align}
f_\pi\propto \sigma_{l,0}\,,\qquad f_K\propto \frac12  \sigma_{l,0}
+\frac{1}{\sqrt{2}}\,\sigma_{s,0}\,,
\end{align}
with the same proportionality constant in the present approximation. Accordingly, it drops out from the ratio and we arrive at 
\begin{align}
\frac{f_K}{f_\pi} \approx \frac{1}{2} + \frac{1}{\sqrt{2}} \frac{\sigma_{s,0}}{ \sigma_{l,0}} \approx 1.1917\,,
\label{eq:fK-fpi}
\end{align}
where on the right hand side we took the expectation value computed on the lattice, $f_{K^\pm}/f_{\pi^\pm}= 1.1917(37)$ see \cite{FlavourLatticeAveragingGroupFLAG:2021npn}. In the present approximation, the two condensates are connected by 
\begin{align}
	\sigma_{s,\textrm{0}} =&\, \frac{1}{\sqrt{2} }
	\sigma_{l,{0}} +
	\frac{\sqrt{2}}{h_\phi} \Delta m_{sl}\,, 
	\label{eq:sigmas-sigma}
\end{align}
valid at $k=0$. Here, $\Delta m_{sl}= M_s(0) - M_l(0)$, defined in \labelcref{eq:Deltamsl,eq:ApproxMq}, is the difference of the strange and light constituent quark masses at vanishing momentum. 
Inserting \labelcref{eq:sigmas-sigma} into \labelcref{eq:fK-fpi}, leads us to 
\begin{align}
	\frac{f_K}{f_\pi} \approx 1 + \frac{1}{2} \frac{\Delta m_{sl}}{ m_l} = 1.1917\,. 
	\label{eq:fK-fpi-sigmas}
\end{align}
With a value of
\begin{align} 
	\Delta m_{sl} = \qty{134.19}{\MeV}\,,
\label{eq:DeltamfRG}
\end{align}
we match the above lattice expectation value for $f_K/f_\pi$. This completely fixes the strange quark sector, and hence the matter sector. The difference $\Delta m_{sl}$ can also be extracted directly from the respective values of the constituent quark masses obtained from lattice simulations. Its value, $\Delta m_{sl}^\textrm{lat} \approx \qty{135}{\MeV}$ sustains its derivation from the ratio of $f_\pi$ and $f_K$ \cite{Aoki:2013ldr}, while not being susceptible to the approximations in the relation. 

In the present work we do not follow the evolution of the strange condensate as the off-shell fluctuations from the strange mesonic sector are irrelevant for the QCD dynamics considered here. Effectively, we only need the strange condensate in the infrared: the strange constituent quark mass is directly proportional to the condensate, and leads to a decoupling of the strange quark contributions in the infrared. This is well accommodated by fixing the running value of $\Delta m_{sl}$ to its infrared value \labelcref{eq:DeltamfRG} and use the relation \labelcref{eq:sigmas-sigma} with this infrared value for all cutoff scales. This overestimates the strange current quark mass in the ultraviolet for $k\gg k_\chi$. In this regime the quark mass gap is dominated by the running cutoff scale and we have 
\begin{align} 
	\frac{\Delta m_{sl}}{k} \ll 1\,.
\label{eq:mslkEstimate}	
\end{align}
The implications of such an approximation for UV-observables like the strange current quark mass is discussed in \Cref{sec:strangeUV}. In short, the effects of this approximation of the full running are negligible for the system and observables considered here.

\subsection{Glue sector} \label{sec:GluonicSector}

For the purely gluonic interactions, we resort to the efficient expansion scheme that has been setup and used in \cite{Braun:2014ata,Fu:2019hdw, Braun:2020ada}. One key ingredient is the flexibility of functional approaches, which allows us to use external results as input. This input can either be obtained from other functional computations or from lattice results. The respective systematics is discussed in detail in \Cref{sec:Lego}. 

In short, one can outsource part of the computation without loss of reliability. The external input comes with its own systematic and, in the case of lattice results, statistical error. Consequently, the statistical and systematic errors of such a mixed approach then depend on the quantitative precision and statistical and systematic errors of the external input, as well as the intrinsic systematic error of the given expansion order of the fRG computation. This scheme has been extended, tested and used in Dyson-Schwinger equations in \cite{Gao:2020fbl, Gao:2020qsj, Gao:2021wun,Gao:2021vsf} and has been formalised in \cite{Lu:2023mkn}, for related earlier work see also \cite{Fister:2011uw, Fischer:2011mz, Fischer:2012vc, Fischer:2014ata}.

The most straightforward use of external input is to simply substitute correlation functions in the loops 
by those obtained from other computations. In the past decades the external input often consists of low order lattice correlation functions, mostly in Yang-Mills theory in the vacuum. Respective lattice results with a low statistical error in the continuum limit have been available for roughly two decades, for their use in functional approaches see the review \cite{Dupuis:2020fhh}. Evidently, if the systematic and statistical errors of the input are significantly smaller than that of the approximation used in the computation at hand, one does not have to consider them in the error analysis.

Furthermore, one can also use the given input as an expansion point for different internal parameters, e.g. the number of flavours $N_f$ and the respective quark masses, or the number of colours $N_c$. For example, the present work uses that one can expand input at fixed $N_f = 2$ and add corrections for an inclusion of a strange quark to obtain a setup with $N_f = 2+1$.
Similarly, one can also expand QCD about a given temperature and density of chemical potential, the canonical choice being all temperatures at a vanishing chemical potential. Then, correlation functions for different chemical potentials can be obtained in \textit{difference flows} or \textit{difference DSEs}. Note that difference functional equations can also be used with sliding expansion points.

\subsubsection{Gluon and Ghost propagators}
\label{sec:GluonGhost}

All explicit computations are carried out in the Landau gauge, 
\begin{align} 
	\xi=0\,,
	\label{eq:LandauGauge}
\end{align}
which is also consistent with the external two flavour input \cite{Cyrol:2017ewj}. The scalar part of the the ghost and gluon propagators, $G_A=G_{s,A}, G_c=G_{s,c}$ are defined in terms of the wave functions, 
\begin{align} \nonumber 
	G_{A,k}(p)=&\, \frac{1}{Z_{A,k}(p)} \frac{1}{p^2 + k^2\,r_A(p^2/k^2))} \,,\\[2ex]
	G_{c,k}(p)= &\, \frac{1}{Z_{c,k}(p)} \frac{1}{p^2 + k^2 \,r_x(p^2/k^2)} \,. 
\label{eq:Gh+glZs}
\end{align}
Note that we have absorbed the mass gap of the gluon propagator in the wave function $Z_A(p)$, which carries the complete gluonic dynamics, both the logarithmic scaling in the ultraviolet and the gluon mass gap, signalled by $1/Z_A(p\to 0) =0$. Indeed, after proper normalisation it is the dressing of the gluon propagator, that enters the loop integrals, 
\begin{align}
	\frac{1}{Z_A(p)} = p^2 G_A(p)\,. 
	\label{eq:GluonDressing}
\end{align} 
\Cref{eq:GluonDressing} governs the physics and infrared differences in the propagator are suppressed and do not inform observables. A prominent example if the confinement-deconfinement phase transition, which is caused by the gluon mass gap and the critical temperature is proportional to it, see \cite{Braun:2007bx, Fister:2013bh, Herbst:2015ona}. The parametrisation of the propagator in terms of the dressing without an explicit gluon mass scale also entails, that the avatars of the strong coupling, \labelcref{eq:StrongCouplingsGlue,eq:StrongCouplingQuarkGluon}, carry the confining dynamics with $\alpha_i(p\to 0)=0$ for $i=A^3,A^4, A q \bar q$, see \Cref{fig:alpha} in \Cref{sec:ResultsAvatars}.  

The ghost and gluon wave functions $Z_c$ and $Z_A$ are taken as input at $k=0$ from the functional precision computation in $N_f=2$ flavour vacuum QCD \cite{Cyrol:2017ewj}. These results agree with the lattice data available in the regime $p\lesssim \qty{5}{\GeV}$, e.g.~\cite{Sternbeck:2012qs}. For $p\gtrsim \qty{5}{\GeV}$ there is no lattice data available for comparison, and the fRG data in \cite{Cyrol:2017ewj} are the only non-perturbative data available, which approaches the correct perturbative behaviour for large momenta. The inclusion of gluon data in the current work is done analogously to \cite{Braun:2014ata, Fu:2019hdw}.

The strange quark contributions are computed in terms of \textit{difference flows} between two and 2+1-flavour QCD for the gluon anomalous dimension $\eta_A=-\partial_t \log Z_A$ defined in \labelcref{eq:etaPhi}. Explicitly, this means we add a correction on top of the input gluon anomalous dimension $\eta_{A,\rm{input}}$,
\begin{align}
	\eta_{A,k} = \eta_{A,\rm{input}}(p=k) + \eta_{A, k}^{(s)}(p=0)\,,
\end{align}
where $\eta_{A,k}^{(s)}$ is the contribution from the strange-quark loop.
Thus, the dressing of the $2+1$ flavour gluon propagator computed in this work is one of the predictions within the current fRG approach, and is shown in \Cref{fig:GlueProp}. 

Here, we do not consider the back-coupling of the correction onto the $N_f=2$-part of the flow, which has already been shown to be negligible in \cite{Fu:2019hdw}. The ghost does not obtain a direct contribution from the added strange-quark and the induced corrections on $\eta_c$ through the other couplings are subleading. Additionally, the ghost only couples back into the flow of $g_{A^3}$ and thus the effect of such corrections on the observables constructed from quarks and mesons is negligible.

Similarly, one can use difference flows for the extension of the current results to finite temperature. For the gluon propagators this amounts to implementing 
\begin{align}
	\eta_{A}(T,\mu) = \eta_{A,\rm{input}} + \Big[\eta_{A}(T,\mu) - \eta_{A}(0,0)\Big]\,. 
\end{align}
As argued before, the advantages of the representation of the right hand side is, that one can use quantitative external input or a more advanced approximation for the computation of $\eta_{A,\rm{input}}$, while the difference can computed in a less advanced approximation without the loss of accuracy. For example, the latter statement holds true for the use of less complete momentum-dependences for the difference flows, if the difference merely amounts to a thermal and chemical potential dependent screening mass. In \cite{Fu:2019hdw} This procedure has been shown to be quantitatively reliable.

\subsubsection{Quark-Gluon interactions}
\label{sec:Ie}

It has been discussed in detail in \cite{Fu:2019hdw}, that the current approximation allows for semi-quantitative accuracy. Key to full quantitative precision is the inclusion of all relevant tensor structures of the quark-gluon vertex, for a detailed discussion see the fRG and DSE-analyses in \cite{Mitter:2014wpa, Cyrol:2017ewj, Gao:2021wun}. 

In the latter work it has been shown that out of the eight transverse tensor structures of the quark-gluon vertex, collected in \labelcref{eq:Quark-Gluon} in \Cref{app:quarkGluonBasis}, 
there are only three relevant ones. The others can be dropped without a sizeable change of any correlation function or observable. In order of descending importance, the three most relevant tensor structures are: the chirally symmetric classical one, ${\cal T}^{(1)}$ in \labelcref{eq:Quark-Gluon}, a further chirally symmetric one ${\cal T}^{(7)}$ in \labelcref{eq:Quark-Gluon}, and a chiral symmetry breaking one, ${\cal T}^{(4)}$ in \labelcref{eq:Quark-Gluon}. The third one is proportional to the quark mass function and hence only provides sizeable contributions below $k_\chi$. This already explains its smallest importance. Moreover, the rise of the ${\cal T}^{(1)}$-dressing triggers chiral symmetry breaking and is dominantly important, whereas the contributions from ${\cal T}^{(7)}$ lower the strength of chiral symmetry breaking. Finally, those originating in ${\cal T}^{(4)}$ enhance the strength of chiral symmetry breaking. 

There are only few computations in functional QCD where the full quark-gluon vertex is considered, see the reviews \cite{Eichmann:2016yit, Fischer:2018sdj, Dupuis:2020fhh, Fu:2022gou} for a rather complete list. In most other works only the classical tensor-structure, ${\cal T}^{(1)}$ in \labelcref{eq:Quark-Gluon}, is considered and the lack of the others is compensated by an infrared modification of its dressing or rather of $\alpha_{Aq \bar q}$. 

In the DSE-approach, the direct contributions to chiral symmetry breaking from the non-classical parts of the quark-gluon vertex dominate the system and dropping them leads to a qualitative decrease of chiral symmetry breaking or even the lack thereof. This is due to the quark gap equation featuring only the quark-gluon self energy diagram, and the only non-trivial vertex being the quark-gluon vertex.
Compensating the missing dynamics by an enhancement of the coupling strength $\alpha_{Aq \bar q}$ of the classical tensor structure requires enhancement factors larger than two. However, including the two most relevant tensor structures in the computation already provides quantitative results both in the vacuum and at finite temperature and density, see \cite{Gao:2020qsj,Gao:2021wun}.

In comparison, the flow equation of the quark propagator also contains a tadpole diagram with the full four-quark scattering vertex. A comparison of the two hierarchies entails that within the fRG approach to QCD, part of the quark-gluon dynamics is carried by the tadpole diagram with the four-quark scatterings. The primary contribution of these interactions lies within the scalar-pseudoscalar channel. Accordingly, we expect a far smaller modification factor, which may be positive (enhancement) or negative ('dehancement'). Indeed, in 2+1 flavour QCD, neglecting the direct contribution from non-classical tensor structures and other four-quark interaction channels results in a slight reduction of the strength of chiral symmetry breaking. Hence a small infrared enhancement of about 3\% is needed, and has already been seen in \cite{Fu:2019hdw}. On the other hand, in two-flavour QCD a small 'dehancement' of about 1\% is needed, see \Cref{app:TwoFlavourQCD}. 

We proceed by discussing the technical implementation of this phenomenological infrared enhancement of the classical tensor structure. We follow the approach put forward in \cite{Braun:2014ata, Fu:2019hdw} and substitute
\begin{subequations}
	\label{eq:IRenhancement}
\begin{align}
		\partial_t g_{A \bar qq}(\rho) \to\, & g_{A \bar q q}(\rho) \partial_t \zeta_{a,b}(k)
		 +\zeta_{a,b}(k)\partial_t g_{A \bar q q}(\rho) \,,
\label{eq:IRenhancementFlow}
\end{align}
where the infrared modification is given by
\begin{align}
		\zeta_{a,b}(k) =1 + a \frac{(k/k_e)^2}{\exp[(k/k_e)^2]-1}\,. 
		\label{eq:enhancementfunction}
\end{align}
\end{subequations}
In 	\labelcref{eq:enhancementfunction}, $a>0$ corresponds to an enhancement of $g_{A \bar qq}(\rho)$, while $a<0$ introduces a 'dehancement'. The modification scale $k_e$ is chosen to be $\qty{2}{\GeV}$, in accordance with data from \cite{Cyrol:2017ewj}, and the enhancement factor $a$ is adjusted with the constituent quark mass. $a$ depends on the truncation used and will be indicated with the result. The strength of the enhancement, encoded in $a$, is of order $10^{-2}$ and thus rather small, as indicated before.

We close with the remark, that the qualitative difference in terms of the infrared modification between DSE and fRG approaches within the discussed approximations originates in the very different resummation schemes both offer, while having rather similar approximations to the effective action. Indeed, this fact offers a further systematic error control: if the results within theses two functional approaches agree quantitatively, one obtains a highly non-trivial self-consistency check.

\subsubsection{Avatars of the strong coupling and gauge consistency}
\label{sec:GaugeConsistency} 

Gauge consistency enforces all avatars of the strong coupling to agree at large momenta, as discussed below \labelcref{eq:AgreeAvatars} in \Cref{sec:PureGlue}. For QCD this condition entails 
\begin{align}
g_{i}(p) = g_s(p)+O_i( g_s^3) \,,\qquad i= A^3,A^4,c\bar cA,q\bar q A\,, 
	\label{eq:AgreeAvatarsQCD}
\end{align}
In \cite{Cyrol:2017ewj}, the couplings at the initial scale $k=\Lambda_\textrm{UV}$ have been chosen such that \labelcref{eq:AgreeAvatarsQCD} was achieved at $k=0$. This is the optimal way to ensure gauge consistency, but a numerically costly one as it involves the solution of a multidimensional logarithmic fine-tuning task. 

In the present truncation we approximate the momentum-dependent couplings at finite $k$ with $g_{i,k}(0)$, see \labelcref{eq:Localg}, to wit 
\begin{align}
	g_{i,k} = g_{i,k}(0)\,.
	\label{eq:defofgs}
\end{align}
These couplings satisfy the \textit{modified} STIs (mSTIs) instead of the STIs at $k=0$, see e.g.~\cite{Cyrol:2017ewj,Pawlowski:2022oyq} for respective discussions. However, at one loop, the respective $\beta$-functions agree due to perturbative one-loop universality: in the absence of further scales the flows can be rewritten as integrals of total momentum-derivatives and the results do not depend on the regulators. This can be also extracted from the mSTIs. Beyond one loop the $g_{i,k}=g_{i,k}(0)$ do not agree anymore, but the coefficient of the $\beta$-functions do. It is this property that finally leads to \labelcref{eq:AgreeAvatarsQCD}. In practice this amounts to ensuring 
\begin{align} 
	\frac{\beta_i}{\alpha_i} =\frac{\beta_{\alpha_s}}{{\alpha_s}} \,,\quad \textrm{with}\quad \beta_i = \partial_t \ln g_{i,k} \,, 
	\label{eq:beta_iEqual}
\end{align}
within the regime governed by the perturbative STIs, 
\begin{align} 
	\Lambda_\textrm{UV} \geq k \geq \qty{5}{\GeV}\,,\qquad \textrm{with}\qquad \Lambda_\textrm{UV} \approx \qty{20}{\GeV}\,,
	\label{eq:STIRange}
\end{align}
and minor differences between the $N_f=2$ and $N_f=2+1$ flavour case, see \labelcref{eq:LambdaUV2-2+1} in \Cref{sec:initialSetup}. This uniquely fixes the set of all avatars of the strong coupling, $\{g_i\}$ with $i=A^3, A^4, c\bar cA,q\bar q A$. This amounts to solving a four-dimensional fine-tuning problem, and we use we use the value of the ghost-gluon coupling at the initial scale from \cite{Cyrol:2017ewj} as the coupling $\alpha_s$, all others are compared to. This is a natural choice, as the ghost-gluon coupling is least sensitive to the modifications of the STI. Note that this implicitly sets the absolute scale and the self-consistency of this scale setting is checked with the result for the 2+1 flavour gluon propagator, that is compared with lattice results, see \Cref{sec:ResultsGluonProp}.

\section{Systematic error estimates and apparent convergence in functional QCD}
\label{sec:ApparentConvergence}

In the present Section we discuss several important aspects of systematic error estimates and convergence of the fRG approach to QCD within a vertex expansion scheme. We emphasise that most of the results concerning the systematics are generic and also apply to other functional approaches, for a related DSE analysis see \cite{Lu:2023mkn}. Still, there are some important properties that are fRG-specific and speed up the convergence of a given approximation scheme. This concerns in particular the finite momentum range of the loops which arises from the regulator insertion. Accordingly, we will mostly concentrate on QCD and the fRG approach here. 

In \Cref{sec:Expansion+Scales} we provide a brief overview on the properties of the vertex expansion scheme, augmented with a systematic expansion of the local scattering sector of resonant channels at low energies. In \Cref{sec:SymPointflows} we argue that the system of vertex flows on symmetric points is closed with a good accuracy. This is important for both, the reduction of full momentum dependences without the loss of quantitative precision as well as the use of cutoff-dependent vertices, and the not momentum-dependent ones: the cutoff dependence captures the symmetric point momentum dependence well, contrary to that of scattering channels and exceptional momenta. In \Cref{sec:ResonantChannels} we discuss how resonant interaction channels are accommodated without a loss of quantitative precision, and \Cref{sec:optimisation} we discuss the evaluation of regulator dependences and optimised flows. In \Cref{sec:DivCouplings} we discuss the treatment of large or diverging couplings, and in \Cref{sec:Lego} we evaluate and stress the importance of systematic error evaluations in closed subsystems. All these different properties and checks combine to the principle of \textrm{apparent convergence} that allows access to the convergence estimates of the present systematic expansion scheme.

\subsection{Expansion scheme and infrared scales} 
\label{sec:Expansion+Scales}

The present expansion scheme is based on a systematic vertex expansion in terms of the RG-invariant vertex functions $\bar \Gamma^{(n)}$ and propagators $\bar G$, see \Cref{sec:RG-invariant expansion}. We emphasise that in the parametrisation of the effective action with RG-invariant fields, the RG-invariant vertices and propagators agree with the vertices and propagators $\Gamma^{(n)}, G$ as all the wave functions are trivial, $\bar Z_\Phi(p)\equiv 1$. However, in the present Section on the systematics and convergence of the expansion schemes, we prefer to keep this distinction for the sake of generality. 

The RG-invariant propagators are reminiscent of the classical propagators. For example, in vacuum, the full quark wave function $Z_q(p)$ in \labelcref{eq:Quarkbilinear} is absorbed in the RG-invariant fields $Z_q^{1/2} q,Z_q^{1/2} \bar q \to \bar q, q$, and the respective RG-invariant quark propagator reads 
\begin{align}
	\bar G_{q,k}(p) = \frac{1}{{\textrm{i}} \slashed p + M_{q,k}(p)+R_q(p)}\,,
	\label{eq:barGq}
\end{align}
for all cutoff scales. Notably, the propagator $\bar G_q$ still encodes the physical decoupling of the quark dynamics below the constituent quark mass, while the momentum-dependent rescaling freedom captured by the full renormalisation group has been used to absorb the wave function. The other ingredients are the RG-invariant vertex dressings \labelcref{eq:barlambda}, 
\begin{align} 
	\bar \lambda_{\Phi_{a_1}\cdots \Phi_{a_n}}^{(i)}(p_1,...,p_n)\,,
	\label{eq:barLambdaSystematics}
\end{align}
of the tensor structures within an expansion of the renormalisation-group invariant $n$-point vertex in \labelcref{eq:CompleteBasisGnRGinv,eq:RGinvGn}.

Hence, the discussion of convergence properties of the underlying systematic vertex expansion can be done by discussing the properties of the flow equations for the RG-invariant vertex dressings. These flows consist out of one-loop diagrams with propagators $\bar G$ with simple dispersions. The diagrams are proportional to products or rather contractions of the vertices, and the properties of vertex flows are related to the properties of the vertices. Moreover, the loop momentum $q$ of the diagrams is restricted to radial momenta $q^2\lesssim k^2$, while the angular loop integration averages over all angular configurations of the vertices. 

This expansion is augmented with a systematic low-energy expansion of the full scattering potential in the scalar-pseudoscalar sector of QCD: For cutoff and momentum scales $p,k\lesssim 500$\,MeV the pion dynamics is dominating the dynamics of QCD. In this regime the multi-scattering events of the pion resonance field $\boldsymbol{\pi}$ are local and can be expanded in terms of momenta over the cutoff scale
$\frac{p^2}{k^2}$. Below the pion mass scale, $k\lesssim 140$\,MeV, the pionic scattering events start dying out. In the present work we consider vertices of quarks and gluons that include all scatterings with the emergent scalar and pseudoscalar composites $(\sigma_l, \boldsymbol{\pi})$ with $k$-dependent wave functions (LPA'). It is well-known that the higher orders of such a derivative expansion are irrelevant in most regimes and are only of subleading importance in scaling regimes of the theory, see e.g.~the review \cite{Dupuis:2020fhh} and references therein. Even there they are responsible for only $\lesssim 5\%$ of the critical exponents. In particular their omission has little impact on the size of the critical regime.

\subsection{Closing the system of symmetric point flows} 
\label{sec:SymPointflows}
 
We proceed with an evaluation of the convergence of the vertex expansion. In the absence of a small parameter we only aim for apparent convergence similar to investigations of continuum scaling in lattice formulations. 
To begin with, let us assume that the vertices have a mild angular dependence: we concentrate on the flow of dressing functions $\bar\lambda_{\Phi^n}^{(i)}(p_1,...,p_n)$ for symmetric point configurations with 
\begin{align} \label{eq:SymPoint}
	\bar \lambda_{\Phi^n}^{(m)}(\bar p)\,,\quad \textrm{with}\quad \frac{|p_i p_j|}{\bar p^2} = \frac{|p_r p_s |}{\bar p^2} \,,
\end{align}
for all $i\neq j,r\neq s$ with $i,j,r,s\in(1,...,n)$. 
Then, the vertices in the flows of the $\bar \lambda(\bar p)$ are themselves well approximated by a symmetric-point approximation with 
\begin{align} 
\bar \lambda_{\Phi^n}^{(m)}(p_1,...,p_n) \approx \bar \lambda_{\Phi^n}^{(i)}(\bar p)\,. 
\end{align}
This originates in the fact that symmetric points are particularly smooth configurations and the angular average in the loop projects on average configurations which reduces the vertices to those close to the symmetric point. For a detailed point of view on the special role of the symmetry point configuration see e.g.,~\cite{Eichmann:2014xya, Eichmann:2015nra}. Moreover, the fact that loop momenta are restricted to $q^2\lesssim k^2$, so the infrared suppressed momentum regime, adds a further smoothening. In turn, for large symmetric point momenta $\bar p^2\gg k^2$, the flow only gives subleading contributions anyway. 

In summary, the flows of symmetric point vertices build a closed system up to subleading contributions, if the angular dependences are small. Accordingly, we have to separately discuss strong angular dependences as they occur for resonant interaction channels.

\subsection{Momentum-local flows via emergent composites}
\label{sec:ResonantChannels}

An emergent strong angular dependence and in particular a resonance is already signalled in the symmetric point approximation. As an illuminating example we consider QCD in the chiral limit: there the pion channel leads to a singularity in the pseudoscalar channel in the broken phase of the theory. Roughly speaking, the infrared cutoff scale serves as an external tuning parameter such as the temperature and QCD enters the phase with spontaneous chiral symmetry breaking for $k_\chi \approx \qty{0.5}{\GeV}$. Below this scale the pseudoscalar channel is massless. This poses several problems: to begin with, it invalidates the simple idea of successively integrating out momentum shell degrees of freedom with $q^2\approx k^2$: for $k\leq k_\chi$ the system has a massless mode and the dynamics always takes place on all scales $q^2 \lesssim k^2$. Evidently, this requires approximations, that have to take into account this regime and momentum expansions fail to be quantitative. 

This asks for an approximation scheme in which also resonant interaction channels of vertices are infrared regularised with the cutoff scale $k$. This requirement can be summarised concisely with 
\begin{align} 
	\textrm{Im} \,p_\textrm{sing,n} \geq k_\textrm{gap}\,,
	\label{eq:PhysGap}
\end{align}
for all $n$-point correlation functions. Here, $p_\textrm{sing,n}$ is the location of the closest singularity to the Euclidean axis and $k_\textrm{gap}$ is the spectral infrared gap in the theory. Supposedly, this gap is introduced by the cutoff \labelcref{eq:CutoffTerm}. Typically it is directly related to the cutoff scale $k$, which is discussed in detail in \cite{Pawlowski:2005xe, Pawlowski:2015mlf, Ihssen:2023xlp}. 

In practice we have to guarantee two properties: 
\begin{itemize} 
\item[(1)] Resonant interaction channels are gapped. 
\item[(2)] Momentum locality: all gaps are roughly at the same scale.
\end{itemize} 
Obviously, the first property is instrumental for guaranteeing the second one. (1) is readily achieved in the fRG approach with emergent composites as discussed in \Cref{sec:EmergentMesons}: in the presence of the emergent composites we can introduce standard cutoff terms for these fields, see \labelcref{eq:CutoffTerm}. These cutoff terms suppress the infrared fluctuations in the resonant channels, effectively arranging for \labelcref{eq:PhysGap}. In conclusion, the fRG approach with emergent composites reduces the need for sophisticated orders of the approximation in order to obtain a small systematic error. 

The second property of \textit{momentum locality}, \cite{Christiansen:2015rva}, is a chiefly important one for the systematics of expansion schemes of theories with different degrees of freedom as it underlies the Wilsonian idea of integrating out momentum shells: let us illustrate its relevance with an extreme example: assume for the moment that we take cutoff scales for quarks, $k_q$, and gluons and ghosts, $k_A$, that differ by more than order of magnitude, $k_q\gg 10 k_A$.
This entails that we first integrate out the gluons, and then the quarks. Accordingly, the full glue effective action has to be encoded in the quark flow at each step. The flow of the latter is given by the quark diagram. Evidently, this requires a far more elaborate Ansatz for the effective action for the same quantitative precision.

\subsection{Optimisation and regulator-dependence}
\label{sec:optimisation}

In conclusion \textit{any} optimisation of fRG flows has to strive for momentum locality or accommodate the physics effects. Indeed, functional optimisation as put forward in \cite{Pawlowski:2005xe}, minimises, amongst other optimisations, also the momentum flow. This is tantamount to momentum locality. For further discussions and a recent application within low-energy effective theories in the phase structure see \cite{Pawlowski:2015mlf} and \cite{Ihssen:2023xlp}, respectively. This also entails, that extreme choices for the regulators $R_A,R_c,R_q,R_\phi$ or rather for the respective physical cutoffs $k_{\Phi_i}$ may induce momentum flows between very different momentum scales as well as the transfer of dynamics between these scales. For example, part of the system may already be in the confining phase with hadrons and glueballs (in the example above the glue part), while another part of the system (in the example above the matter system) is still lingering in the perturbative high energy phase. In conclusion this leads us to the property 
\begin{align} 
	k_{\textrm{phys},\Phi_i} \approx k_\textrm{phys}=k\,, 
	\label{eq:Identicalks}
\end{align}
where $k_{\textrm{phys},\Phi_i}$ is the physical cutoff scale in the subsystem defined by the field $\Phi_i$. In presence of emergent composites for all potentially resonant channels, this scale is nothing but $\textrm{Im}\, p_{\textrm{sing},\Phi_i}$, the distance of the first singularity to the Euclidean frequency axis for the propagator of the field $\Phi_i$, assuming a diagonal (or symplectic in the case of fermions) propagator matrix. In the presence of non-trivial backgrounds the propagator matrix may have non-diagonal entries and the above statements apply to the eigenvalues. Note also that within fully optimised flows the regulators also encode the decoupling which is typically only encoded in the propagator: for cutoff scales 
\begin{align}
k< p^{\textrm{phys}}_{\textrm{sing},\Phi_i}\,,
\end{align}
the regulator of the respective field has to vanish. Here, $p^{\textrm{phys}}_{\textrm{sing},\Phi_i}$ is the distance in the $\Phi_i$ channel in the full physical theory at $k=0$. 

Practically, the above optimised setup requires the solution of flows for complex frequencies including the realtime axis. Accordingly, in the present applications (and most others) we use a proxy for $\textrm{Im}\, p_{\textrm{sing},\Phi_i}$ and simply take the Euclidean curvature masses, see \cite{Helmboldt:2014iya}. They can be extracted from the scalar parts of the renormalisation group invariant propagators $\bar G$, 
\begin{align}
	\frac{1}{k^{d_{\Phi_i}}_{\Phi_i}}= \min\limits_{p,\Phi_\textrm{EoM}} \bar G_{s,\Phi_i}[p,\Phi_\textrm{EoM}]\,, 
\label{eq:Physcutoffscales}
\end{align}
for the definition of the $G_s$ see \labelcref{eq:DefofProps} in \Cref{app:PropagatorsVertices}. In \labelcref{eq:Physcutoffscales} we have $d_{\Phi_i} =2$ for $\Phi_i=A,c,\bar c,\phi$ and $d_{\Phi_i} =1$ for $\Phi_i=q,\bar q$. In the present work we arrange for the different cutoff scales to agree approximately: instead of fine-tuning \labelcref{eq:Identicalks} to hold as an identity, we identify the cutoff scales in the regulators. For similar shape functions this leads to approximately agreeing curvature masses \labelcref{eq:Physcutoffscales} and hence to \labelcref{eq:Identicalks}. In the infrared regime, where the fields decouple successively, the full curvature masses deviate more and more, but it is precisely the decoupling which makes this effect subleading. Finally, we vary the shape functions in order to check the regulator independence of the results. 

The present approximation scheme, based on a systematic vertex expansion, also hosts a derivative expansion for the meson self-couplings and the quark-meson coupling of the emergent composite scalar-pseudoscalar field. This additional expansion as well as the respective systematic error is well-tested in many low-energy effective theories. It has been analysed with the functional renormalisation group, further functional diagrammatic techniques as well as with chiral perturbation theory. In order to appreciate the importance of the latter systematics discussed in abundance in the literature, we remark that the present first principles QCD flows naturally approach chiral perturbation theory ($\chi$PT) at low cutoff scales: for cutoff scales below $k\lesssim 500$\,MeV the pion is the only fully dynamical degree of freedom left, which is the pillar of $\chi$PT. The convergence of the derivative expansion, including its optimisation, has been well studied in the literature, \cite{ Morris:1993qb, Morris:1997xj, Canet:2002gs, Pawlowski:2005xe, Balog:2019rrg, DePolsi:2020pjk, DePolsi:2021cmi, DePolsi:2022wyb, Delamotte:2024xhn, Litim:2001dt, Bervillier:2007rc, Litim:2010tt}. Here we resort to functional optimisation, whose practical version can be summarised in 
\begin{align}
R_\textrm{opt}:\	\min_{R_\bot } \| G_k(p) - G_{k=0}(p)\| \,,
\label{eq:funOp}
\end{align}
where $\{R_\bot \}$ is the set of regulators that feature the same physical gap: $k_{\Phi_i} = k_\textrm{phys}$ for all fields. In the present quantitative approach to QCD, the operator norm in \labelcref{eq:funOp} has to incorporate the need for smooth momentum derivatives, both for accommodating the full momentum dependence but also the quest for convergence of the derivative expansion in the hadronic sector of QCD. Technically this asks for the use of Sobolev norms in 
\labelcref{eq:funOp}, practically this entails that optimal regulators are smoothened versions of the flat or Litim regulator \cite{Litim:2001up}. This has been already derived in \cite{Pawlowski:2005xe} and has been confirmed by now in many applications. Further interesting developments concern the use of symmetry principles such as conformality in critical regimes \cite{Delamotte:2024xhn} as well as numerically adapted regulators \cite{Zorbach:2024zjx}. 

This is one of the reasons for introducing a framework that accommodates general smooth analytic regulators. Another physically important reason is the fact that non-analytic regulators such as the sharp cutoff and the flat cutoff distort the analytic structure of correlation functions in the complex plane for $k\neq 0$. While the complex structure is formally restored in the limit $k\to 0$, this may only hold approximately in approximations, which strongly suggest the use of smooth regulators. 

In summary we arrive at the following systematics: the optimisation of regulators maximises the physics content of a given approximation and leads to the most rapid convergence of the underlying systematic expansion scheme. The details of the optimised regulators depend on the expansion scheme and the variation of the regulator within the constraints of the validity range of the expansion scheme provide an important self-consistency check for the systematics as well as offering an important ingredient of the systematic error estimate.

\subsection{Large couplings \& self-consistency}
\label{sec:DivCouplings} 

Resonant channels, as discussed in \Cref{sec:ResonantChannels}, provide a main source for a rapid growth of couplings or rather the RG-invariant vertex functions and are accommodated in the fRG approach with emergent composites. Indeed, this approach can be used to also incorporate uniformly growing vertex dressings, supplied with a respective infrared regularisation. This leaves us with the discussion of large but finite dressings. A specific intricacy of this discussion is the precise definition of \textit{large}. For example, a (dimensionless) dressing function may have a value of the order 1 - 10, to be specific, but dropping the respective diagrams in the flow of all correlation functions has only a marginal impact. Indeed, this is well-known and studied in vacuum QCD for all but one four-quark dressing (the scalar-pseudoscalar one) as well as five out of eight tensor structures of the quark-gluon vertex, see in particular \cite{Mitter:2014wpa, Cyrol:2017ewj}. This information has been used here and also in \cite{Fu:2019hdw}.  

This is already a well-working example for the general systematics in the present vertex expansion scheme and beyond: In the vertex expansion scheme we assume the subleading nature of higher order vertices on the right hand side of the flow equations. Let us assume that we only take into account 
$n$-point correlation functions with $n\leq N_\textrm{max}$ on the right hand diagrammatic side of the flows. Then the underlying assumption is schematically written as 
\begin{align} 
	\textrm{Flow}_{\bar \lambda_n}[\{ \bar \lambda_{m\leq n+2}\} ] \approx 	\textrm{Flow}_{\bar \lambda_n}[\{ \bar \lambda_{m\leq N_\textrm{max}}\} ]\,,\quad \forall n\,.
\label{eq:VertexExpansionSys}
\end{align}
Here, $n$ is a general $n$-point function including also $n>N_\textrm{max}$. In particular, this also entails that the flow of the dressings $\bar\lambda_{n>N_\textrm{max}}$ is only generated from diagrams with $\bar\lambda_{n\leq N_\textrm{max}}$. 

Importantly, the closed diagrammatic one-loop form of the flow equation allows for a complete self-consistency check of the expansion scheme: we can use the integrated flow of the dressings $\bar\lambda_{n>N_\textrm{max}}$ and feed it back into the flows of the dressings $\bar\lambda_{n\leq N_\textrm{max}}$. If this only leads to subleading changes of the results for these couplings, dropping $\bar\lambda_{n>N_\textrm{max}}$ in the flows is \textit{self-consistent} and the small difference between the results with and without this feedback is another part of the systematic error estimate. We emphasise again that such a \textit{complete} self-consistency check is only possible due to the closed form of the functional equations for correlation functions with a \textit{finite} number of loops in full propagators and vertices: in other resummation schemes one typically has to discard whole diagram classes with infinitely many diagrams. Importantly, this closed form is present for fRG and DSE approaches, and systematic error estimates in their combined use as well as the reduction of the respective systematic error estimates within a comparison of the results are also related to this property. 

Finally, we also discuss the structural nature of such checks at the example of the flow of the mesonic four-point functions, the scattering vertex of four mesonic composites. If we assume their subleading nature, that is $N^\phi_\textrm{max}=3$, we would drop the fish diagram in its flow equation, the only remaining flows being the contribution from the quark loop and the contribution from the mesonic loop with four mesonic three point functions. On the equations of motion, the latter diagram is only non-vanishing in the broken phase for $\phi_\textrm{EoM}\neq 0$. Such an approximation offers semi-quantitative results in the presence of large explicit quark masses but fails in the chiral limit and at phase transitions. The underlying reason is the divergence of all the diagrams in the broken phase due to the diverging correlation length. However, the fish diagram encodes the resummation of diagrams that is also present in the respective BSE or 2PI approaches (in next-to-leading order). Then, the above self-consistency check enforces the enlargement of the system with $N^\phi_\textrm{max}\geq 4$ and 
the same analysis with $N^\phi_\textrm{max}=4$ signals the onset of convergence. In the massless limit or at the chiral phase transition full convergence requires the inclusion of the full mesonic potential, which is one of the achievements of the present work.

\subsection{Systematic error checks in subsystems: the \LEGO-principle}
\label{sec:Lego} 

\begin{figure*}
	\centering%
	\begin{subfigure}[t]{.48\linewidth}
		\includegraphics[width=0.65\linewidth]{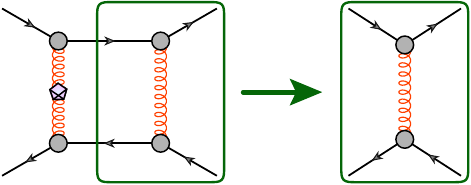}
		\caption{One-gluon exchange coupling as a building block of a diagram for the four-quark vertex.\hspace*{\fill}}
		\label{fig:LegoBlockglue}
	\end{subfigure}%
	\hspace{0.03\linewidth}%
	\begin{subfigure}[t]{.48\linewidth}
		\includegraphics[width=0.65\linewidth]{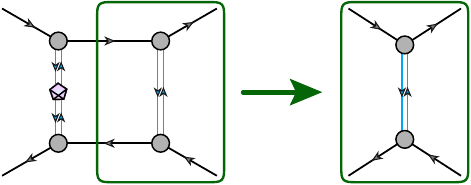}
		\caption{One-meson exchange coupling as a building block of a diagram for the four-quark vertex.\hspace*{\fill}}
		\label{fig:LegoBlockmesonic}
	\end{subfigure}
	\caption{Invariant building blocks from the \LEGO-principle, exemplified at the level of the four-quark interaction. \hspace*{\fill}}
\end{figure*}

In the self-consistency discussion in the previous Section, we have already used that the systematic error estimates in the fRG approach to QCD, and more generally in the fRG approach to QFTs, are also informed by the respective analysis in subsystems.
The underlying systematics is also tightly related to the closed form of the diagrammatic equations. This closed form, and in particular the one-loop form, makes the whole approach modular. This entails, that large subclasses of diagrams only involve propagators and vertices in a given sub-sector. A key example used later is the flow of the four-quark vertex depicted in \Cref{fig:diag_quark_4pt} in \Cref{app:flows}. Its diagrams can be separated in three classes, the quark-gluon diagrams (ultraviolet), the quark-meson-gluon diagrams (interface) and the quark-meson diagrams (infrared). This system will be studied in detail in \Cref{sec:Interface} and shows a well-separated modular form in terms of momentum scale as well as featuring a small class of interface diagrams. A further important example is the flow of the gluon propagator, where the sub-sectors are the diagrams in the pure glue sub-sector and the diagram in the matter sector. Here, the interface are the quark-gluon couplings. We call this modular structure the \LEGO-principle.

In terms of the scale separation of modules and the interface couplings we can isolate sub-sectors which are well-defined (effective) field theories themselves, whose interfaces can be seen as external forces in the flow, leading to a driven system. Then, these sub-sectors can be studied separately in dependence of the 'driving force', and in combination we get a comprehensive systematic error analysis. For example, we may reduce the flows for the matter part of QCD by setting the dressings of all vertices with gluon and ghost legs to zero. In the present fRG approach to QCD with emergent composites this leaves us with a system of flow equations for quark-meson correlation functions. At finite temperature and density this subsystem also includes a gluonic background. The convergence properties of this system can be studied comprehensively as follows: For cutoff scales $k\gtrsim \qty{1}{\GeV}$ the dynamics of the matter system is completely dominated by the quark-gluon diagrams and the systematic error is dominated by the systematic error in this part. For the diagrammatic part of the flow of the four-quark vertex depicted in \Cref{fig:diag_quark_4pt} in \Cref{app:flows}, this is the quark-gluon box, see \Cref{fig:LegoBlockglue}. In this figure we have isolated an important diagrammatic building block, the gluon exchange (green frame). It can be parametrised as the avatar of the quark-gluon coupling $\alpha_{q\bar q A}$, multiplied by the classical gluon propagator with the regulator gap. Note that only the combination is well-defined while we can shift momentum dependences from the quark-gluon coupling to the propagator and back. For $k\lesssim \qty{1}{\GeV}$ the gluon is decoupling successively, and the gluon exchange diagrams get rapidly subdominant. It is superseded by the mixed box diagram with a gluon exchange and a meson exchange with a new building block, the meson exchange. Finally, in the deep infrared the purely mesonic (pion) diagrams dominate depending on two meson exchanges, see \Cref{fig:LegoBlockmesonic}. In this figure we have isolated another diagrammatic important building block, the mesonic exchange (green frame). Similarly to the parametrisation of the gluon exchange in terms of the quark-gluon coupling and the classical gluon propagator, it can be parametrised as the Yukawa coupling squared, $h_\phi^2$, multiplied by the classical mesonic propagators with a sum of the explicit mass function and the regulator gap. 

We illustrate this consecutive decoupling of diagrams in \Cref{sec:Legoatwork} at the example of the quark-meson sector, see in particular \Cref{fig:Slambdaq}. Accordingly, we can perform a complete systematic error analysis of the matter sector within the quark-meson system in the absence of gluon diagrams: we study the convergence of this subsystem, which simply defines a (non-local) low-energy effective theory of QCD with the ultraviolet cutoff $\Lambda_\textrm{UV}\approx \qty{1}{\GeV}$. This analysis is comprehensive if we allow for a rather general initial condition at the initial cutoff scale $k=\Lambda_\textrm{UV}$. The respective systematic error estimate can be directly used in the full system as part of the full systematic error.

In the present QCD system we can use the plethora of results and systematic error analyses of both, the pure Yang-Mills sector, \cite{Cyrol:2016tym, Huber:2016tvc, Huber:2017txg, Corell:2018yil, Huber:2020keu, Pawlowski:2022oyq}, as well as the low-energy matter sector of QCD, see e.g.~\cite{Dupuis:2020fhh} for a review, but also \cite{Schaefer:2004en, Kamikado:2012cp, Tripolt:2013zfa, Mitter:2013fxa, Pawlowski:2014zaa, Jiang:2015xqz, Zhang:2017icm, Tripolt:2017zgc, Resch:2017vjs, Yin:2019ebz, Otto:2019zjy, CamaraPereira:2020xla, Otto:2020hoz, Otto:2022jzl}. Moreover, we can also utilise structural results within other theories, including even non-relativistic condensed matter and statistical systems. In combination this limits the systematic error very efficiently.

\subsection{Apparent convergence}
\label{sec:ApparentConvergencesub}

In preceding Sections, we discussed various aspects of systematic errors inherent in functional approaches, focusing in particular on the \LEGO-expansion. In \Cref{sec:SymPointflows} we have argued that the system of symmetric point flow equations is closed with a very good degree of quantitative precision if the angular dependence of general dressings is sufficiently small. Moreover, its systematic error can be checked in a self-consistent way at each flow step. In \Cref{sec:ResonantChannels} we have shown, that the fRG approach with emergent composites can be used to accommodate resonant interaction channels or more generally, to qualitatively reduce the angular dependence of vertex dressings. A novel ingredient is the computation of full field dependences of emergent composites within a derivative expansion for mesonic multi-scatterings. This does not only gives us access to these scattering but also allows us to evaluate the convergence of Taylor expansions used so far. In \Cref{sec:optimisation} we have discussed, how sufficiently optimised regulators lead to a most rapid convergence of any systematic expansion scheme and how this can be used in QCD for optimal convergence and a further self-consistency check for the systematic error estimate. In \Cref{sec:DivCouplings} we have discussed the important self-consistency check of the vertex expansion, fundamentally rooted in the closed one-loop form of the flow. Finally, in \Cref{sec:Lego} we have argued, that we can directly utilise the wealth of results obtained in closed subsystems of the current full QCD system, including the respective systematic error estimates. An additional important systematic error estimate comes from the combined use of fRG and DSE results in first principles QCD. Moreover, first principles results from lattice simulations serve as important benchmark checks. 

In summary this leads us to the powerful notion of \textit{Apparent Convergence} (AC) in functional approaches: we increase the truncation order in a given systematic expansion scheme and follow the combined systematic error estimate, now also including the convergence of results from one order of the expansion scheme to the next. If this combined error estimate is converging towards zero and is sufficiently small, we call this \textit{apparent convergence}. In our opinion, the functional QCD approach with apparent convergence offers a reliable first principles approach for predictions in QCD. In particular, it allows for quantitative QCD predictions in challenging regimes of great interest such as QCD at finite density, scattering processes and realtime evolutions, where lattice simulations are obstructed by the sign problem. Indeed, these two approaches should be seen as complementary and a combined use in the sense above is instrumental for resolving relevant open problems such as the quest for the critical end point in the QCD phase diagram or rather the onset of new phases and their nature.

\section{Computational setup}
\label{sec:SystematicsQCD}

With increasingly complex truncations of the QCD effective action, it becomes inevitable to use a fully automated process for deriving flow equations and computing results. In this Section, we aim at providing a comprehensive overview of the technical tools employed. These tools range from Mathematica packages developed specifically for functional approaches \cite{Cyrol:2016zqb,Pawlowski:2021tkk,Huber:2011qr,Huber:2019dkb} to high-performance libraries. When combined, they form a highly efficient framework for computing functional flows and represent one of the main advances of this work. We transition from analytical expressions of full flows to a more flexible numerical formulation, allowing for easy and automatic modification of the approximation or regulator functions. This approach also offers the additional benefit of enabling control over numerical accuracy and, consequently, computational speed.

Upon initial reading, this Section may be skipped in favour of proceeding directly to the results in \Cref{sec:Results}. 

We begin by outlining the derivation of functional flows in \Cref{sec:automatedGen}. Next, we discuss field and momentum discretisation, as well as their numerical treatment, in \Cref{sec:implementation}. Momentum dependencies are inherently linked with smooth regulator functions, which are addressed in \Cref{sec:regulators}. Finally, we give an outlook of how the present implementation might be extended even further in \Cref{sec:outlookNum}.

\subsection{Automated generation of functional equations}
\label{sec:automatedGen}
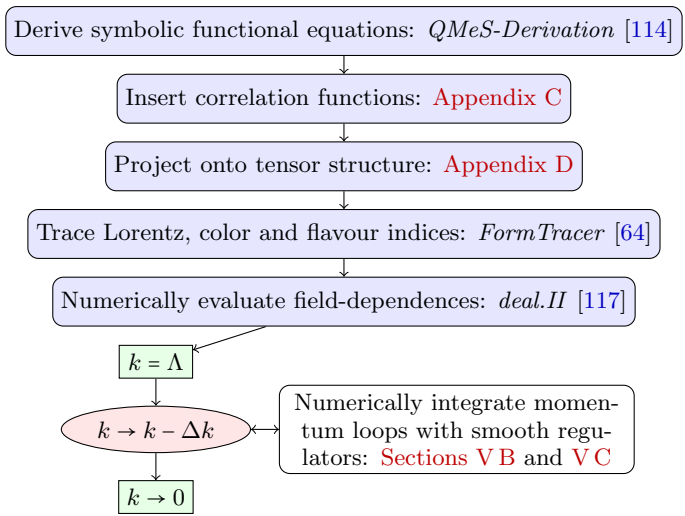
\begin{figure}
	\begin{tikzpicture}
		\node [roundbox] at (0,0.9) (start) {Derive symbolic functional equations: \textit{QMeS-Derivation} \cite{Pawlowski:2021tkk}};
		\node [roundbox] at (0,0) (step1) {Insert correlation functions: \Cref{app:Diagrammatics}};
		\draw [->] (start) -- (step1);
		\node [roundbox] at (0,-0.9) (step2a) {Project onto tensor structure: \Cref{app:Projections}};
		\draw [->] (step1) -- (step2a);
		\node [roundbox] at (0,-1.8) (step2) {Trace Lorentz, color and flavour indices: \textit{FormTracer} \cite{Cyrol:2016zqb}};
		\draw [->] (step2a) -- (step2);
		\node [roundbox] at (0,-2.7) (step3) {Numerically evaluate field-dependences: \textit{deal.II} \cite{dealII95}};
		\draw [->] (step2) -- (step3);
		\node [process] at (-2.5,-3.5) (step4) {$k =\Lambda$};
		\node [integration] at (-2.5,-4.4) (step5) {$k \to k - \Delta k$};
		\node [process] at (-2.5,-5.3) (step6) {$k \to 0$};
		\draw [->] (step3) -- (step4);
		\draw [->] (step4) -- (step5);
		\draw [->] (step5) -- (step6);
		\node [roundbox, fill=white,text width=4.5cm] at (1.5,-4.4) (int) {Numerically integrate momentum loops with smooth regulators: \Cref{sec:implementation,sec:regulators}};
		\draw [<->] (step5) -- (int);
	\end{tikzpicture}
	\vspace{1mm}
	\caption{Workflow chart for the computation of the effective action $\Gamma_{k \to 0}$, see \labelcref{eq:effAct}. Details on single steps can be found in the linked appendices, as well as the cited software packages. \hspace*{\fill}}
	\label{fig:workflow}
\end{figure}

The computation of the full effective action within a given approximation scheme \labelcref{eq:effAct} requires the derivation of flow-equations for all couplings. To ensure a consistent, automated, and easily reproducible workflow, especially in anticipation of even more complex truncation schemes in the future, we provide a detailed account of all steps and software packages utilised in the following. 

We use the Mathematica module \textit{QMeS-Derivation} \cite{Pawlowski:2021tkk}, which computes functional derivatives of the right-hand side of the Wetterich equation \cite{Wetterich:1992yh} and other functional relations such as DSEs and modified Slavnov-Taylor identities. This module enables us to derive the flow equations with correct momentum routing, prefactors, and signs. Additionally, the results of this step can be depicted diagrammatically, see \Cref{app:flows}. Currently, the generalised RG-transformation terms $\dot \phi$ in the composite operator flow \labelcref{eq:GenFlow} are not generated automatically in \textit{QMeS}, and we have to include them manually. 
Alternatively, the derivation of flow equations can also be performed with \textit{DoFun}, \cite{Huber:2011qr,Huber:2019dkb}. 

Subsequently, we insert the explicit expressions for vertices and propagators into the diagrammatic/symbolic representation of the flow equations. These expressions, derived directly from the Ansatz \labelcref{eq:effAct}, are summarised in \Cref{app:Diagrammatics}. It is only here, that approximations enter the system.

The flow equation of any full n-point function is then projected onto the respective coupling or dressing which closes all Lorentz-, colour- and flavour-indices and selects the momentum structure of a given coupling.
Finally, the trace over all indices is taken using the Mathematica module \textit{FormTracer} \cite{Cyrol:2016zqb}. Conventions and explicit projectors used in the derivation of flows are detailed in \Cref{app:Projections}.

A depiction of the workflow, beginning with the derivation of the equations and concluding with their numerical evaluation, is presented in \Cref{fig:workflow}. A brief discussion of the numerical evaluation of the flow and the loop-momentum integration follows in the subsequent Sections.

\subsection{Numerical evaluation of the equations}
\label{sec:implementation}

Once the flows are derived, the resulting set of partial differential equations is solved numerically. The current framework incorporates a full dependences on mesonic fields, as well as full momentum dependences of the couplings. Both are continuous quantities and require discretisation, necessitating an efficient computational implementation.

For the mesonic field dependences, we adopt the fluid-dynamical approach to the fRG, initiated in \cite{Grossi:2019urj}, which is discussed in \Cref{sec:fluiddynamics}.
The integration of full momentum loops is carried out at each step of the RG-time evolution $k \to k - \Delta k$. This is elaborated on in \Cref{sec:momentumIntegration}.

\subsubsection{Implementation of the full field dependence}
\label{sec:fluiddynamics}

The effective action of the matter part \labelcref{eq:Gammamatter} includes complete mesonic field dependences. In this work, we use continuous Galerkin methods, for more details on fluid dynamical techniques in the fRG see \cite{Grossi:2019urj, Grossi:2021ksl, Ihssen:2022xkr, Ihssen:2023xlp, Koenigstein:2021rxj, Koenigstein:2021syz, Steil:2021cbu}.

In this approach, the flow of the effective mesonic potential $V_l(\rho_l)$ is treated as a convection-diffusion equation in terms of its first derivative $u= m_\pi^2 = \partial_{\rho_l} V_l(\rho_l)$. In the fluid-dynamical picture the time direction corresponds to the RG-time and the spatial variable corresponds to $\rho_l$, compare \labelcref{eq:rho-rhos}. Consequently, we employ a 1D numerical grid to capture the mesonic field dependence. In future projects, additional condensates or resonant channels, such as diquarks or vector mesons, will necessitate the implementation of a 2D or even higher dimensional numerical grid.

Schematically, the flow of the mesonic potential or rather its $\rho_l$-derivative $u$ is given by
\begin{align}\label{eq:easyQCDflow}
	\partial_{t}u(\rho_l, p)
	=
	\partial_{\rho_l} & \Big[\eta_{\phi,k}(p) u(\rho_l, p)\rho_l + \frac{1}{{\cal V}_4} \mathrm{Flow}[\Phi_c](p)\Big]\,, 
\end{align}
with the space-time volume ${\cal V}_4$ and constant field $\Phi_c$. We consider general regulators, enabled by the fact, that all integrals and sums are computed numerically. 

The computation of the full field dependence is of importance for quantitative precision in the chiral limit and even for a qualitative access to the phase structure at high densities. In this regime, non-analyticities appear in the RG-flow of the potential and in the potential itself at $k \to 0$, see e.g. \cite{Grossi:2021ksl} for an investigation in the quark-meson model. Thus, higher derivatives of the potential - which correspond to higher mesonic scattering orders - cannot be neglected.
Presently, we use the finite element library \textit{deal.II} \cite{dealII95}. For time-stepping the highly stiff flow of $m_{\pi,k}^2(\rho_l)$, we use the \textit{IDA} solver from the \textit{SUNDIALS} library \cite{gardner2022sundials,hindmarsh2005sundials}. It implements a highly efficient BDF scheme with variable order up to 5, see also \cite{Ihssen:2023qaq}, where BDF methods have been found to be an excellent match to convexity-restoring fRG-flows. 

Presently, we do not take any further fully field-dependent quantities into account.
For example, a naive inclusion of a full field dependence of the Yukawa coupling $h_\phi$ seems to at least require the introduction of relative cutoff scales, similar to \cite{Ihssen:2023xlp} and lies beyond the scope of this work.

\subsubsection{Momentum integrals and grids}
\label{sec:momentumIntegration}

A further new and important improvement is the computation of complete momentum dependences of propagators and couplings. This improvement is paid for with a significant additional computational cost, as momentum integrals must be numerically evaluated at each step of the RG-time integration.

In the present symmetric point configuration, \labelcref{eq:SymPoint}, the 4-dimensional momentum integrals reduce to an integration over their absolute value $x = p^2 / k^2$ and a cosine between external and internal momentum.
To evaluate them, we employ multi-dimensional Legendre quadrature integrals of fixed order. We set the upper integration boundary $x_\mathrm{max}$ such that the error incurred by this truncation remains below $10^{-4}$. To determine the value of $x_\mathrm{max}$, we use quadrature integrals of extremely high order ($\sim 10^4$) to accurately isolate the error resulting from truncating a tail.
The Legendre quadrature orders used are:
\begin{itemize}
	\item \underline{$x$-integral:} order $64$ (no significant change from $32$)
	\item \underline{angular integral:} order $16$ (no sig. change from $8$).
\end{itemize}
Due to the highly regular structure of the integrands, we also find no significant difference when employing more sophisticated adaptive integration routines, such as those provided by the \textit{cubature} \cite{cubature} package or quasi-Monte-Carlo methods \cite{Borowka:2018goh}. The utilisation of fixed-order quadratures yields integrals that are at least ten times faster compared to adaptive routines, primarily due to their straightforward parallelisation and avoidance of error-checking overhead. Despite this, one can still easily obtain numerical error estimates by slightly increasing quadrature orders.

Parallelisation within this setup is both straightforward and efficient: In one and two dimensions, CPU processing surpasses GPU algorithms due to the substantial overhead associated with communication and memory access/transfer for GPUs. Consequently, we use the \textit{oneTBB} library for efficient CPU-side reduction over quadrature points. For three-dimensional and higher calculations, where GPU overhead becomes negligible, we employ the \textit{CUDA} API to harness the computational power of GPUs. In the present work, this is used for the system of avatars of the strong coupling at the symmetric point. This already gives a sizeable performance boost relative to a purely CPU calculation and this option will prove crucial for future extensions built upon our current truncation scheme -- for full momentum dependences, the calculation can be massively parallelised on one or more GPUs.

For the calculation of full momentum dependences, e.g. for $M_l(p)$, we utilise logarithmic momentum grids with 64 grid points. The integration of the flow can be again performed in parallel for every momentum-grid point.

\subsection{Regulators for rapid convergence}
\label{sec:regulators}

The cutoff term \labelcref{eq:CutoffTerm} implements the successive integration of momentum shells in the fRG, discussed in \Cref{sec:GenflowsQCD}. The precise form of these momentum shells is in one-to-one correspondence to that of the regulators detailed in \labelcref{eq:RegPhi} in \Cref{app:regs}. Their general structure is given by 
\begin{align} 
	R_{i}(p) = {\cal T}^{\ }_{i}(p) \, k^{d^{\ }_{i}} r_{i}(x)\,,\qquad x=\frac{p^2}{k^2}\,,  
\label{eq:RegStructure} 
\end{align}
where ${\cal T}^{\ }_{i}(p)$ is the dimensionless tensor structure of the kinetic term of the field $i=A_\mu, c,\bar c, q,\bar q, \phi$. The factor $k^{d_{i}}$ carries the dimension of the kinetic operator with $d_i=2$ for $i=A_\mu, c,\bar c, \phi$ and $d_q=1$. The last factor, the shape functions $r_i$, carries the details of the momentum shell for the momentum of the field $\Phi_i$. 

In the simplest case $r_i=1$, the regulators in \labelcref{eq:RegStructure} reduce to mass-type terms that carry the tensor structure of the kinetic term: they introduce a 'kinetic' mass gap.
While this is simply a mass term for the ghosts and mesonic composites, it is a chiral mass-type term for the quarks and it carries the Lorentz structure of the kinetic term for the gauge field, for more details see \Cref{app:regs}.
In consequence, regulators as defined in \labelcref{eq:RegStructure} maintain global symmetries of the respective sector of the theory, such as chiral symmetry. This regulators are behind the functional Callan-Symanzik equation, see \cite{Symanzik:1970rt}. More recently, they have been considered for realtime functional flows, as they keep all space-time symmetry, see \cite{Fehre:2021eob, Braun:2022mgx, Horak:2023hkp}. 

However, in order to have a UV regularisation of the flows for large momenta $x\gg 1$, the regulators and hence the shape functions should decay rapidly. The shape functions we consider in this work have an exponential decay and so have the regulators. 

The importance for an optimised convergence of the present approximation scheme has been discussed in \Cref{sec:ResonantChannels,sec:optimisation} and we will put this to work in the following. With the present setup we aim at the resolution of QCD in the vacuum and at finite temperature and density. Specific interesting phenomena concern for example its phase transitions, the different emergent degrees of freedoms as well as the resolution of competing order effects at large densities. Attaining rapid convergence in the present approximation scheme requires thus a class of momentum-local operators, that also include optimal regulators, which obey the following criteria:

\subsubsection{Analyticity} 
\label{sec:Analyticity} 

This property is crucial for finite-temperature results, where non-analytic regulators may lead to large oscillations due to the discrete nature of the Matsubara frequencies \cite{Fister:2015eca}. Moreover, non-analytic regulators such as the sharp cutoff or the flat regulator lead to momentum non-localities such as the occurrence of $\sqrt{p^2} (p^2)^n$ terms in momentum-dependent dressings. For example, the flow of the bosonic two-point function, $\textrm{Flow}^{(2)}(p)$, with the sharp and flat regulators have the following small momentum expansions, 
\begin{align} \nonumber 
 \textit{sharp:}\ \ \textrm{Flow}^{(2)}(p) = &\, c_0 + c_1 \sqrt{p} +O(p^2)\,,\\[1ex]
\textit{flat:}\ \ \textrm{Flow}^{(2)}(p) = &\, c_0 + c_2 p^2 +c_3 \sqrt{p^2} p^2+ O(p^4)\,. 
\end{align}
Consequently, the two-point functions for $k\neq 0$ carry these non-localities that only disappear for $k\to 0$. Their presence spoils the momentum structure in the complex frequency and momentum plane, and hence the reconstruction of spectral properties such as the pole mass. Moreover, within given truncations remnants may survive at $k=0$. In the context of reconstructions of spectral functions, this has been studied in \cite{Bonanno:2021squ}, a comprehensive analysis in the context of realtime flows can be found in \cite{Braun:2022mgx}. 

From the numerical point of view, non-analytic regulators introduce non-analyticities in the integrands of the loop integrals in $\textrm{Flow}^{(n)}$, which are not well-suited for any numerical recipe. 

We close with the remark, that the preservation of analyticity is of primary importance for momentum locality and momentum dependences, like the spectral properties of QCD. In particular, non-analytic regulators are not well suited for the partial use of the derivative expansion in the low-energy sector of QCD. Specifically, we have found in this work, that using the flat or Litim regulator significantly impacts the mesonic masses through $Z_{\phi,k}(p=0)$. This effect immediately vanishes if a smooth approximation of the flat regulator is chosen.

\subsubsection{Convexity restoration}
\label{sec:Convexity} 

The full effective action $\Gamma_k[\Phi]+\Delta S_k[\Phi]$ is convex as it is a Legendre transform. This entails that the full mesonic effective potential $V_{k}(\rho) + k^2 \rho$ is convex. In regimes with spontaneous symmetry breaking, the non-trivial expectation value of $\rho_0$ at $k=0$ is obtained from the effective potential $V_\textrm{eff}(\rho) = V_{k=0}(\rho)$ via 
\begin{align} 
\lim_{\epsilon\to 0} \left[ \frac{\partial V_\textrm{eff}(\rho_l)}{\partial {\rho_l}} - \epsilon \right]_{\rho= \rho_0}=0 \,,
\label{eq:EoMSpontSym}
\end{align}
where $\epsilon$ induces a small (linear) explicit symmetry breaking. 
\Cref{eq:EoMSpontSym} is the functional analogue of the standard definitions via expectation values with 
\begin{align}
	\rho_0 = \frac{1}{{\cal V}_4} \int_x \langle \hat \rho(x)\rangle \,,\qquad \textrm{or} \qquad \phi_0 =\lim_{\epsilon\to 0} \, \langle \hat\phi\rangle\,,  
	\label{eq:EoMExpt}
\end{align} 
where ${\cal V}_4$ is the dimension of Euclidean space-time. The fields $\hat\phi,\, \hat \rho$ in \labelcref{eq:EoMExpt} are the field operators, whose expectation values are the mean field in the effective action, $\rho = \langle \hat\rho\rangle ,\,\phi= \langle \hat \phi\rangle$. A non-vanishing $\rho_0$ in \labelcref{eq:EoMExpt} signals a macroscopic occupation of the field zero mode, as it does in \labelcref{eq:EoMSpontSym}. 

For $\rho_0\neq 0$, \labelcref{eq:EoMSpontSym} entails that the effective potential features a flat regime for $\rho<\rho_0$. In the fRG approach, this flat regime is obtained as follows: in the broken regime, the effective potential $V_k(\rho)$ is not convex at intermediate $k$ and has non-trivial minima $\rho_{0,k}$. 
In the limit $k\to0$, the $\rho_{0,k}$ converge towards the physical expectation value $\rho_0$, while the potential $V_k(\rho)$ develops flat regimes for $\rho< \rho_{0,k}$. 

The flow equation has a convexity restoring property, c.f. \cite{Litim:2006nn}: the propagator is getting close to a repulsive singularity for small momenta and $\rho< \rho_{0,k}$. In this regime, the flow grows strong and its contributions are convexity-restoring. 

In particular, in the present QCD application the mesonic mass functions \labelcref{eq:rawMass} become negative for $\rho\leq \rho_0$ ($m^2_\pi$ is negative for $\rho\leq \rho_0$, $m^2_\sigma$ follows suit for smaller $\rho$). It can be shown that in the absence of approximations, this singularity is sufficiently repulsive so that it cannot be reached. The negative mass functions are pushed towards vanishing ones and the rate is always such that $k^2+m_\phi^2 > 0$ and the singularity is never reached \cite{Litim:2006nn}. 

One of the novel ingredients of the present approximation is the inclusion of the full effective potential which accommodates the convexity-restoring property, where e.g.~Taylor expansions in $\rho$ (with $\rho<\rho_{0,k}$) and perturbation theory or related resummation schemes fail. Moreover, even if the Taylor expansion is restricted onto the flowing minimum $\rho_{0,k}$ or to a fixed $\rho_0 = \rho_{0,k\to0}$, in the case of $m^2_\phi(\rho_0)\to 0$, it becomes far less stable. For a detailed discussion see \cite{Pawlowski:2014zaa}. 

Numerically convenient inclusions of the convexity restoring property require specific properties for the regulator or rather the shape function in the limit $p^2\to 0$: in the case of integrable singularities, in the present case defined via 
\begin{align} 
	\int_p \partial_t R_\phi(p) G_\phi^2(p,\phi) <\infty\,,
\label{eq:IntegrableSings}
\end{align}
the repulsive property of the singularity is only implemented via the successive RG-steps in the $t$-direction, which implicitly amounts to taking $\phi$-derivatives of \labelcref{eq:IntegrableSings}. 
Evidently, this involves a combination of $t$- and $p$-integration steps which is numerically challenging. This intricacy is present for most standard regulator choices, including the exponential shape functions $r_k(x) = \frac{x}{e^{x}-1}$, $e^{-x}$ and variants thereof with higher order polynomials in the exponent. 

Note that the non-analytic sharp and flat regulators lead to non-integrable cuts in the integrand of \labelcref{eq:IntegrableSings}. While this readily incorporates convexity restoration, it is precisely their non-analyticity that leads to the cut and clashes with the required analyticity property discussed in \Cref{sec:Analyticity}. 

Accordingly, computationally convenient analytic regulators have to introduce poles of sufficiently high order in \labelcref{eq:IntegrableSings}. In the present case of vacuum QCD in $d=4$ this explicitly means a requirement of order four and higher for the singularity. This has been investigated recently in \cite{Zorbach:2024zjx}, where classes of regulators with non-integrable singularities for \labelcref{eq:IntegrableSings} have been suggested and studied comprehensively, introducing the \textit{principle of strongest singularity}.

\subsubsection{An optimised class of regulators} 
\label{sec:ClassRegulators} 

In the present work, we do not use one of the classes studied in \cite{Zorbach:2024zjx} and concentrate more on the combination of flatness, analyticity and smoothness. The latter two properties are related but not the same. While not discussed further he, we add that smoothness is chiefly important for frequency-dependent regulators at finite density, see e.g.~\cite{Fister:2015eca, Cyrol:2016tym}. Details are deferred to \Cref{app:regs}, here we summarise the main properties. 

First of all, momentum locality requires closely related physical cutoff scales for all fields as discussed in \Cref{sec:optimisation}, see \labelcref{eq:Identicalks} and below. While it is possible to satisfy this constraint, it is computationally demanding and we refrain from a full optimisation. Instead, we use identical regularisations for the scalar part $G_s$ of all propagators, see \labelcref{eq:ScalarProp} in \Cref{app:regs}. We recall it here for the sake of convenience, 
\begin{align}\nonumber 
	G_{i,s} =&\, \frac{1}{p^2\left[1 + \frac{k^2}{p^2} r_i(x)\right] + m^2_{i} }\,,\qquad \textrm{for}\qquad i= A,c,\phi\,,\\[1ex] 
	G_{q,s} = &\, \frac{1}{ p^2\left[1 + \frac{k}{p} r_q\right]^2 + m^2_q }\,.
	\label{eq:ScalarPropMain}
\end{align}
where we have used regulators carrying the kinetic tensor structure as indicated in \labelcref{eq:RegStructure}. Consequently, identical regularisations for all propagators amounts to using identical shape functions $r_i=r$ for $i=A,c,\phi$. For $i=q$, we infer from	\labelcref{eq:ScalarPropMain} a simple relation between the shape function $r$ and $r_q$. In summary, we are led to 
\begin{align}\nonumber 
	r_A(x)=&\,r_c(x)=r_\phi(x)= r(x)\,,\\[1ex] 
	r_q(x) = &\,\frac{1}{k} \left( \sqrt{p^2 +k^2 r(x)} -p\right) \,.
	\label{eq:rPhiEqualMain}
\end{align}
As discussed in the previous Section, a numerically convenient implementation of the convexity restoring property \cite{Litim:2006nn} of the flow is achieved with regulators that lead to non-integrable poles in the flow equation for negative curvature masses. Again, this can be investigated by concentrating the scalar part of the propagators provided in \labelcref{eq:ScalarPropMain}: we require that the scalar parts \labelcref{eq:ScalarPropMain} of all propagators develop an $n$th order pole at $p=0$ and $m^2_i + k^2 = 0$, 
\begin{align}
	G_{i,s}(p^2\to 0,m^2_i\to -k^2) = \textrm{const.} \,\left(\frac{1}{p^2}\right)^n\,, 
	\label{eq:GsSingn} 
\end{align}
for all fields, $i=A,c,q,\phi$. For sufficiently large $n$, this implements the convexity restoring behaviour discussed above, leading to numerically accessible flows within double precision. In the present work, we chose $n=8$ which is sufficient to flow very far into the infrared, and employ a general class of shape functions given by
\begin{align}
	r_n(x) = e^{-f_n(x)}\,, \qquad f_n(x) = \frac{P_n(x)}{Q_{n-1}(x)}\,.
	\label{eq:RationalExpShapeMain}
\end{align}
Here, $P_n(x), Q_{n-1}(x)$ are polynomials of order $n$ and $n-1$ respectively, and their coefficients are chosen such that equation \labelcref{eq:GsSingn} is fulfilled. A detailed discussion and explicit expressions can be found in \Cref{app:regs}.

\begin{figure}
	\centering
	\includegraphics[width=0.5\textwidth]{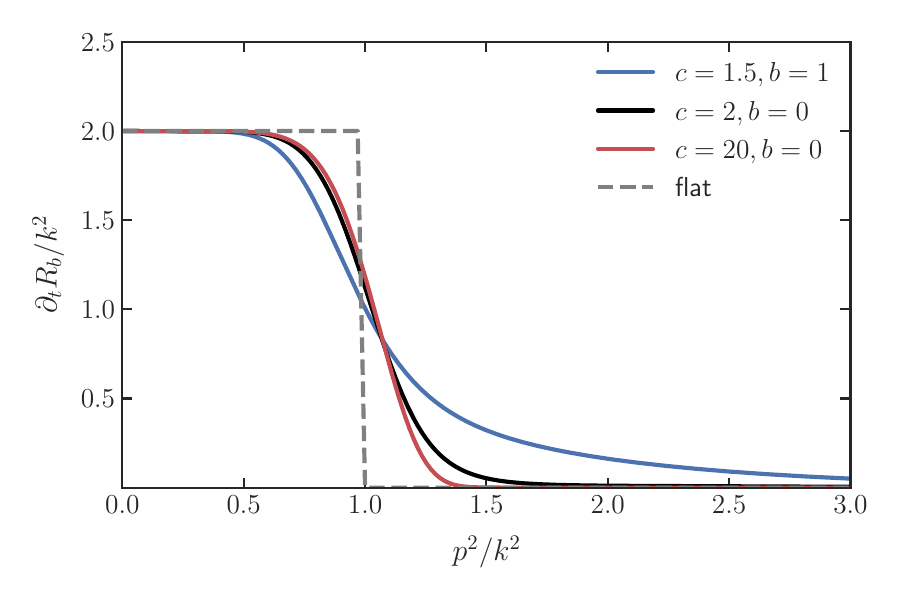}
	\caption{We show $\partial_t R_k(p)/k^2$ for smooth shape functions $r(p^2/k^2)$ with exponential decay, that approximate the shape function of the flat regulator. In this work, we adopt the rational exponential regulator, defined in \labelcref{eq:RationalExpShapeMain} and detailed in \Cref{app:CompRegs}: All depicted regulators induce $8$th order poles, $c$ determines the rapidity of the exponential decay for large $p^2/k^2$, while $b$ is tuned for avoiding substructures. Our main results are obtained for $c=~2$ (\textit{black straight line}). For our error analysis we also compute results within the same regulator class, but with $c=1.5$ (\textit{blue straight line}) and $c=20$ (\textit{red straight line}), which respectively elongate or truncate the regulator tail. For comparison, we also depict $\partial_t R_k(p)/k^2$ for the flat regulator (\textit{grey dashed line}). 
		\hspace*{\fill}}
	\label{fig:regs}
\end{figure}

The resulting class with a fixed order $n=8$ has two free parameters $b$ and $c$. In \Cref{fig:regs} we show the regulators used in the present work with 
\begin{align} 
 (c,b) \in\left\{\ (1.5,1)\,,\ (2,0)\,,\ (20,0)\ \right\}\,.
\end{align}
Our optimised result uses $(c,b)=(2,0)$, while the two other choices are more extreme choices in terms of sharpness $(20,0)$ and lack of decay $(1.5,1)$. Note that the main parameter is $c$, while $b$ is tuned for smoothness of second and higher order derivatives.

\subsection{Full momentum dependences and further developments}
\label{sec:outlookNum}

The setup detailed in this Section allows to systematically improve the current truncation. Moreover, the numerical implementation of all flows allows to check for systematic errors by varying regulators, an important ingredient of a systematic expansion with a controlled systematic error estimate. The present versatile and flexible setup also gives us access to fully momentum-dependent propagators and couplings as we have seen in \Cref{sec:optimisation}. This will be used in the present work to extract fully momentum-dependent meson and quark propagators on the solution of the flow equations. They are currently not fed back into the flow equations, which is left to future work. Still, the extracted momentum dependence gives us direct access to the meson pole masses. We can reconstruct the pole masses by using Pad\'e approximants. This enables us to tune the current quark masses directly with the pion pole mass. Moreover, one may use approximate formulae such as the Pagels-Stokar formula for $f_\pi$, that utilise the quark mass-function and wave function. Its direct computation in the present approach uses the fact that the scalar-pseudoscalar Yukawa coupling is nothing but the Bethe-Salpeter wave function, and $f_\pi$ follows directly from its evaluation on the pion pole. For $m_\pi\neq 0$ this requires realtime fRG methods as prepared with the spectral fRG in \cite{Horak:2020eng, Braun:2022mgx, Horak:2023hkp}. 

Feeding back the frequency and momentum dependences is especially relevant for computations at high $\mu$, where the momentum structure of both, the quark and meson propagators is chiefly relevant and non-trivial: for the quark this originates in the non-trivial Fermi-surface that introduces non-trivial frequency and momentum structures in quark propagators and vertices. The meson propagators may carry non-trivial maxima in momentum space that are signatures of inhomogeneities such as a moat regime \cite{Pisarski:2021qof, Rennecke:2023xhc}. First signatures of the latter have been detected in \cite{Fu:2019hdw}, and we expect that these signatures are corroborated in an improved computation. Moreover, inevitably the flows will carry a minimal momentum transfer, mainly due to the non-trivial Fermi surface of the quarks and diquarks, for a first discussion see \cite{Ihssen:2023xlp}. The results there imply that regulator independence and quantitative reliability for QCD at large densities can only be obtained by a full momentum resolution of the flow of at least the quark- and meson-propagators. 

In a first step beyond the present work we will also consider the full momentum dependence of propagators, a Fierz-complete inclusion of the four-quark scattering vertices, the symmetric point momentum dependences of all couplings as well as the full field dependence of the Yukawa coupling, as well as the strong couplings. We also note that apart from the improved systematics and the full access to the physics and phenomena at large densities it will also allow for a fully self-contained calculation of the gluon propagator. While this seemingly looks like a further major step, it was one of the chief goals of the present work to put forward a comprehensive numerical framework in which all these steps, and more, are readily implemented. In this respect, the current work has to be viewed as part of a combined effort, that draws from many previous works in low-energy effective models in QCD, for a rather complete list see \cite{Dupuis:2020fhh}. In functional QCD it builds specifically on \cite{Mitter:2014wpa, Cyrol:2016zqb, Cyrol:2017ewj, Fu:2019hdw} and earlier work. It is accompanied by \cite{Fu:2022uow, Fu:2024ysj}, that will be used in an extension \cite{FHPT2024} of \cite{Mitter:2014wpa, Cyrol:2017ewj} with a more complete momentum structure of the four-quark vertices. This line of work is accompanied by works on the full momentum structure of four-fermi vertices in the presence of a non-trivial Fermi-surface and ordering \cite{C?W2024} as well as the construction of optimal non-singular tensor bases for four-quark vertices and beyond \cite{?SW2024}.

\section{Results}
\label{sec:Results}
	
We begin this Section by discussing the systematic error estimation procedure and the input parameters used in our computation in \Cref{sec:initialSetup}. We indicate the full table of observables obtained from this set of parameters in \Cref{tab:results}.

This is followed by a discussion of results, focussing on the different components of the truncation:
The gluonic interactions dominating at high energies are discussed in \Cref{sec:GlueResults}. In 
\Cref{sec:Legoatwork} the transition regime is investigated, where the gluonic sector starts to decouple and the low-energy degrees of freedom begin to emerge. In particular we focus on the systematic improvements of the current setup, also in comparison to preceding works \cite{Braun:2014ata, Fu:2019hdw} we build upon.
Finally, we discuss the results for the emergent mesonic sector in \Cref{sec:ResultsMatterSector}. There, we also showcase fully momentum dependent propagators in the matter sector.

The present Section mainly discusses results for $N_f = 2+1$. A summary of $N_f=2$ results is deferred to \Cref{app:TwoFlavourQCD}, including some comparison plots.

\subsection{Initial conditions and systematic errors}
\label{sec:initialSetup} 

The input parameters at $k=\Lambda_\textrm{UV}$ of the present approximation to the effective action of QCD contain those of first principle QCD, i.e. the strong coupling $g_s$ of QCD, and the current quark masses $m_q$. Additionally, we include one phenomenological infrared parameter $a$, which is an infrared rescaling of the running strong coupling by a few percent.

We inherit the value for the initial strong coupling from the quantitative functional two-flavour QCD computation in \cite{Cyrol:2017ewj}, while the current quark masses are tuned such that the pion pole mass in 2+1 flavour QCD has the value 138\,MeV. Also, the ratio of pion and kaon decay constant takes its physical value $1.1914$. Finally, the rescaling $a$ is adjusted such that the constituent quark mass function has the infrared value $M_l(0)=350$\,MeV.

Below, we describe the details of the respective initial conditions in \Cref{sec:InitialConditions} and discuss our systematic error estimates in \Cref{sec:RegulatorErrors}.

\subsubsection{Initial conditions}
\label{sec:InitialConditions}

The initial conditions are set at the initial UV-cutoff scale $k=\Lambda_\textrm{UV}$,
\begin{align} 
	\Lambda^{N_f=2+1}_\textrm{UV} = \qty{20.137}{\GeV}\,. 
	\label{eq:LambdaUV2-2+1}
\end{align}
The underlying scale setting originates from the two-flavour QCD input \cite{Cyrol:2017ewj}, where we choose $\Lambda^{N_f=2}_\textrm{UV} = \qty{20}{\GeV}$. The slight adjustment of $\Lambda_\textrm{UV}$ in 2+1-flavour QCD accommodates the changes in the flow due to the strange quark.
\begin{table}[b]\centering
	\begin{tabular}{|>{\centering}m{0.3\linewidth} || >{\centering\arraybackslash}m{0.25\linewidth} |}
		\hline & \\[-1ex]
		Coupling & Value \\[1ex]
		\hline& \\[-1ex]
		$\alpha_{A^3,\Lambda_\textrm{UV}}$& 0.210 \\[1.5ex]
		\hline& \\[-1ex]
		$\alpha_{c\bar c A,\Lambda_\textrm{UV}}$ & 0.235
		\\[1.5ex] 
		\hline& \\[-1ex]
		$\alpha_{A^4,\Lambda_\textrm{UV}}$   & 0.216 \\[1.5ex]
		\hline& \\[-1ex]
		$\alpha_{l\bar l A,\Lambda_\textrm{UV}}$  & 0.227
		\\[1.5ex]
		\hline& \\[-1ex]
		$\alpha_{s\bar s A,\Lambda_\textrm{UV}}$    & 0.227
		\\[1.5ex]  
		\hline
	\end{tabular}
	\caption{Strong couplings at the initial scale $\Lambda$ for $N_f = 2+1$. Their determination is outlined in \Cref{sec:GaugeConsistency}.
		\hspace*{\fill}	\label{tab:alphas} }
\end{table}
In short, in the UV the gluonic sector dominates the dynamics and the RG-flow can be matched to perturbation theory or lattice results, see also \Cref{fig:pertAlpha}. The relevant input parameters in this regime are the gluonic couplings, compare \labelcref{eq:AgreeAvatars}. These couplings are also implicitly contained in the input data \cite{Cyrol:2017ewj}, i.e. the gluon and ghost propagators, but the present cutoff-dependent couplings differ due to the mSTIs. 
Their determination at $k=\Lambda_\textrm{UV}$ is also described in \Cref{sec:GaugeConsistency}.

We take $\alpha_{c\bar c A}$ directly from the input in \cite{Cyrol:2017ewj} and then adjust \labelcref{eq:beta_iEqual} at the initial cutoff scale \labelcref{eq:LambdaUV2-2+1}, leading to the values summarised in \Cref{tab:alphas}.

In the perturbative regime, the flow of the four-quark couplings is proportional to $\alpha_s^2$ and rapidly tends towards zero for large $k$. Moreover, its value at smaller perturbative cutoff scales $k\gtrsim \qty{5}{\GeV}$ is dominated by the contributions of quark-gluon box diagrams. Accordingly, we have to choose 
\begin{align} 
 \Lambda_\textrm{UV}^2	\lambda_i \ll c_i \alpha_{l\bar l A}^2\,, 
	\label{eq:lambda=0}
\end{align} 
for all four-quark couplings to ensure the strong suppression of $\lambda_i$ at higher scales.
In \labelcref{eq:lambda=0}, $c_i \alpha_{l\bar l A}^2$ has been defined as the absolute value of the respective quark-gluon box diagram contribution to the flow of $\lambda_i$.
In the present work we only consider the scalar-pseudoscalar coupling $\lambda_q$, in a complete basis also called $\lambda_\pi$ or $\lambda_{\sigma-\pi}$. 
The flow of $\lambda_q$ is depicted in \Cref{fig:diag_quark_4pt} in \Cref{app:flows}. The flows of all other four-quark couplings are structurally identical but are projected differently.
The coefficient $c_q$ in \labelcref{eq:lambda=0} is given by 
\begin{align}
c_q = 9.4\cdot 10^{-3} \,, \quad c_q \alpha_{l\bar l A}^2 = c_q\cdot 0.046656 \approx 4.4\cdot 10^{-4}\,, 
\label{eq:boundlambdaq}
\end{align}
where we have used \Cref{tab:alphas}. In the deep UV, the initial effective potential of the mesonic composites is quadratic, 
\begin{align}\label{eq:UVPot}
	V_\Lambda(\rho) = m_{\phi, \Lambda_\textrm{UV}}^2\,\rho \,.
\end{align}
Moreover, on the equations of motion for $\phi$ we obtain
\begin{align}
\lambda_q(p) = \frac{1}{2} h_\phi^2 G_{\phi_l}(p) \,,\quad G_{\phi_l}(p) =\frac{1}{p^2 + m_\phi^2 + R_{\phi_l}}\,,
\end{align} 
with the mesonic propagator $G_{\phi_l}$ defined in \Cref{app:PropagatorsVertices}. As long as the combination of the square of the Yukawa coupling and the meson propagator satisfies 	\labelcref{eq:lambda=0}, the results at $k=0$ are independent of this choice. This has been checked thoroughly in \cite{Gies:2002hq, Braun:2008pi, Braun:2009gm, Braun:2014ata, Rennecke:2015eba, Fu:2019hdw} for the present one-channel approximation and in \cite{Mitter:2014wpa, Cyrol:2017ewj} including all channels, for details we refer to these works. In our explicit computations, we have chosen 
\begin{align} 
	h_{\phi;\Lambda_\textrm{UV}}=1 \,,\quad \frac{m_{\phi,\Lambda_\textrm{UV}}^2}{\Lambda_\textrm{UV}^2} = 10^4\,, \quad \to \quad \Lambda_\textrm{UV}^2	\lambda_q = 5\cdot 10^{-5}\,, 
\end{align}
in line with \labelcref{eq:lambda=0} and \labelcref{eq:boundlambdaq}.

In the perturbative regime with $k\gtrsim \qty{5}{\GeV}$, the RG-invariant Yukawa coupling settles at $h_\phi \approx 9 - 12 $, while the mass function $m_\phi$ of the scalar-pseudoscalar composite drops with $1/\alpha_{l\bar l A}^2$. 
This has been baptised a 'quasi fixed point' running in \cite{Gies:2002hq} and has also been seen in \cite{Gies:2002hq, Braun:2008pi, Braun:2009gm, Braun:2014ata, Rennecke:2015eba, Fu:2019hdw, Mitter:2014wpa, Cyrol:2017ewj}. Its persistence in the fully momentum-dependent approximations \cite{Mitter:2014wpa, Cyrol:2017ewj} corroborates the fact that even in the ultraviolet, the $t$-channel exchange is well described by an exchange of a massive exchange boson with a classical dispersion.
In turn, within the deep infrared, full QCD runs into $\chi$PT as discussed before in \Cref{sec:optimisation} and is dominated by pion exchange processes of a point-like pion. 

\begin{table}[t]\centering
	\begin{tabular}{|>{\centering}m{0.24\linewidth} ||>{\centering}m{0.3\linewidth}|| >{\centering\arraybackslash}m{0.4\linewidth} |}
		\hline & & \\[-1ex]
		Observables & Value & Parameter in $\Gamma_{\Lambda_\textrm{UV}}$\\[1ex]
		\hline& & \\[-1ex]
		$m_{\pi,\textrm{pol}}$ [MeV]     &  138(9) & $c_{\sigma_l} = \qty{4.67}{\GeV^3}$  \\ [1ex]
		\hline & & \\[-1ex] 
		$f_K/f_\pi $   & $1.1914$    & 
		  $\Delta m_{sl} = \qty{134.2}{\MeV}$  \\[1ex]
		\hline & & \\[-1ex]
		$\alpha_{l\bar l A ,\Lambda_\textrm{UV}}$   &    & $  \alpha_{l\bar l A,\Lambda_\textrm{UV}} = 0.227$
		\\[1ex]
		\hline& &\\[-2ex]
		\hline & & \\[-1ex]
		$m_l$ [MeV]      &  $350$  &   $a= 0.0251 \,\quad b=\qty{2}{\GeV}$ \\[1ex]  
		\hline& &\\[-2ex]
		\hline & & \\[-1ex]
		$f_\pi$ [MeV]      &  $97.2^{+4.0}_{-2.2}$ & \noindent\rule{1cm}{0.4pt}
		\\[1ex]
		\hline& &\\[-1ex] 
		$m_s$ [MeV]      &  $485.0^{+0.0}_{-0.3}$  &  \noindent\rule{1cm}{0.4pt}
		\\[1ex]  
		\hline& &\\[-1ex] 
		$m_{\pi,\textrm{cur}}$ [MeV]      &  $138$ & \noindent\rule{1cm}{0.4pt}
		\\[1ex]
		\hline& &\\[-1ex] 
		$m_\sigma$ [MeV]      &$388.1^{+0.0}_{-1.1}$  & 
		\noindent\rule{1cm}{0.4pt} \\[1ex]  
		\hline& &\\[-1ex] 
		$\sigma_{0,l}$ [MeV]      & $69.^{+1.2}_{-0.2}$ & 
		\noindent\rule{1cm}{0.4pt} \\[1ex]  
		\hline
	\end{tabular}
	\caption{IR-Observables and corresponding parameters of the initial UV action for $N_f=2+1$. The parameters in in $\Gamma_{\Lambda_\textrm{UV}}$ are chosen to adjust the masses on the physical equations of motion \labelcref{eq:eomQCD}. The quark mass is adjusted using an enhancement of the quark-gluon vertex which is indicated in \labelcref{eq:IRenhancement}. 
	The $\pm$ error bounds correspond to the systematic error linked to the choice of regulator, recall \Cref{sec:RegulatorErrors}. The error on the pion pole mass is due to the extrapolation to the complex plane \labelcref{eq:padePole}.
	\hspace*{\fill}	\label{tab:results} }
\end{table}

We integrate our results down to an RG-scale $k_{\mathrm{min}}=~\qty{0.02}{\MeV}$, where no changes in any observables are discernible anymore. 

The current setup has one physical parameter for $N_f=2$ and two physical parameters for $N_f=2+1$: They correspond to the current quark masses and are set by the coefficients $c_{\sigma_l}$ (light current quark mass) and $c_{\sigma_s}$ (strange current quark mass) of the linear terms in $\sigma_l, \sigma_s$ that induce explicit chiral symmetry breaking. 
In both cases, one phenomenological parameter remains: the infrared scaling factor $1+a$ of the quark-gluon coupling $\alpha_{\bar q q A}$, which compensates for the missing tensor structures of the quark-gluon vertex. 

We fix the coefficient $c_{\sigma_l}$ through the pion pole mass and the coefficient $c_{\sigma_s} $ by the ratio of pion and kaon decay constants.
The phenomenological parameter introduced in \labelcref{eq:IRenhancement} is used to adjust the strength of chiral symmetry breaking through the value of the constituent quark mass at vanishing momentum. 

In the present approximation, it introduces a $\lesssim 3$\% enhancement of the quark-gluon coupling $\alpha_{l\bar lA}$ in the infrared. This small remaining phenomenological parameter will be removed in a forthcoming work, where all tensor structures and momentum dependences of the dressings will be considered. 

All initial parameters, resulting observables and the corresponding quark and meson pole masses at $k\to 0$ are summarised in \Cref{tab:results}. We comment on the results for the pole masses in the following sections.
Furthermore, we show results for $N_f=2$ in \Cref{app:TwoFlavourQCD}.

We also indicate the chiral symmetry breaking scale, which we define as the largest cutoff scale with $\langle\sigma\rangle(k<~k_\chi)\neq 0$ in the chiral limit and is given by
\begin{align}
	k_\chi^{N_f=2+1} & = \qty{388}{\MeV}\,.
\label{eq:ChiralSymmbScale}
\end{align}
A similar $k_\chi$ has already been found in previous QCD computations with the fRG, \cite{Braun:2014ata, Mitter:2014wpa, Cyrol:2017ewj, Fu:2019hdw}. Notably, \labelcref{eq:ChiralSymmbScale} is significantly lower than the chiral symmetry breaking scale in most low-energy effective theory calculations. This indicates the incompleteness in terms of QCD fluctuations and suggests to use to augment them with QCD information such as flows of specific couplings. These energy effective theories are called QCD-assisted models and have a qualitatively enhanced reliability and regime of validity. For a recent example see \cite{Fu:2023lcm}. 

For illustration, we also show the evolution of the order parameter $\sigma_0$ in \Cref{fig:sigmarunning} in \Cref{app:plots}, both for the chiral case defining $k_\chi$, and at the physical point. Furthermore, in \Cref{tab:resultsChiral} we also present our results in the chiral limit.
\begin{table}[t]\centering
	\begin{tabular}{|>{\centering}m{0.3\linewidth} | >{\centering\arraybackslash}m{0.3\linewidth} |}
		\hline & \\[-1ex]
		Observable & Value \\[0.5ex]
		\hline& \\[-1ex]
		$(f_K/f_\pi)_{\chi}$ & $1.2168^{+0.0006}_{-0.0007}$ \\[1ex]
		\hline& \\[-1.5ex]
		$f_{\pi,\chi}$ [MeV] & $93.2^{+3.5}_{-3.1}$	\\[1ex]
		\hline& \\[-1.5ex]
		$m_{l,\chi}$ [MeV]  & $311.6^{+0.3}_{-0.1}$	\\[1ex]
		\hline& \\[-1.5ex]
		$m_{s,\chi}$ [MeV]&  $446.7^{+0.3}_{-0.2}$ \\[1ex]
		\hline& \\[-1.5ex]
		$m_{\sigma,\chi}$ [MeV]&  $214.7^{+5.4}_{-9.3}$ \\[1ex]
		\hline& \\[-1.5ex]
		$\sigma_{l,0,\chi}$ [MeV] & $67.1^{+1.2}_{-0.0}$	\\[1ex]
		\hline
	\end{tabular}
	\caption{IR-Observables for $N_f = 2 +1$ in the chiral limit. The corresponding initial parameters are indicated in \Cref{tab:alphas} and \Cref{tab:results}. In contrast to the latter table we use $c_{\sigma_l} = 0$ but keep $\Delta m_{sl} = \qty{134.2}{\MeV}$.
		\hspace*{\fill}	
		\label{tab:resultsChiral} }
\end{table}
%

\subsubsection{Regulator dependence and systematic error estimates}
\label{sec:RegulatorErrors}
\begin{table}[t]
	\begin{tabular}{|>{\centering}m{0.25\linewidth}||>{\centering}m{0.2\linewidth}|>{\centering\arraybackslash}m{0.2\linewidth} |}
		\hline & &  \\[-1ex]
		Regulator & $a$ & $c_\sigma$ [GeV$^3$] 
		\\[2ex]\hline&& \\[-1ex]
		$(c,b) = (1.5,1)$ & 0.0182 & 4.54
		\\[1ex]	\hline& & \\[-1ex] 
		$(c,b) = (2,0)$ & 0.0251 & 4.67 
		\\[1ex]	\hline& & \\[-1ex] 
		$(c,b) = (20,0)$ &0.0265 & 4.69
		\\[1ex]	\hline
	\end{tabular}
	\caption{Changes in parameters for different regulators.\hspace*{\fill}	}
	\label{tab:RegulatorChanges}
\end{table}
The combination of approximation scheme and regulator choice is one of the main sources for systematic errors in the fRG approach to QCD, recall \Cref{sec:optimisation}. Due to our current implementation with numerical integration of momentum loops, we are in a position to easily check the regulator dependence of the present approximation scheme. 
We limit ourselves to a class of regulators that are sufficiently flat for small momenta, see \labelcref{sec:ClassRegulators}. On the basis of this, we vary the decay from steep to mild ones, where steepness is limited by future use at finite temperature. The construction is detailed in \Cref{sec:regulators} and \Cref{app:regs}. Within this class we chose three different regulators, indicated in \Cref{fig:regs}, to obtain results with an accurate systematic error incurred by the RG-time integration. This procedure allows us to perform a general scan of the present setting within the constraints coming from an easy accessibility to convexity restoration and at finite temperature. Note that these constraints are also related to optimisation as discussed before and are not mere convenience. 

Changing the specific regulator requires a readjustment of the enhancement factor and $c_{\sigma_l}$, which is indicated in \Cref{tab:RegulatorChanges} for completeness. Although the regulator changes quite strongly, we observe the change in the enhancement factor to be around $15\%$.
The resulting error from the change of regulator and readjusting of parameters is indicated in all plots as error bounds.

\begin{figure*}
	\centering%
	\begin{minipage}[t]{.48\linewidth}
		\includegraphics[width=\linewidth]{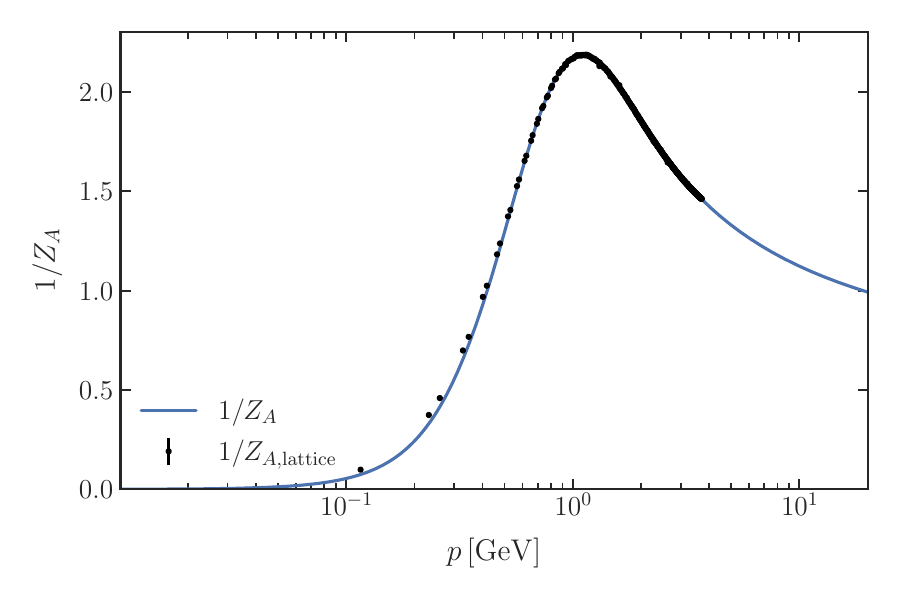}
		\caption{Dressing $1/Z_A(p) = p^2 G_A(p)$ in \labelcref{eq:GluonDressing} of the scalar part of the gluon propagator in the Landau gauge in comparison to lattice computations for $N_f = 2+1$ flavours from \cite{Zafeiropoulos:2019flq}. The systematic error estimate is that obtained by varying the regulator, see \Cref{sec:RegulatorErrors}. It is indicated as an error band but is barely discernible. 
			\hspace*{\fill}}
		\label{fig:GlueProp}
	\end{minipage}%
	\hspace{0.03\linewidth}%
	\begin{minipage}[t]{.48\linewidth}
		\includegraphics[width=1.02\linewidth]{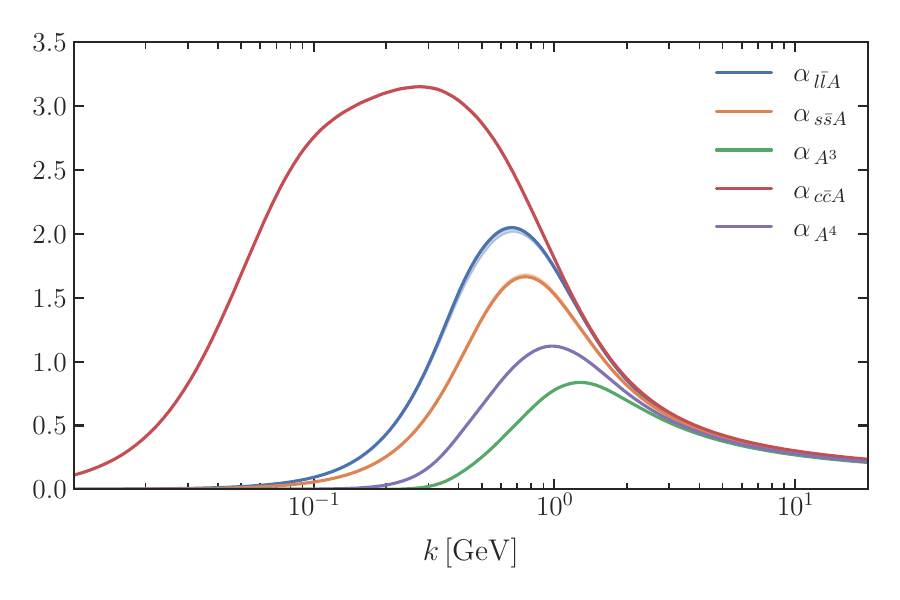}
		\caption{RG-scale dependence of the strong couplings $\alpha_{i,k}$ for $N_f=2+1$ flavours. These couplings are the RG-invariant strengths of the one-gluon exchange diagrams as shown in \Cref{fig:LegoBlockglue} that control and drive the glue dynamics. 		
			We show also a double-logarithmic plot of the strong couplings in \Cref{fig:alphasZoomed}, where the behaviour at small $k$ is clearly visible -- in particular the zero crossing of the three-gluon coupling.	
			\hspace*{\fill}
		}
		\label{fig:alpha}
	\end{minipage}
\end{figure*}
%
\subsection{Gluonic correlation functions}
\label{sec:GlueResults}

We start our discussion of the results with the gluonic correlation functions: The result for the 2+1 flavour gluon propagator is evaluated in \Cref{sec:ResultsGluonProp} concerning two aspects, its agreement with other quantitative results from functional approaches and lattice simulations, and its confirmation of the absolute momentum scale which is taken over from the two-flavour results of the two-flavour input propagator. In \Cref{sec:ResultsAvatars} we compare the avatars of the strong coupling and discuss the decoupling of the glue sector from the low-energy dynamics for $k\lesssim \qty{1}{\GeV}$.

\subsubsection{Gluon propagator}
\label{sec:ResultsGluonProp}

To compute the gluon and ghost propagators in $N_f=~2+1$ flavours, we apply the \LEGO-principle put forward in \Cref{sec:Lego}: Given that the contribution of the strange quark to the gluonic dynamics is at most subleading, we simply add the strange quark loop to the RG-flow without feedback into the two-flavour part. 
As discussed in \Cref{sec:GluonGhost}, it is the gluon dressing \labelcref{eq:GluonDressing}, that carries the dynamics and it depicted. A comparison of the respective result for the gluon dressing with corresponding lattice data in $N_f = 2+1$ is depicted in \Cref{fig:GlueProp}. The excellent agreement between both solutions affirms our approach, and has been also previously observed in \cite{Fu:2019hdw}. We point out the small error band on the fRG data due to the regulator dependences, see \Cref{sec:regulators} and \Cref{sec:RegulatorErrors}. The lattice data is continuum-extrapolated and taken from \cite{Zafeiropoulos:2019flq}.

Given that our computation lacks the full momentum dependence of the glue sector, we identify $p = k$ in the fRG data for comparison. Physical scales are introduced to the system by aligning the peak of the gluon dressing \labelcref{eq:GluonDressing} with the lattice data. Away from the peak, the momentum dependence of the gluon dressing matches well with the lattice data. The minor deviations in the deep infrared are well understood and originate mostly from the non-perturbative gauge-fixing~\cite{Cyrol:2016tym, Cyrol:2017ewj, Maas:2019ggf, Pawlowski:2022oyq}. 

\begin{figure*}
	\centering%
	\begin{minipage}[b]{.48\linewidth}%
		\begin{subfigure}[t]{\linewidth}%
			\includegraphics[width=1.0\linewidth]{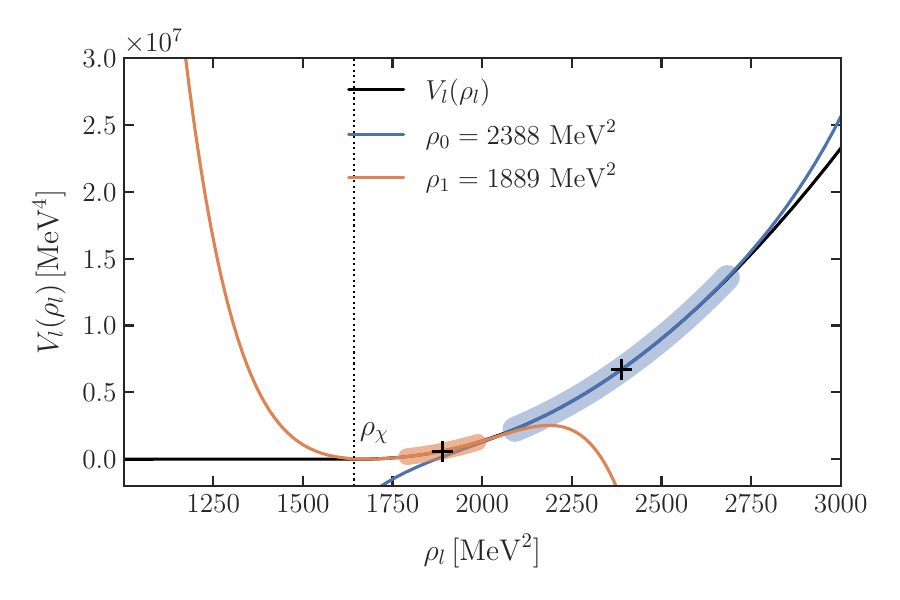}%
			\caption{Full effective potential $V_l(\rho_l)$ and Taylor expansion potentials of order $N=5$ for two expansion points: the expectation value $\rho_0$ at the physical point and one close to the chiral limit, $\rho_1 = \rho_\chi\cdot1.15$, including their validity regime, see also \Cref{fig:taylor_rad}. 
				\hspace*{\fill}
			}
			\label{fig:MesonPot}
		\end{subfigure}%
	\end{minipage}
	\hspace{0.03\linewidth}%
	\begin{minipage}[b]{.48\linewidth}
		\begin{subfigure}[t]{\linewidth}
			\includegraphics[width=\linewidth]{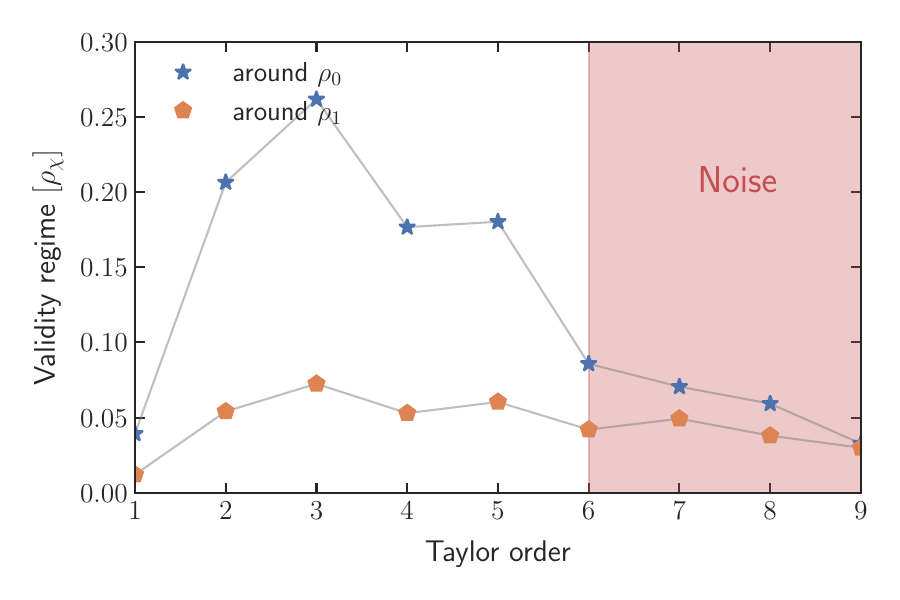}
			\caption{Convergence properties of the Taylor expansion. The validity regime has been defined by the deviation of the Taylor expansion from the full result by less than $2\%$. 
				For larger $N$ numerical noise makes it difficult to define a Taylor expansion within our setup. 
				\hspace*{\fill}
			}
			\label{fig:taylor_rad}
		\end{subfigure}%
	\end{minipage}
	\caption{Full effective potential of mesonic scatterings $V_l(\rho_l)$ as a function of the light condensate $\rho_l$ \labelcref{eq:rho-rhos}. In \Cref{fig:MesonPot} it is shown for $\rho_l$'s including the expectation value in the chiral limit, $\rho_\chi$ and the physical expectation value. This figure also shows the Taylor expansion potentials of order $N=5$ and their validity regimes for two expansion points: the expectation value $\rho_0$ at the physical point and one close to the chiral limit, $\rho_1 = \rho_\chi\cdot1.15$. The validity regimes are estimated by the interval, for which the Taylor expansion deviates less than $2\%$ from the full result. In \Cref{fig:taylor_rad} we show the size of this validity regime for different $N\leq 10$, indicating the radius of convergence. In \Cref{tab:TaylorValues}, we provide explicitly the coefficients of the expansion around $\rho_0$ shown above.\hspace*{\fill}}
	\label{fig:MesonPot+ConvergenceTaylor}
\end{figure*}
%

\subsubsection{Avatars of the strong coupling and glue decoupling }
\label{sec:ResultsAvatars}

We show the RG-scale running of the strong couplings in \Cref{fig:alpha}. Again, we find a remarkably narrow error band on the RG-scale dependence. In the RG-invariant formulation, as introduced in \Cref{sec:RG-invariant expansion}, these couplings are avatars of the strong interaction and can be directly linked to the importance of corresponding diagrams.
Within the present approximation, all avatars of the strong coupling are computed self-consistently.

In comparison to prior studies \cite{Braun:2014ata, Fu:2019hdw}, we have lifted most of the approximations concerning the gluonic sector. In particular, the self-consistent computation of the four-gluon coupling $g_{A^4}$ is a novelty of the present setup. Its relative agreement with $g_{A^3}$, which also does not directly couple to the matter sector, corroborates this approximation on $g_{A^4}$ made in \cite{Braun:2014ata, Fu:2019hdw}.

\subsection{\LEGO-principle at work}
\label{sec:Legoatwork}

Our results provide a detailed insight in the \LEGO-principle at low energies, which is a the heart of the \LEGO-expansion. The building blocks of the current RG-invariant expansion scheme are the RG-invariant dressings $\bar\lambda^{(i)}_n(p_1,...p_n)$ and the RG-invariant normalised propagators $\bar G_{\Phi_i}(p)$. The latter are split into their scalar parts $\bar G_{i,s}(p)$ with $i=A,c,\bar c , q,\bar q, \phi$, \labelcref{eq:ScalarProp}, and the respective tensor structure, \labelcref{eq:DefofProps}. The relative strength of diagrams depends on the loop momentum integral over products of these building blocks and the combinatorial factors computed from the contraction of the product of the tensor structures of vertices and propagators. 

The systematic error estimate of the \LEGO-expansion in QCD is based on the fact, that QCD has two well-separated subsystems: the pure glue system and the matter sector, which includes the emergent scalar-pseudoscalar mesonic fields, constituting the quark-meson system. The interface of these subsystems are only a few connecting diagrams (the \LEGO studs) in a reduced set of flows of propagators and vertices in either sector. Moreover, the flow of the entire system is dominated by that of the glue sector for momenta and cutoff scales $k,p\gtrsim \qty{1.5}{\GeV}$. In turn, for $k,p\lesssim \qty{0.5}{\GeV}$ the matter, or rather the pion sector, is dominating the flows. Hence, the convergence properties and systematic error estimate of the subsystems can be done separately, taking the impact of the respective other sector into account by evaluating the potential influence of the connecting diagrams, the diagrammatic studs.

\subsubsection{QCD Subsystems} 
\label{sec:QCDsubsystems}

Accordingly, the key to the convergence and systematic error analysis of functional QCD is the evaluation of these diagrammatic studs built up from the interface couplings. Therefore, we separate the subsystems for analysis and collect the couplings in three sets: the glue self-couplings $\{\lambda_\textrm{glue}\}$, the matter self-couplings $\{\lambda_\textrm{mat}\}$, and the interface couplings $\{\lambda_\textrm{inter}\}$.

The dynamics of the pure glue system dominates the flows and hence the dynamics of QCD for $k\gtrsim \qty{1.5}{\GeV}$. The set of self-couplings in the present approximation are the pure glue avatars of the strong coupling, 
\begin{align} 
	\{\lambda_\textrm{glue}\}	=\left\{ \alpha_{A^3}\,,\,\alpha_{A^4}\,,\, \alpha_{c \bar c A}\right\}\,. 
	\label{eq:lambdaGlue}
\end{align} 
see \labelcref{eq:StrongCouplings} and \Cref{fig:alpha}. Further self-couplings and the convergence of the subsystem have been considered in particular in \cite{Mitter:2014wpa, Cyrol:2016tym, Cyrol:2017qkl, Corell:2018yil}. 

For cutoff scales $k\lesssim \qty{1.5}{\GeV}$, the dynamics of the QCD system is successively taken over by that of the matter subsystem, which fully dominates for scales $k\lesssim \qty{0.5}{\GeV}$, naturally settling in the regime described by $\chi$PT. The set of self-couplings in the matter sector in the present approximation is given by 
\begin{align} 
	\{\lambda_\textrm{mat}\}	=\left\{h_\phi(\rho_0)\,,\, \lambda_{\phi,n}(\rho_0) \right\}\,, \quad \lambda_{\phi,n}(\rho) = \partial_\rho^n V(\rho) \,. 
	\label{eq:lambdaMatter}
\end{align} 
The set of interface couplings $\{\lambda_\textrm{inter}\} $ of the glue subsystems to the matter subsystem in the present approximation only contains the quark-gluon coupling, 
\begin{align}
	\{\lambda_\textrm{inter}\} = \left\{\alpha_{l\bar l A}\right\}\,,
	\label{eq:lambdaInter}
\end{align}
which was defined in \labelcref{eq:StrongCouplingQuarkGluon}.

To wrap up, \labelcref{eq:lambdaGlue,eq:lambdaMatter,eq:lambdaInter} classify all couplings considered in the present approximation into the three subsystems of glue, matter and interface couplings.

\subsubsection{Effective potential of the chiral order parameter}
\label{sec:EffPot} 

The subset $\{ \lambda_{\phi,n} \}$ is simply the set of all meson self-scattering events contained in the full effective meson potential in QCD, which is taken fully into account for the first time in the present work. 
\begin{table}[b]
	\begin{tabular}{|>{\centering}m{0.3\linewidth}|>{\centering\arraybackslash}m{0.25\linewidth} |}
		\hline &\\[-1ex]
		Coefficient $c_n$ & Value[MeV$^{2n}$] \\[1ex]
		\hline&   \\[-1ex] 
		$c_1$ & $1.9046\cdot10^{5}$\\[1ex]
		\hline& \\[-1ex]
		$c_2$ & $27.148$\\[1ex]
		\hline& \\[-1ex]
		$c_3$ & $-3.1020\cdot10^{-3}$\\[1ex]
		\hline& \\[-1ex]
		$c_4$ & $7.8183\cdot10^{-5}$\\[1ex]
		\hline& \\[-1ex]
		$c_5$ & $2.5407\cdot10^{-6}$\\[1ex] 
		\hline
	\end{tabular}
	\caption{Coefficients of the 5th order Taylor expansion around $\rho_0 =  \qty{2245.7}{\MeV^2}$ as shown also in \Cref{fig:MesonPot+ConvergenceTaylor}.\hspace*{\fill}}
	\label{tab:TaylorValues}
\end{table}

A fully field-dependent effective potential of mesonic scattering processes $V(\rho)$ allows unprecedented quantitative access to observables in the chiral limit or critical regions. Moreover, we can access the convergence radius of Taylor expansions in the fRG approach. This is a significant methodological improvement in comparison to the proceeding works \cite{Braun:2014ata, Fu:2019hdw}, which can be also used for reliability estimates of these expansions. 

\Cref{fig:MesonPot} depicts the result for the full effective potential of the light meson condensate, $V_l(\rho_l)$. Additionally, we indicate two Taylor expansions for a fixed expansion order $N=5$ at the physical point and close to the chiral limit. These are obtained by straight-forwardly taking high-order numerical derivatives of our results at the expansion points. The regime in which their relative error to the full potential is $<2\%$ is also highlighted.

Of course, the validity regime defined by a $2\%$ deviation depends on the order of expansion. We illustrate this in \Cref{fig:taylor_rad}. As can be seen, for 
\begin{align}
	N_\textrm{min}^\textrm{phys} = 5\,,\qquad N_\textrm{min}^\chi= 4\,, 
\end{align}
the expansions are able to fully capture the possible range of the effective potential while maintaining precision at the expansion point. Moreover, the convergence radius shrinks with the distance to the chiral EoM, where it vanishes. We provide our coefficients of a fifth-order Taylor expansion around the physical minimum in \Cref{tab:TaylorValues}. For completeness, we also show the time evolution of the effective potential in \Cref{fig:timeevol} in \Cref{fig:ZQkvsp}.

The rapid convergence of the respective Taylor expansion in vacuum QCD at the physical point has been tested in \cite{Mitter:2014wpa, Braun:2014ata, Rennecke:2015eba, Cyrol:2017ewj, Fu:2019hdw}, that of the Taylor expansion of $h_l(\rho)$ has been tested in vacuum QCD in \cite{Mitter:2014wpa, Braun:2014ata, Rennecke:2015eba, Cyrol:2017ewj}, a comprehensive study of the double convergence in low-energy QCD effective theories in the phase structure of QCD has been put forward in \cite{Pawlowski:2014zaa}. 

The inclusion of the latter work within an effective theory draws already from the \LEGO-principle: The convergence properties and systematic error estimate of an expansion in mesonic self-interactions can safely be done within the quark-meson sector without considering full QCD. The flow of the full mesonic effective potential in QCD only couples indirectly to the glue sector and its impact is controlled by the interface, the diagrammatic studs. This allows us to rely on systematic error estimates and the evaluation of convergence properties in low-energy effective theories with functional approaches, for a compilation of low-energy QCD fRG-studies see \cite{Dupuis:2020fhh}.

\begin{figure}
	\centering
	\includegraphics[width=0.5\textwidth]{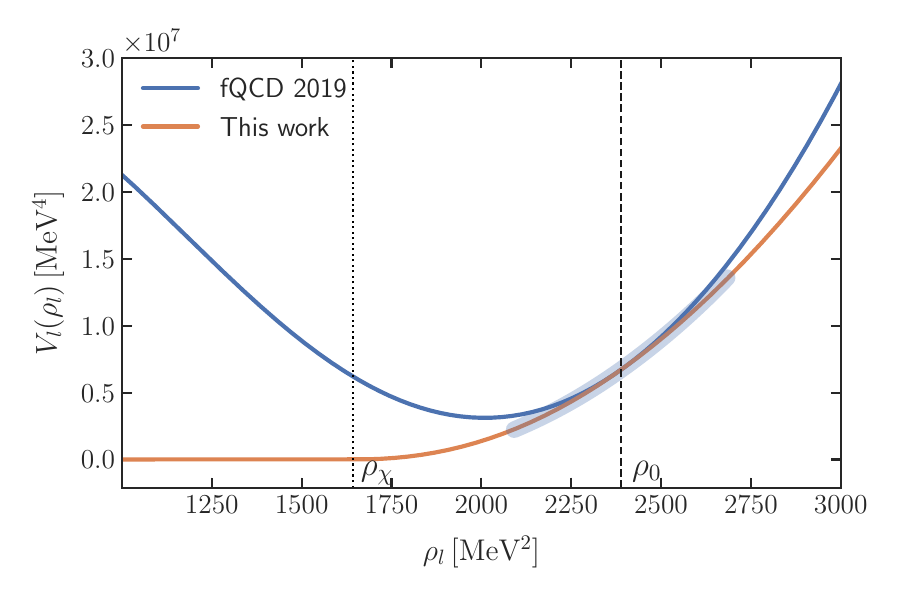}
	\caption{Effective potential $V_l(\rho_l)$ of the light quark condensate (orange) in comparison to the Taylor expansion potential from \cite{Fu:2019hdw} (blue). The $\rho_l$-range includes the expectation value in the chiral limit $\rho_\chi$ (dotted vertical line), as well as the expectation value at the physical point $\rho_0$ (dashed vertical line). We also indicate the validity regimes of the Taylor expansions as determined in \Cref{fig:MesonPot} by transparent blue bands. We emphasise that the validity regime around $\rho_\chi$ vanishes, while that around the physical point neither incorporates $\rho_\chi$ nor $\rho_{s,0}$. \hspace*{\fill}
	}
	\label{fig:VeffComparison2019}
\end{figure}

In \Cref{fig:VeffComparison2019} we compare our full potential to the Taylor-expanded result from \cite{Fu:2019hdw}. Here, it can be seen that our estimate for the validity of the expansion fits quite well. In particular, the potential is almost perfectly identical around the physical EoM. However, the Taylor-expansion deviates slightly faster than our validity estimate would suggest. This can be explained by the reduced precision due to a running expansion point and other errors picked up during the flow.
Clearly, the Taylor expansion can also not be used to extrapolate to the chiral limit, as has been already indicated in \Cref{fig:MesonPot+ConvergenceTaylor}.
\begin{figure*}
	\centering
	\begin{minipage}[b]{.48\linewidth}
		\begin{subfigure}[t]{\linewidth}
			\centering
			\includegraphics[width=1.\linewidth]{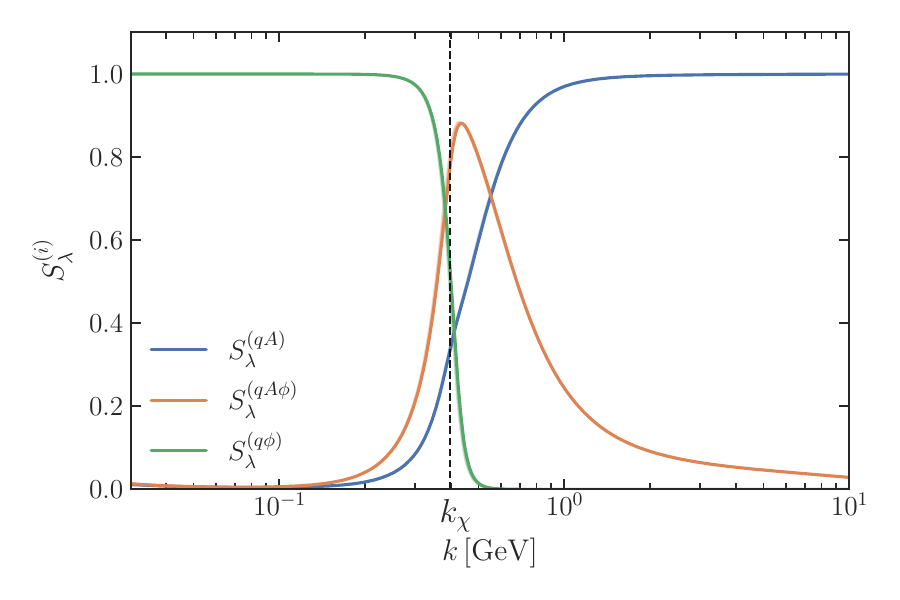}
			\subcaption{Relative strengths $S_\lambda^{(i)}$ with $i=qA,qA\phi,q\phi$, defined in \labelcref{eq:Strength}, of the diagrammatic contributions to the four-quark flow in the scalar-pseudoscalar channel. \hspace*{\fill}}
			\label{fig:Slambdaq}
		\end{subfigure}%
	\end{minipage}
	\hspace{0.02\linewidth}%
	\begin{minipage}[b]{.48\linewidth}
		\begin{subfigure}[t]{\linewidth}
			\centering
			\includegraphics[width=1.0\linewidth]{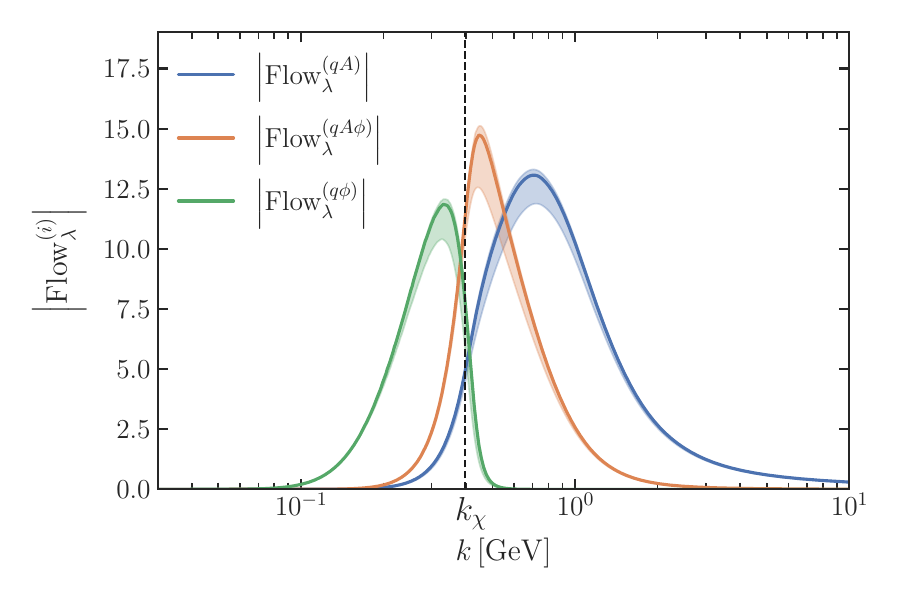}
			\subcaption{Absolute strengths $\left |\textrm{Flow}^{(i)}_{\lambda}\right|$ of the diagrammatic contributions to the scalar-pseudoscalar four-quark channel. \hspace*{\fill}}
			\label{fig:lambdaq}
		\end{subfigure}%
	\end{minipage}
	\caption{Strengths of the diagrammatic contributions to the flow of the $t$-channel dressing of the scalar-pseudoscalar four-quark dressing: quark-gluon box ($qA$), mixed quark-gluon-meson box, ($qA\phi$), and quark-meson box ($q\phi$). The error-bands indicate the systematic error incurred by the regulator-dependence. For a discussion see \Cref{sec:RegulatorErrors}. For the resulting strength of he coupling, see also \Cref{fig:lambda} in \Cref{app:plots}. \hspace*{\fill}}
	\label{fig:mixdiags}
\end{figure*}
%

\subsubsection{Interface}
\label{sec:Interface}

As mentioned before, the set of interface couplings \labelcref{eq:lambdaInter} only contains the quark-gluon coupling $\alpha_{l\bar l A}$. A diagrammatic depiction of its flow is found in \Cref{fig:diag_quark_gluon_3pt} in \Cref{app:flows}. Further interface couplings and their importance for gauge consistency have been considered in particular in \cite{Cyrol:2017ewj}. For $p\gtrsim \qty{5}{\GeV}$, the value of $\alpha_{l\bar l A}$ agrees with that of the other avatars of the strong coupling. On the other hand, for smaller values, all the avatars start to differ due to non-trivial terms in the respective STIs. At low momenta and cutoff scales, all avatars of the strong coupling decay and the glue sector decouples. 

The diagrammatic studs of the two subsystems in the present approximation enter the glue subsystem via the flow of the dressing of the gluon propagator and the pure glue couplings in terms of a quark loop with external gluons, see \Cref{fig:diag_gluon_3pt,fig:diag_gluon_2pt} in \Cref{app:flows}. The diagrammatic studs in the present approximation enter the matter subsystem via the flowing constraint for the scalar-pseudoscalar channel of the four-quark vertex, \Cref{fig:diag_quark_4pt} in \Cref{app:flows}. These are the flow of the Yukawa coupling and the wave function of the quark propagator, both derived from \Cref{fig:diag_quark_2pt} in \Cref{app:flows}. 

\begin{figure*}
	\centering
	\begin{minipage}[t]{.48\linewidth}
		\begin{subfigure}[t]{\linewidth}
			\centering
			\includegraphics[width=\linewidth]{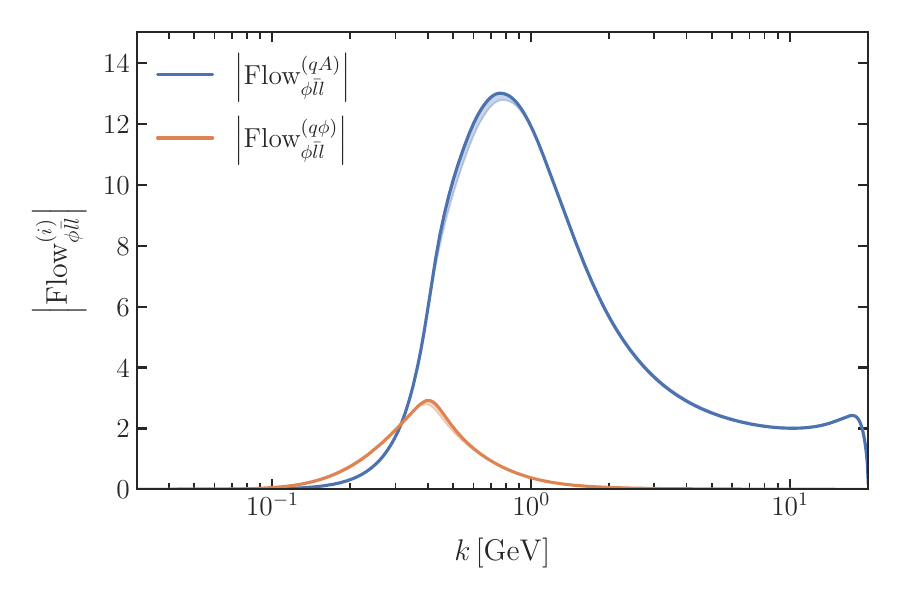}
			\subcaption{Absolute strengths $\left |\textrm{Flow}^{(i)}_{\phi\bar qq}\right|$ of the diagrammatic contributions to the Yukawa coupling $h_\phi$. \hspace*{\fill}}
			\label{fig:Shphi}
		\end{subfigure}%
	\end{minipage}
	\hspace{0.02\linewidth}%
	\begin{minipage}[t]{.48\linewidth}
		\begin{subfigure}[t]{\linewidth}
			\centering
			\includegraphics[width=1.01\linewidth]{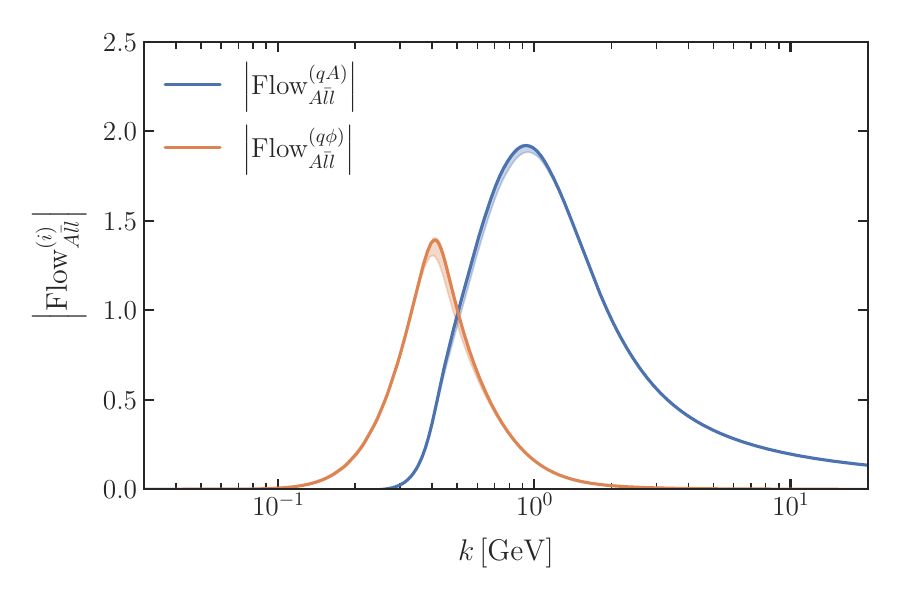}
			\subcaption{Absolute strengths $\left |\textrm{Flow}^{(i)}_{A\bar qq}\right|$ of the diagrammatic contributions to the quark-gluon coupling $g_{\bar q qA}$. \hspace*{\fill}}
			\label{fig:SqbarqA}
		\end{subfigure}
	\end{minipage}
	\caption{Absolute strength of the diagrammatic quark-gluon and quark-meson contributions to the flow of the Yukawa coupling and the quark-gluon coupling. Note that the pertaining diagrams do not have mixed meson-gluon contributions, see \Cref{fig:diag_quark_2pt} and \Cref{fig:diag_quark_gluon_3pt} in \Cref{app:flows}. The error-bands indicate the systematic error incurred by the regulator-dependence. For a discussion see \Cref{sec:RegulatorErrors}. \hspace*{\fill}}
	\label{fig:couplingflows}
\end{figure*}

One key information for the present analysis is the relative strength of the contributions to the diagrammatic part of the flow $\textrm{Flow}_{\lambda}$ of a given dressing $\lambda\in\{\lambda_\textrm{glue}, \lambda_{\textrm{mat}},\lambda_\textrm{Inter}\}$ from the sets \labelcref{eq:lambdaGlue,eq:lambdaMatter,eq:lambdaInter}. These different parts are labelled by $\textrm{Flow}^{(qA)}_{\lambda}$ (quark-gluon diagrams), $\textrm{Flow}^{(q\phi)}_{\lambda}$, (pure matter sector) and $\textrm{Flow}^{(qA\phi)}_{\lambda}$, (mixed). Their relative strength is then measured by 
\begin{align}
	{\cal S}_\lambda^{(i)} = \frac{\left |\textrm{Flow}^{(i)}_{\lambda}\right|}{\sqrt{ \sum_j \left(\textrm{Flow}^{(j)}_{\lambda}\right)^2}}\,,\qquad i,j= qA\,,\, q\phi\,,\,qA\phi\,,
	\label{eq:Strength}
\end{align} 
and is depicted in \Cref{fig:mixdiags,fig:couplingflows} for $\lambda=\lambda_q, h_\phi, \alpha_{q\bar q A}$. 

The second key information is the absolute strength of the flow at a given cutoff scale $k$: if the combined flow of a dressing, measured in the dressing itself, tends towards zero, it may be dominated by one of the contributions above, but is irrelevant. 

We put this to work within an evaluation of the flow of the scalar-pseudoscalar channel of the four-quark vertex depicted in \Cref{fig:diag_quark_4pt} in \Cref{app:flows}, see also \cite{Fu:2019hdw}. The different building blocks are the quark, gluon and meson propagator, and the couplings $\alpha_{q\bar q A}, h_\phi$. The full analysis necessarily also includes that of the quark-gluon coupling and the Yukawa coupling. However, the Yukawa coupling is quite stable for all cutoff scales $k$. As discussed before, this entails that the momentum-dependence of the scalar-pseudoscalar $t$-channel is well described by the exchange of a quasi-particle with a constant coupling strength. Moreover, this offers a further important example for using QCD information for improving low energy effective theories (LEFTs) towards QCD-assisted LEFTs: Setting up a Quark-Meson model within the LPA' approximation allows us to use a constant RG-invariant Yukawa coupling in a first good approximation to QCD at low scales. In any case, this implies that the full analysis reduces to that of the four-quark flow and the quark-gluon coupling. 

For $k\gtrsim \qty{1.5}{\GeV}$ the quark-gluon box diagrams dominate the flow of the four-quark vertex. Moreover, its strength at a given scale $k\gtrsim \qty{1.5}{\GeV}$ is dominated by the flow at $k$ itself. This stabilises the system considerably as the error proliferation from larger cutoff scales is efficiently decoupled, and the systematic error in this regime is dominated by the quark-gluon coupling. 

We also remark that the closed quark-loop contributions to the gluon propagator and the pure glue couplings are merely shifting the characteristic scale $\Lambda_\textrm{QCD}$ from that in pure Yang-Mills to that in QCD. Thus, the convergence and systematic error estimate of the glue subsystem is the same as in pure Yang-Mills. 

\begin{figure*}
	\centering
	\begin{minipage}[t]{.48\linewidth}
		\includegraphics[width=\linewidth]{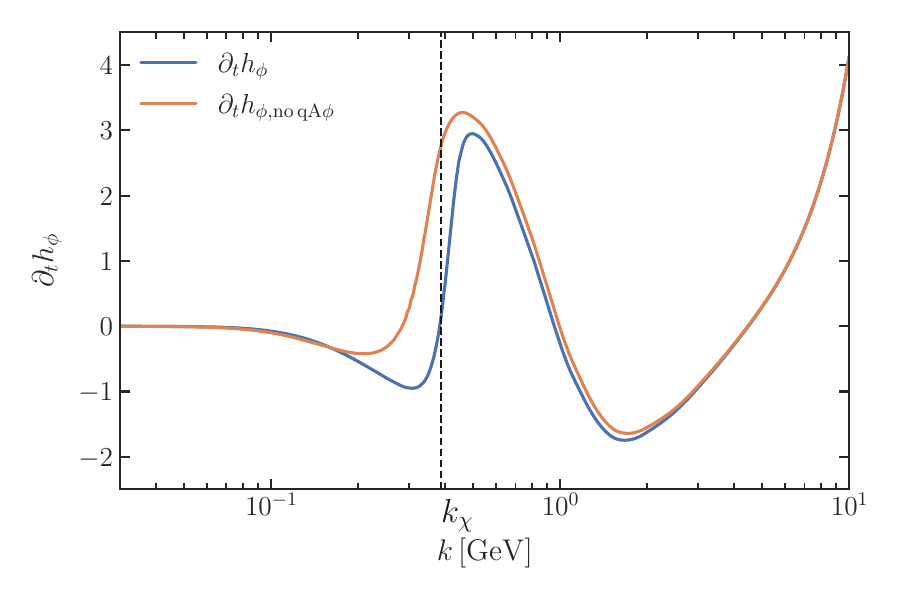}
		\caption{The full flow of $h_\phi$ with and without the mixed quark-gluon-meson diagrams. Note the sizeable discrepancy arising around the chiral symmetry breaking scale $k_\chi$ due to the absence of the mixed quark-gluon-meson diagrams in the flow of the scalar-pseudoscalar four-quark coupling. See also \Cref{fig:hphi} in \Cref{app:plots} for the resulting $h_\phi$.\hspace*{\fill}}%
		\label{fig:TransitionReg}%
	\end{minipage}
	\hspace{0.02\linewidth}%
	\begin{minipage}[t]{.48\linewidth}
		\centering%
		\includegraphics[width=\linewidth]{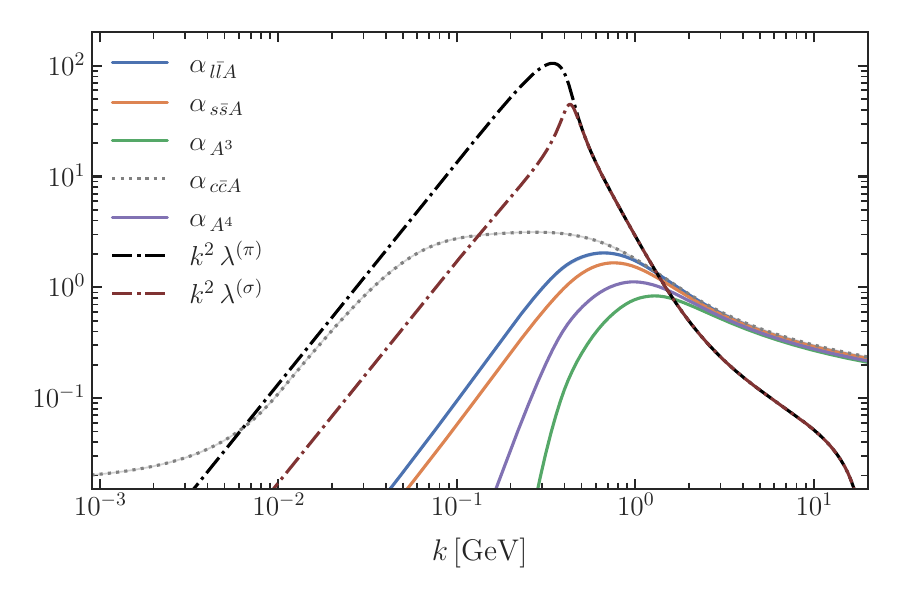}%
		\caption{RG-scale dependence of the strong couplings $\alpha_{i,k}$ for $N_f=2+1$ flavours on a double-logarithmic scale in comparison to the running of the parts of the scalar-pseudoscalar four-fermi channel $\lambda^{(\pi)} = {h_\phi^2}/{2 G_\pi}$ and $\lambda^{(\sigma)}~=~{h_\phi^2}/{2 G_\sigma}$.\hspace*{\fill}}
	\end{minipage}
\end{figure*}
In the regime $\qty{0.5}{\GeV} \lesssim k \lesssim \qty{1.5}{\GeV}$, the different flow contributions to the four-quark flow have a similar size, with the quark-meson contributions successively taking over the dynamics, see \Cref{fig:Slambdaq}. The same analysis holds true for the Yukawa coupling and the quark-gluon coupling, see \Cref{fig:couplingflows}. However, contributions to the matter sector from the quark-gluon coupling decouple rapidly and are irrelevant for the systematic error. 

\begin{figure*}
	\centering
	\begin{minipage}[t]{.48\linewidth}
		\begin{subfigure}{\linewidth}
			\centering
			\includegraphics[width=\linewidth]{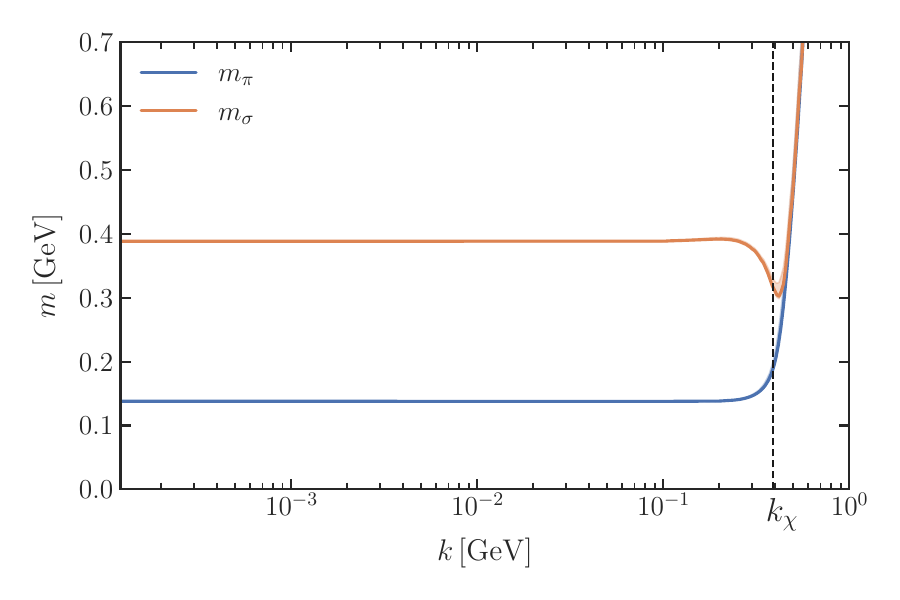}
			\subcaption{Running of the meson masses on the physical EoM. \hspace*{\fill}}
			\label{fig:mesonmassphys}
		\end{subfigure}%
	\end{minipage}
	\hspace{0.02\linewidth}%
	\begin{minipage}[t]{.48\linewidth}
		\begin{subfigure}{\linewidth}
			\centering
			\includegraphics[width=\linewidth]{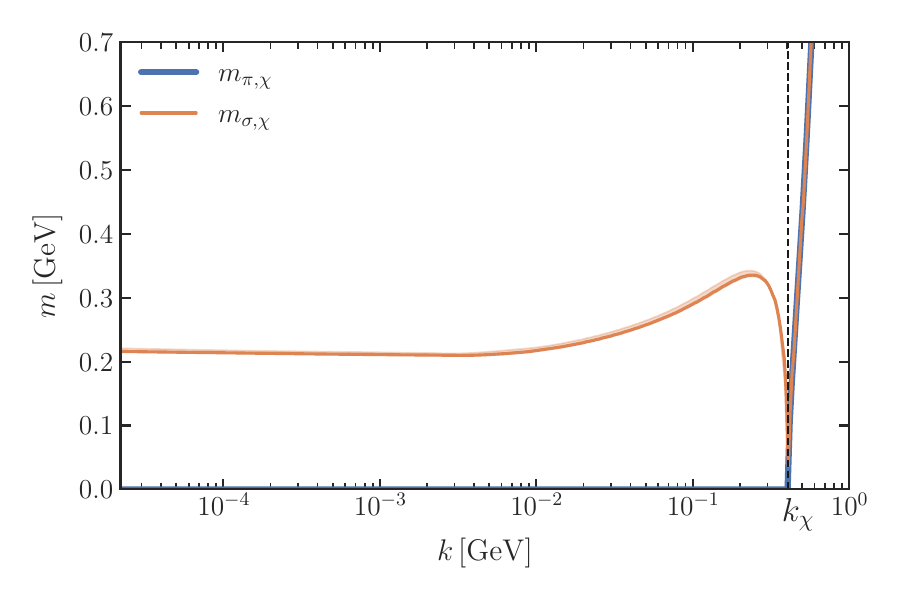}
			\subcaption{Running of the meson masses on the chiral EoM. \hspace*{\fill}}
			\label{fig:mesonMasschiral}
		\end{subfigure}
	\end{minipage}
	\caption{RG-scale dependence of mesonic masses for $N_f = 2+1$. There values at $k=0$ are collected in \Cref{tab:results,tab:resultsChiral}. We have indicated the chiral symmetry breaking scale $k_\chi$, which given in \labelcref{eq:ChiralSymmbScale}. The error-bands indicate the systematic error incurred by the regulator-dependence. For a discussion see \Cref{sec:RegulatorErrors}.\hspace*{\fill}}
	\label{fig:mesonMass}
\end{figure*}
For the explicit evolution of the Yukawa coupling $h_\phi$ and the associated four-quark vertex see also \Cref{fig:hphi} and \Cref{fig:lambda} in \Cref{app:plots}, where these are explicitly shown.

We conclude from the above discussion that the total systematic error of the system is well estimated by the combined systematic errors of the quark-gluon sector at high energies for $k\gtrsim \qty{1.5}{\GeV}$, of the low-energy sector of QCD for $k\lesssim \qty{0.5}{\GeV}$ and of the transition regime for $\qty{0.5}{\GeV} \lesssim k \lesssim \qty{1.5}{\GeV}$. 

The high- and low-energy sectors have been studied thoroughly in the literature. For high energies this has been done with self-consistency evaluations of higher order vertices and with comparisons with other functional approaches -in particular DSE studies- which offer different resummation schemes, as well as perturbative results at asymptotically high energies. In the regime $p\lesssim \qty{4}{\GeV}$, one can also compare the results from functional approaches with lattice results for quark and gluon propagators, as well as a few results for vertices, albeit the lattice results still have large statistical errors. 

On the other hand, the systematic error estimate for the matter sector can be estimated by a combination of that from self-consistency analyses in fRG calculations of low-energy effective theories, see \cite{Dupuis:2020fhh}, benchmark predictions in comparison to experiment, and the fact that the present approach converges towards $\chi$PT for $k\to 0$, giving access to the systematic error estimates of $\chi$PT. 

Together, this leads to a combined conservative systematic error estimate of less than 5\% from the high and low-energy sector of QCD for the QCD correlation functions considered here. The largest part of the systematic error estimate stems from the interface regime with $\qty{0.5}{\GeV} \lesssim k \lesssim \qty{1.5}{\GeV}$: Here we can only draw from the self-consistency estimate discussed in \Cref{sec:ApparentConvergence}, as neither lattice simulations nor other functional approaches offer a direct comparison. 
In this setup, we do not calculate the full interface, which contains further quark-gluon interactions associated with non-classical tensor structures. This expansion has been shown to be stable under the inclusion of higher order contributions, \cite{Mitter:2014wpa, Braun:2014ata, Rennecke:2015eba, Cyrol:2017ewj}.
Here, we use a small enhancement factor to estimate the impact of these further interface studs. Although this enhancement is small, our results are sensitive to the exact value thereof, which leads us to estimate the overall systematic error to be around 10\%, and part of the remaining error is accommodated by the adjustment of the current quark masses and the pion pole mass. 

This concludes our evaluation of the systematic error estimate in the present approximation to QCD. In summary we arrive at a conservative systematic error estimate of 10\% for the present functional first principle study of vacuum QCD, being dominated by the interface error.

\subsection{Momentum dependences in the matter sector}
\label{sec:ResultsMatterSector}

The present setup of the matter sector contains a full mesonic field dependence and also gives access to fully momentum-dependent propagators. Even though the momentum dependence of propagators is extracted from the data and not fed back, this it allows for a direct access to the pole masses and the pion decay constant. 
Furthermore, the resulting full momentum dependence serves as a important consistency check for both the current and previously employed approximation schemes of the effective action.

\subsubsection{Meson masses}
\label{sec:MesonMasses}
\begin{figure*}
	\centering
	\begin{minipage}[t]{.48\linewidth}
		\begin{subfigure}[t]{\linewidth}
			\includegraphics[width=\linewidth]{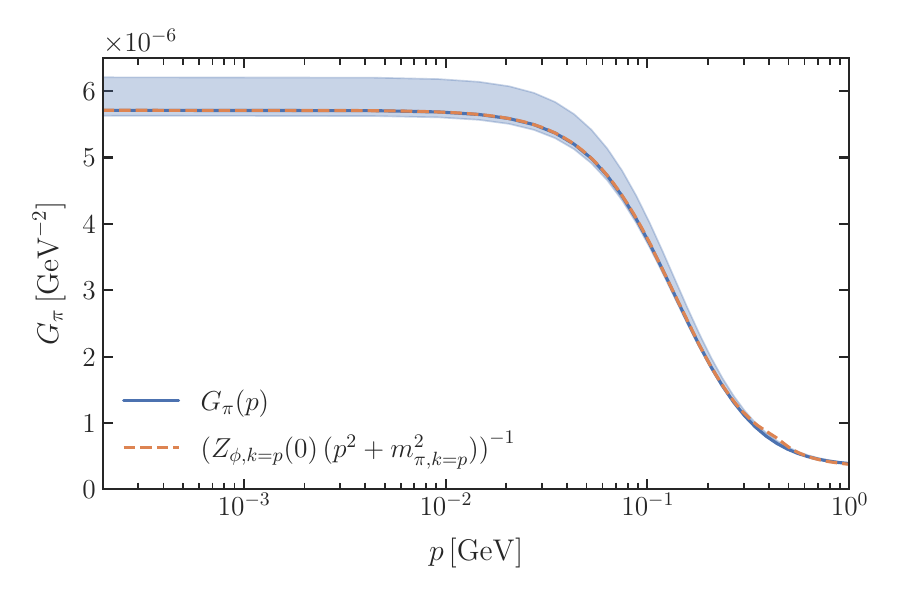}
			\caption{Fully momentum-dependent pion propagator for $N_f=2+1$, compared to the RG-scale-dependent propagator as defined in \labelcref{eq:Gpi}. Whilst the \textit{blue} curve shows a genuine momentum dependence, the \textit{orange} curve shows a RG-scale dependence which is identified as $k=p$. \hspace*{\fill}}
			\label{fig:Gpi}
		\end{subfigure}%
	\end{minipage}
	\hspace{0.02\linewidth}%
	\begin{minipage}[t]{.48\linewidth}
		\begin{subfigure}[t]{\linewidth}
			\centering
			\includegraphics[width=\linewidth]{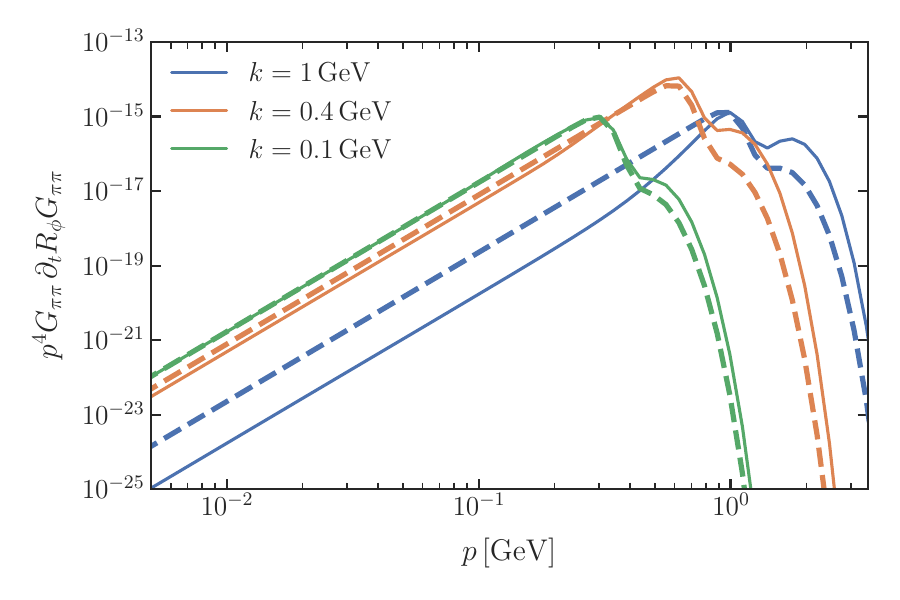}
			\subcaption{Comparison of the loop-term $G_{\pi,k}(p)\,\partial_t R_{\phi,k}(p)$ (solid lines) with the approximation we use in our loops, 
				$\frac{1}{Z_{\phi,k}} \, \frac{\partial_t R_{\phi,k}(p)}{p^2 + m_{\pi,k}^2 +R_{\phi,k}(p)}$ (dashed lines) at three different cutoff-scales $k = {(1,\,0.4,\,0.1)}\unit{\GeV}$.
				The region around the respective $k$ contributes by far the most to the loop integrals due to higher scales being suppressed by $\partial_t R_\phi$ and lower scales due to the four-dimensional integral over loop momenta. As such, we see excellent agreement within the relevant region.
				\hspace*{\fill}}
			\label{fig:Gpiloopapprox}
		\end{subfigure}
	\end{minipage}
	\caption{RG-scale dependence of mesonic masses for $N_f = 2+1$. We have indicated the chiral symmetry breaking scale $k_\chi$ provided in \labelcref{eq:ChiralSymmbScale}. The error-bands indicate the systematic error incurred by the regulator-dependence. For a discussion see \Cref{sec:RegulatorErrors}.\hspace*{\fill}}
	\label{fig:PionProp}
\end{figure*}
The computation of the full mesonic potential gives direct access to the curvature masses in the matter sector, which are summarised in \Cref{tab:results} and \Cref{tab:resultsChiral}.
Additionally, we show the RG-scale-dependent mesonic masses in \Cref{fig:mesonMass}. This plot emphasises the different precision requirements for calculations in the chiral limit. Whereas the mesonic masses on the physical point saturate at scales below the physical pion mass $ k \lesssim m_\pi = \qty{138}{\MeV}$, this is not the case if we have a truly massless mode, such as in the chiral limit or, importantly, at second order phase transitions such as the critical endpoint.

In view of a future application of the present setup to the resolution of the critical endpoint, as well as for an analysis of the mass spectrum of QCD in the chiral limit, we also briefly discuss the results in the chiral limit. There, the determination of the sigma mass $m_\sigma$ requires an increased precision in field space, and we obtain 
\begin{align}
	m_{\sigma,\chi}= 214.7^{+5.4}_{-9.3}\,\textrm{MeV}	\,,
	\label{eq:msigmachi}
\end{align}
see \Cref{tab:resultsChiral}. We have ascertained that the field discretisation is fine enough to avoid sizeable errors for the (right-sided) derivative of $m_\pi^2$ at the chiral EoM $\rho_\chi$. This has been done by increasing the cell count around the EoM by more than an order of magnitude. At this point, the derivative of $m^2_\pi(\rho)$ jumps sharply at the chiral EoM and an increase in precision does not yield changing results. In summary, we corroborate low energy effective theory results in the literature, see \cite{Dupuis:2020fhh}. Technically, this originates in the successive decoupling of gluons and quarks from the off-shell dynamics, effectively leaving us with the off-shell dynamics of an O(4) model with its well-known mass spectrum. Note that the chiral limit is very sensitive to the dynamical details of anomalous chiral symmetry breaking, for a comprehensive analysis see \cite{Resch:2017vjs, Pisarski:2024esv}, for related considerations see \cite{Zwicky:2023krx}. In summary, the chiral limit deserves further investigation within an improved approximation which is deferred to future work. 

We proceed by investigating the full momentum dependence of the pion propagator $G_\pi(p)$ in \Cref{fig:PionProp}. It can be directly constructed from the curvature mass $m_\pi$ of the pion at vanishing cutoff scale and the integrated wave function $Z_\phi(p)$, defined in \labelcref{eq:Zphip}, with the momentum-dependent anomalous dimension $\eta_{\phi_l}(p)$ in \labelcref{eq:AnomalousDimEasier}.
We are thus led to 
\begin{align}
	G_\pi(p) = \frac{1}{Z_\phi(p)}\frac{1}{p^2+m^2_\pi}\,, 
	\label{eq:Gpi}
\end{align}
at vanishing cutoff scale. The full $\eta_{\phi_l}(p)$ is a result of the flow and we do not fully feed it back into the system, but only $\eta_{\phi_l}(0)$ and $\eta_{\phi_l}(k)$. This is explained in \Cref{sec:Wavefuncts}, where we also discuss the resulting systematic error. In future works we will feed back $\eta_{\phi_l}(p)$ into the flow to obtain a fully self-consistent pion propagator. In any case, we can compare this to an approximation of $G_\pi$, given through the identification of $p=k$,
\begin{align}\label{eq:Gpik}
	\frac{1}{Z_{\phi,k}(0)}\frac{1}{k^2 + m_{\pi,k}^2}\,,
\end{align}
which we show in \Cref{fig:Gpi}. The agreement between the results is quite good, with the largest difference being a bump around $p \approx \qty{500}{\MeV}$.

As indicated above, we use the approximation $\eta_{\phi_l}(p) \approx~\eta_{\phi_l}(p=k)$ for the loop propagators, recall \Cref{sec:Wavefuncts}, instead of the full momentum dependence. We can use the fully momentum-dependent result to verify the quality and consistency of this assumption. 

To this end, we compare the full pion propagator as it enters into the flow of $\Gamma_k[\Phi]$ to the one we use in the loop-momentum integration, i.e.
\begin{align}\nonumber 
	p^4\,G_k(p)&\partial_t R_{\phi,k}(p)G_k(p) \\[1ex]
	&=\frac{1}{Z_{\phi,k}(p)} \, \frac{p^4\partial_t R_{\phi,k}(p)}{(p^2 + m_{\pi,k}^2+R_{\phi,k}(p))^2}\,,
\label{eq:FlowGPi_p}
\end{align}
to
\begin{align}\label{eq:FlowGPi_k}
	\frac{1}{Z_{\phi,k}(k)} \, \frac{p^4\partial_t R_{\phi,k}(p)}{(p^2 + m_{\pi,k}^2+R_{\phi,k}(p))^2}\,,
\end{align}
where $Z_{\phi,k}$ is computed $k$-dependently at $p=k$ using the relations below \labelcref{eq:Zk-p}. The above combination is the main way in which the pion propagator enters the flow equations.
Their good agreement, which is visible in \Cref{fig:Gpiloopapprox}, confirms \labelcref{eq:loopapprox}.

The fully momentum-dependent pion propagator can be used to extract the pion pole mass in the complex momentum plane, where the pole of the propagator is situated at. Using a Pad\'e-fit, we extract a pion pole mass
\begin{align}\label{eq:padePole}
	m^{N_f=2+1}_{\pi,\textrm{pole}}	= \qty{138(9)}{\MeV}\,,
\end{align}
see \Cref{tab:results}. For its determination we took the average of fit results for the orders $N=2,\dotsc,10$ and the error is given by standard deviation of these fits.
Due to the reconstruction process of the propagator, the errors are relatively large. This will be improved on in future work.

The approximate agreement with the curvature mass \labelcref{eq:rawMass} with $m_{\pi,\textrm{curv}}=\qty{138}{\MeV}$ is in line with respective computations in the quark-meson model, \cite{Helmboldt:2014iya}. It also serves as additional support of the methodology employed previously in \cite{Fu:2019hdw} based on \cite{Helmboldt:2014iya}. Furthermore, it is suggestive to identify the curvature mass of the $\sigma$-mode, indicated in \Cref{tab:results}, with the lowest lying pole mass in the scalar sector.

\begin{figure*}
	\centering
	\begin{minipage}[t]{.48\linewidth}
		\begin{subfigure}[t]{\linewidth}
			\centering
			\includegraphics[width=\textwidth]{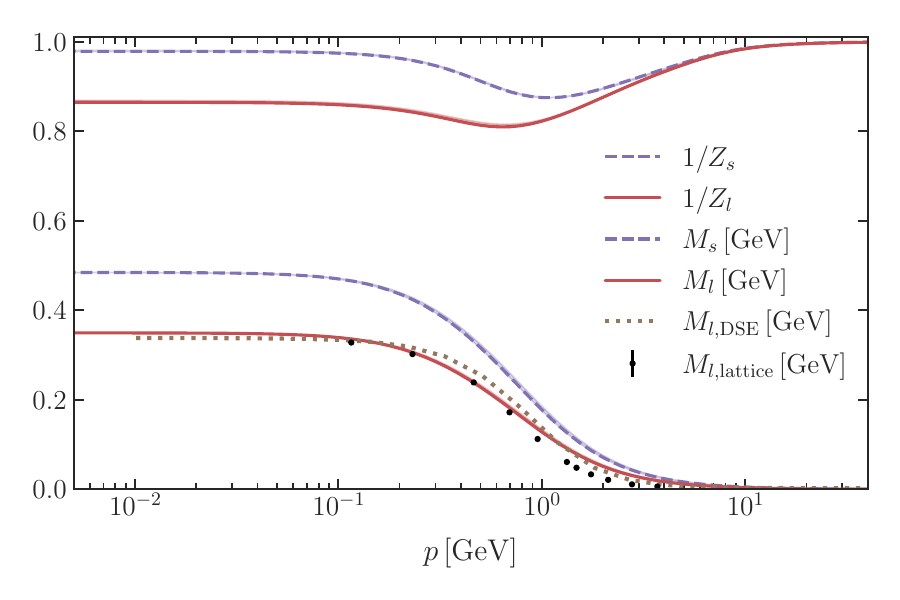}
			\caption{Results for the momentum-dependence of the quark mass function $M_q(p)$ and the quark wave function renormalisation $Z_q(p)$ for both light and strange quarks.
			The DSE data is taken from \cite{Gao:2021wun}, whereas the lattice data is from the lattice study in \cite{Chang:2021vvx}.
			 \vspace{8mm}
			\hspace*{\fill}}
		\label{fig:Mq}
		\end{subfigure}%
	\end{minipage}
	\hspace{0.02\linewidth}%
	\begin{minipage}[t]{.48\linewidth}
		\begin{subfigure}[t]{\linewidth}
			\centering
			\includegraphics[width=1.02\linewidth]{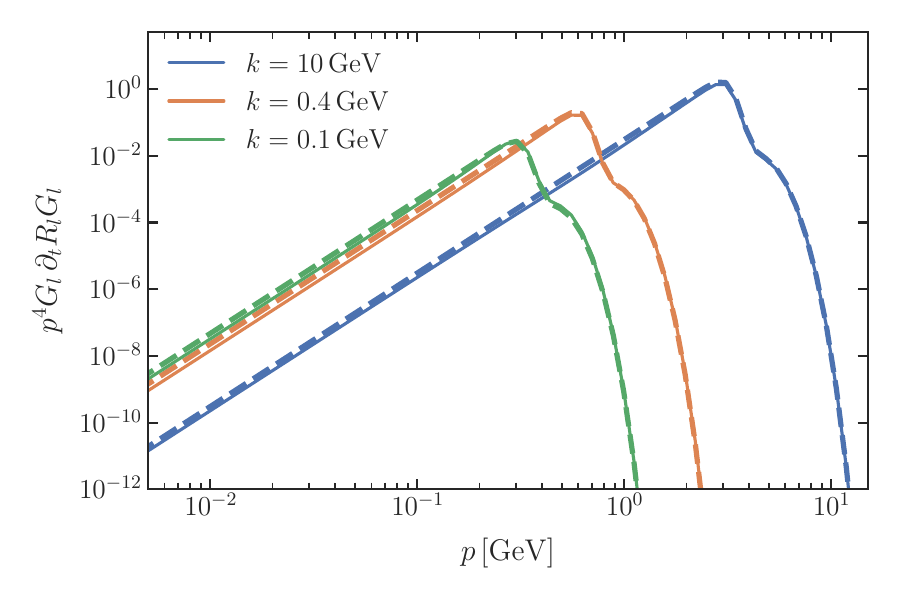}
			\subcaption{Comparison of the loop-term $G_{l,k}(p)\,p\partial_t R_{l,k}(p)$ (solid lines) with the approximation we use in our loops, 
				$\frac{1}{Z_{l,k}} \, \frac{p\partial_t R_{l,k}(p)}{m_{l,k}^2 + \,(p + R_{l,k}(p))^2}$ (dashed lines) at three different cutoff-scales $k = (1,\,0.4,\,0.1)\unit{\GeV}$.
				The region around the respective $k$ contributes by far the most to the loop integrals due to higher scales being suppressed by $p\partial_t R_l$ and lower scales due to the four-dimensional integral over loop momenta. As such, we see excellent agreement within the relevant region.
				\hspace*{\fill}}
			\label{fig:Gqloopapprox}
		\end{subfigure}
	\end{minipage}
	\caption{Momentum dependences of the quark propagator. The error-bands indicate the systematic error incurred by the regulator-dependence. For a discussion see \Cref{sec:RegulatorErrors}. Here, error-bands are so small that they are hardly visible, indicating that the error of the quark mass and wave function has little dependence on the regulator choice.\hspace*{\fill}}
	\label{fig:QuarkMomdep}
\end{figure*}
%

\subsubsection{Quarks and pion decay constant}
\label{sec:resultsfpi}

In the previous Section we found that the different scale-dependent solutions to the mesonic wave function \labelcref{eq:Zatkandp0} and \labelcref{eq:Zphip} yielded very similar results for the momentum-dependent pion propagator, as seen in \Cref{fig:Gpi}. This already entails that our approximation of the meson propagators in loops with $\eta_{\phi_l}(p)\to \eta_{\phi_l}(p=k)$ is quantitatively reliable. 

For the quark wave functions and mass functions, the situation is very different: we discuss this in the chiral limit as this illustrates the underlying properties most lucidly: to begin with, the wave function is only mildly varying between approximately 1 - 1.25 over the full momentum range, and any approximation to the $\eta_q$'s will do. The flow of the mass function, on the other hand, vanishes identically for all cutoff scales $k>k_\chi$ in the chiral limit. For cutoff scales $k< k_\chi$, the flow of the mass function switches on and generates contributions for all momenta, as its flow decays with $1/p^2$ for $p^2 \gg k^2$. This feed back to larger momenta is missing in the local approximation with $M(p) \approx M_{k=\gamma p}(0)$ for all $\gamma$. Indeed, in the chiral limit we have $M_{k=\gamma p}(0) =0$ for $\gamma p>k_\chi$. We illustrate this in \Cref{fig:Mqp-MqkChiralLimit} in \Cref{app:plots}, where $M(p)$ and $M_{k=p}$ are depicted. We close this discussion with the remark that this feedback of momentum contributions is also missing in the scalar case, but there it is a subleading contribution. 

With the above structure in mind, induced by dynamical chiral symmetry breaking, we reconsider the wave function $Z_q$. In the Landau gauge it vanishes identically at one loop order and hence it is approximately zero deep in the perturbative regime. We have discussed in \Cref{sec:QuarkZ}, how this is implemented with gauge consistency in the fRG approach. We conclude that while the flow of the wave function is not vanishing identically, it is approximately zero deep in the perturbative regime due to the Landau gauge. These differences can only play a subleading role in the flow equation, but we still expect sizable deviations of $Z_{k=\gamma k}(0)$ from $Z(p)$ for all $\gamma$.

We depict the momentum-dependent quark dressings $1/Z_{q, k=0}(p)$ with approximate gauge consistency in \Cref{fig:Mq}. They are computed with \labelcref{eq:Zqp}, and details of the procedure and the final light quark wave function are also shown in \Cref{fig:Zq-Procedure}. The momentum-dependent quark dressings are compared with the $k$-dependent ones, computed from \labelcref{eq:Zqp0}, in \Cref{fig:ZQkvsp} in \Cref{app:plots}. As expected, the scale-dependence does not reflect the $p$-dependence of $Z_q$. This begs the question how the approximation $\eta_{q,k}(p) \to \eta_{q,k}(k)$ fares in the loops, even though it is a subleading term in the first place. To that end we compare 
\begin{align}\label{eq:FlowGq_p}
	p^4\,G_l(p)&\partial_t R_{\phi,k}(p)G_l(p) 
	\notag\\[1ex]
	&=\frac{1}{Z_{l,k}(p)} \, \frac{p^4\partial_t R_{l,k}(p)}{(M_{l,k}(p)^2 + (p + R_{l,k}(p))^2)^2}\,,
\end{align}
with
\begin{align}\label{eq:FlowGq_k}
	\frac{1}{Z_{l,k}(k)} \, \frac{p^4\partial_t R_{l,k}(p)}{(m_{l,k}^2 + (p + R_{l,k}(p))^2)^2}\,,
\end{align}
where $Z_{l,k}(k)$ is computed $k$-dependently at $p=k$ using the relations below \labelcref{eq:Zk-p}.
The agreement of the loop-terms is excellent, as shown in \Cref{fig:Gqloopapprox}, confirming again \labelcref{eq:loopapprox}. Technically this originates in the following: even though the wave functions at low momenta differ, as depicted in \Cref{fig:ZQkvsp} in \Cref{app:plots}, the quark masses suppress this difference. In summary, this sustains quantitatively the approximation for $\eta_q(p)$ used in the present work. 

We proceed with the discussion of the momentum dependence of the quark mass functions. First of all, $M_l(p)$ and $M_s(p)$ agree with the momentum-independent ones used in the flow at vanishing momentum, see \labelcref{eq:ApproxMq}, which is one of the reasons behind the excellent agreement of \labelcref{eq:FlowGq_p,eq:FlowGq_k}. Their momentum dependence is obtained by integrating the momentum-dependent flow of the Yukawa coupling \labelcref{eq:Flowhphip} in \Cref{sec:QuarkZ} on the $k$-dependent solution in the present approximation. They are depicted in \Cref{fig:Mq} in comparison to results from quantitative functional computations from \cite{Gao:2021wun} as well as recent lattice results from \cite{Chang:2021vvx}. The fRG results, obtained by reading out the momentum dependence on the flow, agree very well with the results in the literature, which provides further support for the approximation scheme set up here. 

We finally discuss the computation of the pion decay constant in the present approach. It can be computed from the quark mass function $M_l(p)$ using the exact relation
\begin{align}\label{eq:fpi}
	f_\pi =&\, \lim\limits_{P\to -m_\pi^2} \Bigg( \frac{Z_l(P)}{(N_f^2 - 1)\,P^2} 
	\notag\\[1ex]
	&\hspace{.3cm}\times\int_q \textrm{tr}\left[
	T^i \gamma_5 \slashed P G_{l\bar l}(q+P) \Gamma^i(q,P) G_{l\bar l}(q)
	\right]\Bigg)\,.
\end{align}
Here, $\Gamma^i(q,P)$ is the light-quark-pion vertex, 
\begin{align}
	\Gamma^i(q,P) = [\Gamma^{\bar ll\pi}]^i(q+P,-q,-P)\Big\vert_{h_\phi = \hat h_\phi(q+P,-q)}\,. 
\end{align}
These relations imply a relation between the normalised Bethe-Salpeter (BS) wave function $\hat h_\phi$ and the Yukawa coupling in the fRG approach with emergent composites, 
\begin{align}
	f_\pi\, \hat h_\phi(p,q) = M_l(q)Z_l(p) + M_l(p)Z_l(q)\,.
\end{align}
In the explicit computation we use two approximations of the full relations above: Firstly, we do not resolve the angular dependence of $h_\phi$ in this setup, and thus simply approximate
\begin{align}
	f_\pi \, \hat h_\phi(p,q) = 2\,M_l(\bar p)Z_l(\bar p)\,,
\end{align}
where $\bar p = \sqrt{\frac{p^2 + q^2}{2}}$ is the average momentum.
Secondly, we do not resolve the additional tensor structures of the full BS wave function.

Within this approximation, the results for the pion decay constants on the physical point and in the chiral limit in $N_f=2+1$ are given by
\begin{align}
		f_\pi &= 97.2^{+4.0}_{-2.2}\,\unit{\MeV}\,,
		\notag \\[2ex]
		f_{\pi,\chi} &= 93.2^{+3.5}_{-3.1}\,\unit{\MeV}\,.
\label{eq:fpiresults}
\end{align}
We have checked that the influence of implementation details of gauge-consistency described in \Cref{sec:QuarkZ}, and specifically that of the choice of $\Lambda_\textrm{pert}$ in \labelcref{eq:Zqp}, is only of subleading importance for the values of $f_{\pi}$ and $f_{\pi,\chi}$: Varying $\Lambda_\textrm{pert}$ from \qty{5}{GeV} to $\Lambda_\textrm{UV}$ leads to an increase of $\qty{0.5}{\MeV}$ in $f_\pi$ and of $\qty{1}{\MeV}$ in $f_{\pi,\chi}$. 

For the sake of completeness we also report on results for the pion decay constant with the commonly used approximate Pagels-Stokar formula and improvements thereof, see \Cref{app:other_fpi}. The results in \labelcref{eq:fpiresults} amount to an approximately $5\%$ overestimation as compared to its physical value of $f_\pi = 92-\qty{93}{\MeV}$ and that from lattice $N_f = 2+1$ lattice simulations, see \cite{Workman:2022ynf}. Apart from the two approximations discussed above as well as the overall systematic error estimate of 10\%, a further reason for the deviation may be hidden in the momentum tails of $M_q(p)$ and $Z_q(p)$: As we have not fully fed back the quark mass function into their flows, larger $p$ carry a greater error, but the tail is crucial for the determination of $f_\pi$. 
We expect that $f_\pi$ will drop by a few percent in a fully momentum-dependent, gauge-consistent approach to the full BS wave function. 

In future work we plan to use the full momentum dependence to obtain the full Bethe-Salpeter wave function $\Gamma^{\phi\bar qq}(p_1,p_2)$ and consequently the pion decay constant, see \cite{Eichmann:2016yit}. Furthermore, this development would allow us to significantly reduce the systematic error on \labelcref{eq:fpiresults}.

\section{Summary and outlook}
\label{sec:outlookQCD}

In this work we have set up a systematic expansion scheme in the functional renormalisation group approach with emergent composites to QCD. The systematics concerns both the QCD effective action as well as its functional renormalisation group flow. The present approach is built on and uses previous studies, in particular \cite{Gies:2002hq, Pawlowski:2003hq, Braun:2008pi, Braun:2009gm, Mitter:2014wpa, Braun:2014ata, Rennecke:2015eba, Cyrol:2016tym, Cyrol:2017ewj, Cyrol:2017qkl, Corell:2018yil, Fu:2019hdw, Braun:2020ada}. Thus, the current work includes many evolutionary improvements collected over the past two decades. Moreover, it also contains three novel important developments: 
\begin{itemize} 
\item[(i)] A comprehensive systematic error analysis based on the modular nature of the fRG approach, described in \Cref{sec:ApparentConvergence}. The modular structure is summarised in the \LEGO-principle, see \Cref{sec:Lego}. 
\item[(ii)] The introduction of a fully automated computational framework, allowing for unprecedented access and improvement of the fRG approach to QCD. It is detailed in \Cref{sec:SystematicsQCD} and related Appendices. 
\item[(iii)] The inclusion of the full effective potential of the chiral order parameter, and the access to the full momentum dependence of correlation functions as a first application of the automated computational framework (ii), for results see \Cref{sec:EffPot} and \Cref{sec:ResultsMatterSector} in \Cref{sec:Results}. 
\end{itemize} 
We do not repeat the detailed discussions in the respective Sections, and refer the reader for a compilation of the results on correlation functions and observables in 2+1 flavour QCD to \Cref{sec:Results} and additional Appendices. The latter also include a comparison study of two-flavour QCD. 

In summary, in the present work we have made a major step towards fully converged QCD computations with the fRG approach within the phase structure. Importantly, the current setup already allows for QCD computations at finite temperature and density for $\mu_B/T\gtrsim 4$, pushing the quantitative reliability regime beyond the current bound of $\mu_B/T\lesssim 4$.
The endeavour towards pushing this bound out requires the inclusion of full momentum dependences, further tensor structures both in the quark-gluon and four-quark sector, as well as further emergent composites, such as the diquark field and the density mode. All these steps have been already performed and benchmarked in the respective matter and quark-gluon sectors and are currently included in the fully automated computational fRG framework to QCD, making use of the \LEGO-principle. We hope to report on respective results including a prediction for the location of the critical end-point in the near future, thus performing the final step beyond the current estimates \cite{Fu:2019hdw, Gao:2020fbl, Gunkel:2021oya}.

\begin{acknowledgments}
	We thank Jens Braun, Wei-jie Fu, Andreas Geissel, Chuang Huang, Álvaro Pastor, Fabian Rennecke, Jonas Wessely and Shi Yin for discussions. This work is done within the fQCD collaboration \cite{fQCD} and we thank its members for discussions and collaborations on related projects. This work is funded by the Deutsche Forschungsgemeinschaft (DFG, German Research Foundation) under Germany’s Excellence Strategy EXC 2181/1 - 390900948 (the Heidelberg STRUCTURES Excellence Cluster) and the Collaborative Research Centre SFB 1225 (ISOQUANT). It is also supported by EMMI. 
	FRS acknowledges funding by the GSI Helmholtzzentrum f\"ur Schwerionenforschung and by HGS-HIRe for FAIR. FI acknowledges support by the Studienstiftung des Deutschen Volkes. 
	NW acknowledges support by the Deutsche Forschungsgemeinschaft (DFG, German Research Foundation) – Project number 315477589 – TRR 211 and by the State of Hesse within the Research Cluster ELEMENTS (Project ID 500/10.006). 
\end{acknowledgments}
%

\appendix
\begingroup
\allowdisplaybreaks

\section{Flow equations}
\label{app:flows}	

In this Section we complement the flow diagrams in the main text with those relevant to this work. 
\begin{enumerate}
	\item The notation is indicated in \Cref{fig:diag_notation}.
	\item The flow of the effective action is shown in \Cref{fig:FlowGamma}.
	\item The flow of the gluon self-energy correction is shown in \Cref{fig:diag_gluon_2pt}.
	\item The flow of the quark self-energy is shown in \Cref{fig:diag_quark_2pt}.
	\item The flow of the meson self-energy is shown in \Cref{fig:diag_meson_2pt}.
	\item The flow of the three-gluon vertex is shown in \Cref{fig:diag_gluon_3pt}.
	\item The flow of the quark-gluon vertex is shown in \Cref{fig:diag_quark_gluon_3pt}.
	\item The flow of the four-quark vertex is shown in \Cref{fig:diag_quark_4pt}.
\end{enumerate}
\begin{figure*}[t]\centering
	\begin{subfigure}[t]{.4\linewidth}\centering
		\includegraphics[width=\textwidth]{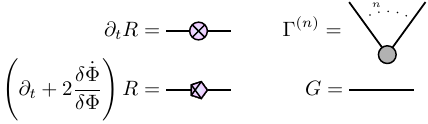}
		\caption{Notation of diagrammatic building blocks.\hspace*{\fill}}
	\end{subfigure}%
	\hspace{0.1\linewidth}%
	\begin{subfigure}[t]{.4\linewidth}\centering
		\includegraphics[width=0.35\textwidth]{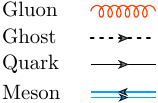}
		\caption{Different propagators appearing in diagrams.\hspace*{\fill}}
	\end{subfigure}%
	\caption{Diagrammatic notation used throughout this work. \hspace*{\fill}}
	\label{fig:diag_notation}
\end{figure*}
\begin{figure*}[h]
	\centering
	\includegraphics[scale=0.65]{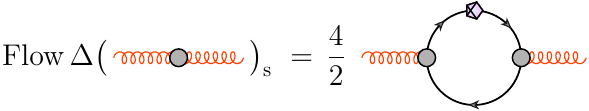}
	\caption{Flow equation of the correction to the gluon self-energy. The subscript 's' denotes that only the strange quark contributes on the right-hand side. The diagrammatic notation is explained in \Cref{fig:diag_notation}.\hspace*{\fill}}
	\label{fig:diag_gluon_2pt}
\end{figure*}
\begin{figure*}[h]
	\centering
	\includegraphics[scale=0.65]{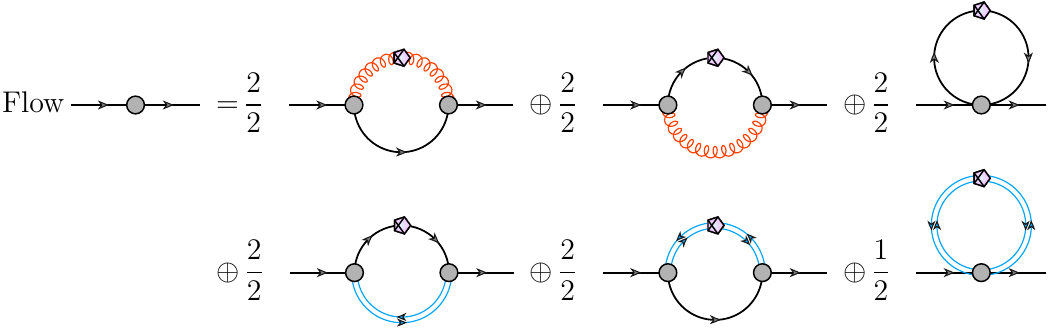}
	\caption{Flow equation of the quark self-energy. The diagrammatic notation is explained in \Cref{fig:diag_notation}.\hspace*{\fill}}
	\label{fig:diag_quark_2pt}
\end{figure*}
\begin{figure*}[h]
	\centering
	\includegraphics[scale=0.65]{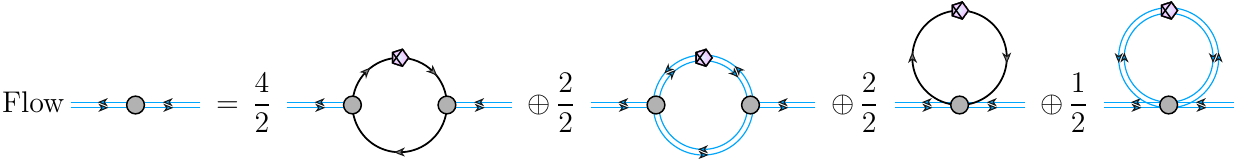}
	\caption{Flow equation of the mesonic self-energy. The diagrammatic notation is explained in \Cref{fig:diag_notation}.\hspace*{\fill}}
	\label{fig:diag_meson_2pt}
\end{figure*}
\begin{figure*}[h]
\centering
\includegraphics[scale=0.65]{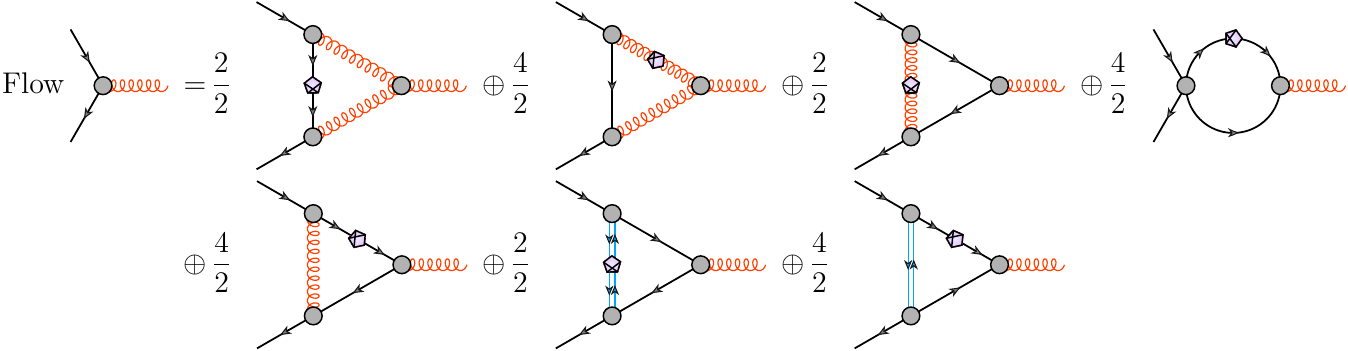}
\caption{Flow equation of the quark-gluon vertex. The diagrammatic notation is explained in \Cref{fig:diag_notation}.\hspace*{\fill}}
\label{fig:diag_quark_gluon_3pt}
\end{figure*}
\begin{figure*}[h]
	\centering
	\includegraphics[scale=0.65]{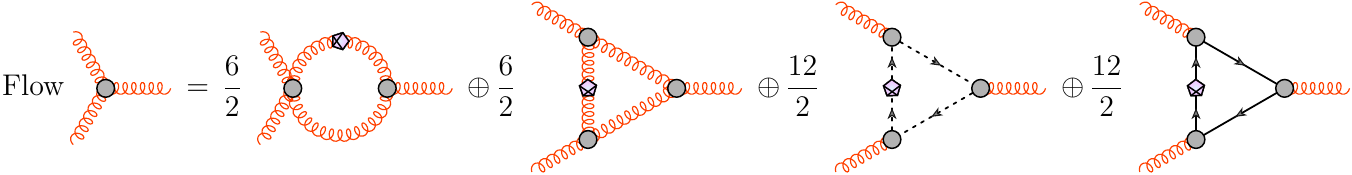}
	\caption{Flow equation of the three-gluon vertex. The diagrammatic notation is explained in \Cref{fig:diag_notation}.\hspace*{\fill}}
	\label{fig:diag_gluon_3pt}
\end{figure*}

\begin{figure*}[h]
	\centering
	\includegraphics[scale=0.65]{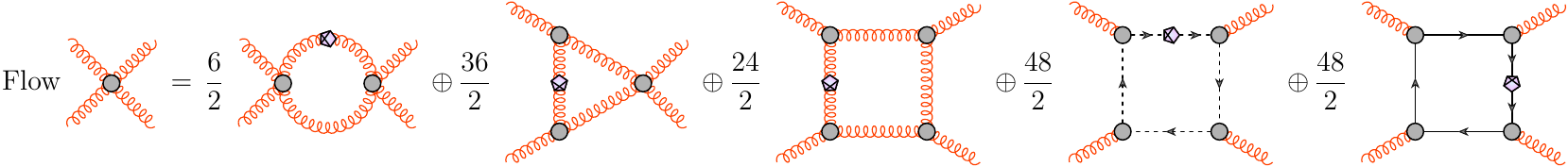}
	\caption{Flow equation of the four-gluon vertex. The diagrammatic notation is explained in \Cref{fig:diag_notation}.\hspace*{\fill}}
	\label{fig:diag_ghost_gluon_3pt}
\end{figure*}
\begin{figure*}[h]
	\centering
	\includegraphics[scale=0.65]{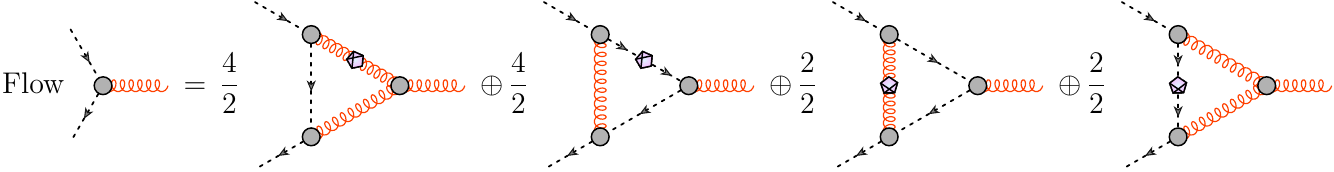}
	\caption{Flow equation of the ghost-gluon vertex. The diagrammatic notation is explained in \Cref{fig:diag_notation}.\hspace*{\fill}}
	\label{fig:diag__gluon_4pt}
\end{figure*}
\begin{figure*}[h]
	\centering
	\hspace*{-20pt}\includegraphics[scale=0.65]{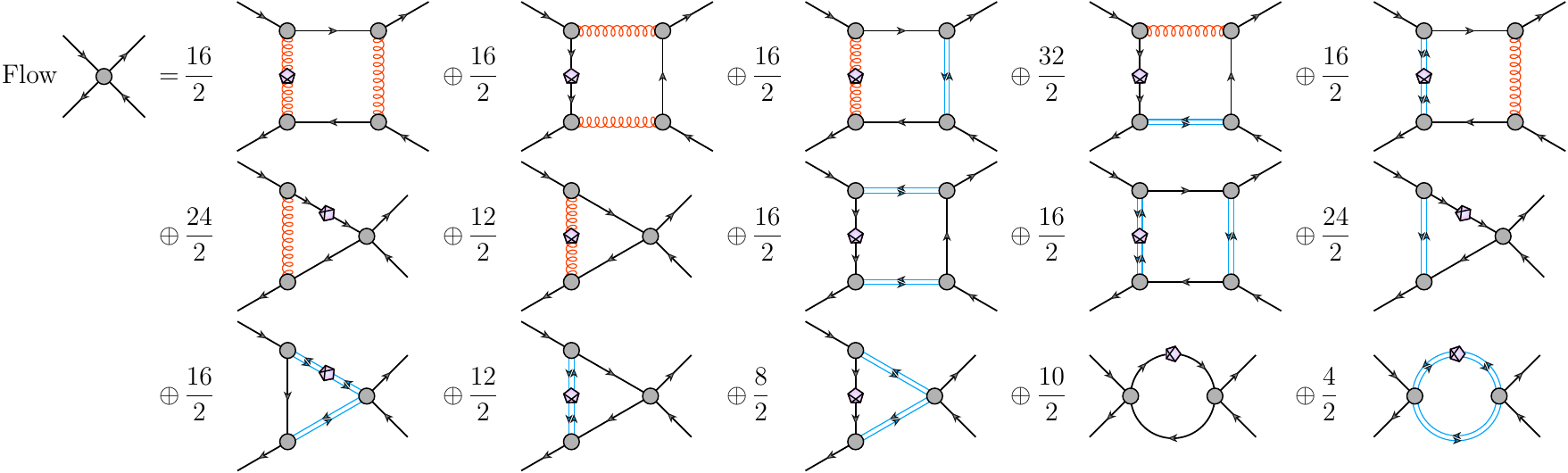}
	\caption{Flow equation of the four-quark vertex. The diagrammatic notation is explained in \Cref{fig:diag_notation}.\hspace*{\fill}\vspace{8cm}}
	\label{fig:diag_quark_4pt}
\end{figure*}
\clearpage

\section{Strange quark sector}
\label{sec:strangeUV}

Here we discuss the limitations of the setup of the strange sector as introduced in \Cref{sec:strangeQuark}. We emphasise again that the current approximation is quantitatively sufficient for the calculation of the dynamics and all observables considered in this work. However, the full RG scale-evolution of the strange-mass functions is not captured quantitatively. This fact can be ignored due to the negligible feed-back into the flows.

A parameter that is tailor-made for this investigation is the ratio of the quark mass functions, $M_s(p)/M_l(p)$: at vanishing momentum it settles at the ratio of the constituent quark masses, 
\begin{align}
\frac{M_s(0)}{M_l(0)}=\frac{m_{s,k=0}}{m_{l,k=0}}\approx 1.386\,,
\label{eq:RatMsMl-0}
\end{align}
in our computation, see \Cref{tab:results}. For $p\to\infty$ it settles at the ratio of the current quark masses, $m_s^0/m_l^0$, that is 
\begin{align}
	\lim_{p\to \infty} \frac{M_s(p)}{M_l(p)}\approx \frac{m_{s,k=\Lambda}}{m_{l,k=\Lambda}}\approx \frac{ m_s^0}{m_l^0} = 27.42(12)\,,
	\label{eq:RatMsMl-Infty}
\end{align}
where the right hand side of \labelcref{eq:massratioCurrent} is taken from lattice simulations \cite{FlavourLatticeAveragingGroupFLAG:2021npn}. 

Both ratios, \labelcref{eq:RatMsMl-0,eq:RatMsMl-Infty}, can be extracted from the $k$-dependence of the quark mass functions: first, we note that the mass ratio of the light and strange current quark masses $m_{l,s}^0$ is related to that of the explicit breaking terms as discussed in Appendix A in \cite{Fu:2019hdw}. This leads us to 
\begin{align}
	\frac{c_{\sigma_ s}}{c_{\sigma_l}}\approx 	\frac{ m_s^0}{m_l^0}= 27.42(12)\,. 
	\label{eq:massratioCurrent}
\end{align}
In principle, the determination of the ratio $c_{\sigma_ s}/c_{\sigma_l}$ requires a computation of the full effective potential $V(\rho, \rho_s)$. 
However, in \cite{Fu:2019hdw}, the relation \labelcref{eq:Vs-V-UV} has been used together with the approximate relation between $V_l$ and $V_s$, 
\begin{align} 
	V_s(\rho_s) \approx  \frac{1}{2}  V_l(2  \rho_s)\,,
	\label{eq:Vs-V}
\end{align}
to approximate the full potential as
\begin{align} 
	V(\rho) \approx  V_l( \rho_l) +\frac{1}{2}  V_l(2  \rho_s)\,.
	\label{eq:VfullVl+Vs}
\end{align}
\Cref{eq:Vs-V} originates from the fact that the light quark effective potential in the UV gets its main contributions from the sum of the up and down quark loops, while the contributions from the mesonic loops sourced by the mesonic self-interaction are strongly suppressed. The contributions from the up and down quark loops are equivalent due to isospin symmetry, $V_d(\rho_l)=V_u(\rho_l)$. Moreover, for cutoff scales far larger than the constituent quark masses all quark loops approach each other as full three-flavour symmetry is restored. We deduce 
\begin{align} 
	V_l(\rho_l) = 2\, V_u(\rho)\,, \qquad V_s(\rho_s ) \stackrel{\frac{k}{k_\chi}\to\infty}{\longrightarrow}  V_u(2 \rho_l)  \,,  
	\label{eq:UVVuVdVs}
\end{align} 
in the ultraviolet. Using the UV relations \labelcref{eq:UVVuVdVs} for all cutoff scales, 
leads us to \labelcref{eq:Vs-V,eq:VfullVl+Vs}. Within this approximation one ignores the infrared differences of the light and strange quark loop contributions as well as the non-trivial infrared dynamics of the light quark condensate triggered by the pions. Both effects can be approximated by using 
\begin{align} 
	V_{s,k=0}(\rho_s ) \approx  \frac12 V_{l,k= k_s}(2 \rho_l)\,,\qquad  k_s\approx \qty{500}{\MeV}\,. 
	\label{eq:Vsks}
\end{align} 
Here, $ k_s$ is the approximate scale where the strange dynamics decouples from that of the light quarks and freezes in. 
In \Cref{fig:csks} we show the ratio \labelcref{eq:massratioCurrent} as a function of the decoupling scale $k_s$. The lattice ratio is obtained with 
\begin{align}
	k_s = \qty{505}{\MeV} \approx m_s \,, 
	\qquad \textrm{with} \qquad  	
	\frac{c_{\sigma_ s}}{c_{\sigma_l}}=  27.4\,, 
	\label{eq:ksRation}
\end{align}
sustaining the arguments above. In summary our analysis entails that our framework has the capacity of determining the ratio of the current quark masses and the current qualitative estimate is in the correct ballpark. We also conclude that its prediction requires an extension of the current approximation that takes into account the effective potential of the strange condensate $\rho_s$ as well as the wave function of the strange quark. Incorporating the full strange dynamics is well within the means of the current computational setup. We defer this step to future works, since it is not necessary to access the observables of interest which are gathered in \Cref{tab:results}.  
\begin{figure}
	\includegraphics[width=0.48\textwidth]{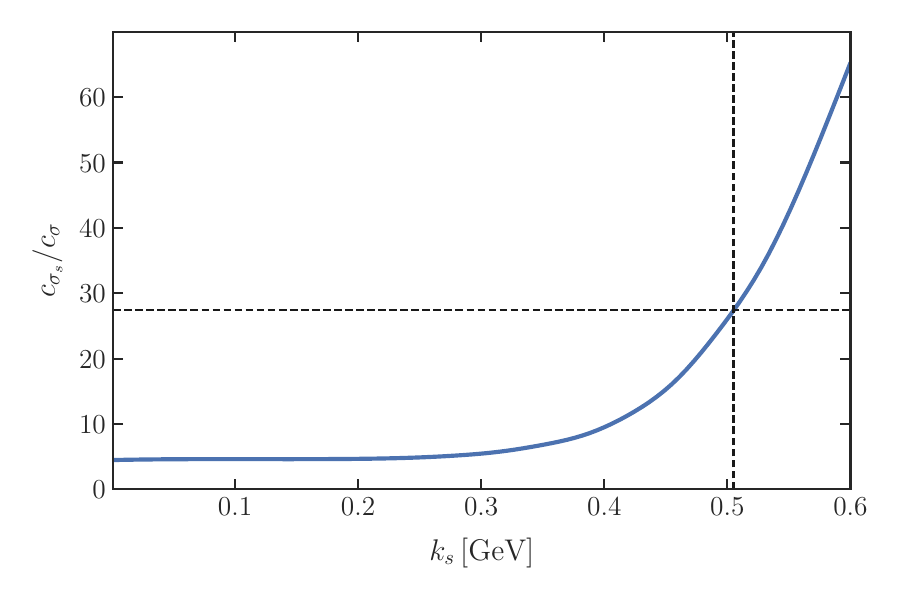}
	\caption{We show $c_{\sigma_s} / c_{\sigma_l}$ as a function of $k_s$ in \labelcref{eq:massratioCurrent}. Clearly, the fraction is strongly dependent on $k_s$, but this is expected as both the strange meson wave function as well as the strange potential itself would effectively stop running around the strange quark mass at $\approx \qty{500}{\MeV}$. The lattice ratio, $c_{\sigma_s} / c_{\sigma_l}\approx 27.4$ is obtained for $k_s= \qty{505}{\MeV}$, see \labelcref{eq:ksRation}. This $k_s$ is indicated as a vertical line and the lattice ratio \labelcref{eq:massratioCurrent} is shown as a horizontal line. These values are indicated by the horizontal and vertical dashed lines.\hspace*{\fill}}
	\label{fig:csks}
\end{figure}

We close this Section with an evaluation of the respective results in \cite{Fu:2019hdw}. There, $V_l(\rho_l)$ has been computed within the fifth order of the Taylor expansion about the flowing minimum $\rho_0$. In terms of the present setup with $k_s$, in \cite{Fu:2019hdw}, \labelcref{eq:Vsks} with $k_s=0$ has been used. This led to the ratio $c_{\sigma_s}/c_{\sigma_l}\approx 27$ for $\Delta m_{sl}= \qty{150}{\MeV}$, and the ratio was rather sensitive to a variation of the value of $\Delta m_{sl}$: For $\Delta m_{sl} \in (\qty{120}{\MeV}\,,\, \qty{150}{\MeV})$, the ratio varied between 14 and 34, while this relatively small variation of $\Delta m_{sl}$ has no sizeable impact on observables such as the chiral condensate. This mirrors the large dependence on $k_s$ shown in \Cref{fig:csks}, as well as the discussion concerning its irrelevance for the observables considered here. In short, the current computations corroborate the findings in \cite{Fu:2019hdw}. 

Given the large sensitivity of the ratio to small scale changes, we expect some changes for the value of the ratio with the current setup. Indeed, emulating the procedure used in \cite{Fu:2019hdw} in the current improved setup by using \labelcref{eq:Vsks} with $k_s=0$, we obtain ${c_{\sigma_ s}}/{c_{\sigma_l}}\approx 5.3$, the limiting value for $k_s\to 0$ in \Cref{fig:csks}. In this limit the strange potential also contains the infrared dynamics of the pions: This dynamics is absent in the strange sector and leads to a further non-trivial $\rho_l$-dependence of the potential $V_l(\rho_l)$, which is absent in $V_s(\rho_s)$. In \cite{Fu:2019hdw}, the choice of $k_s=0$ is effectively compensated by the fact, that $2\rho_{s,0}$ lies far beyond the convergence radius of the Taylor expansion around the expectation value of $\rho_l$, and the Taylor potential shows a stronger $\rho$-dependence. This has been also discussed in detail in \Cref{sec:EffPot}, see in particular \Cref{fig:MesonPot,fig:taylor_rad,fig:VeffComparison2019}. 

In summary, the reliable determination of the current quark ratio requires the computation of the potential $V(\rho,\rho_s)$ and the current results indicate that at the physical point it suffices to use a Taylor expansion about the combined solution of the EoMs, $(\rho_0,\rho{s,0})$. We recall again, that none of the observables computed in the present work shows any sensitivity to this ratio and the discussion here only provides some insight to the properties of the full potential $V(\rho_l, \rho_s)$.

\section{Diagrammatics}
\label{app:Diagrammatics}

For the sake of completeness and transparency, we collect the full set of correlation functions which is encoded in the effective action \labelcref{eq:effAct}. Our conventions are compatible with the \textit{QMeS-Derivation} Mathematica module \cite{Pawlowski:2022oyq} that is used for the derivation of flows in this work. Thus, directly inserting these expressions into the output of \textit{QMeS-Derivation} allows for a full reproduction of our flow equations.

\subsection{Notation}

Here we detail our notation. It is consistent with the external software modules \textit{QMeS} \cite{Pawlowski:2021tkk} and \textit{FormTracer} \cite{Cyrol:2016zqb}, which we use for the derivation of the flow equations, the projections and the traces over field-, Lorentz-, colour- and flavour-indices. The super-index notation used for the super-field \labelcref{eq:Superfield} is based on \cite{Pawlowski:2005xe} and can be summed up by:\\[-2ex]

An index $a$ for the field $\Phi_a$ contains the species of fields, momentum and possibly Lorentz, Dirac, colour, flavour or further internal indices. For convenience we group fermions together with the respective anti-fermions, 
\begin{align}
	(\Phi_a) = \begin{pmatrix}
		\psi_{i} \\ \bar{\psi}_{i} 
	\end{pmatrix} \,, 
	\label{eq:PsiBarPsi}
\end{align}
with $a=(i,i)$ and for quarks, $i$ labels Dirac, colour and flavour indices. Indices can be contracted with the help of a metric
\begin{align}
	\gamma_{ab} = \gamma^{ab} = \begin{cases}
		\begin{pmatrix}
			0 & -1 \\
			1 & 0
		\end{pmatrix}\,\delta_{ab}
		& \text{if $a$ and $b$ are fermionic,} \\[3ex]
		\delta_{ab} & \text{if $a$ and $b$ are bosonic,}   \\[2ex]
		0               & \text{otherwise} \,.
	\end{cases}
\end{align}
Lowering and raising the index of the super-field is done by using the metric
\begin{align}
	\Phi_a = \Phi^b\gamma_{ba} \,,
	\quad \quad
	\Phi^a = \gamma^{ab}\Phi_b \,.
\end{align}
For the QCD superfield in \labelcref{eq:Superfield}, this leads to the following contraction of all fields,
\begin{align}
\Phi_a 	\Phi^a =\int_x \left[ A_\mu^b A_\mu^b+ 2\, \bar c^b c^b + 2 \,\bar q^A q^A +\left(  \sigma^2 +\boldsymbol{\pi}^2\right)\right] \,.
	\label{eq:SuperfildSquared}
\end{align}
Raising and lowering indices, as introduced at the example of the superfield (vector), also applies to general higher rank tensors, e.g.
\begin{align}
	M^{ab}_{\phantom{ab}c} = \gamma^{aa'} \gamma^{bb'} M_{a'b'}^{\phantom{a'b'}c'} \gamma_{c'c} \,.
\end{align}
The convention introduced above is that used in \textit{QMeS}, and the Wetterich equation reads 
\begin{align}
	\partial_t \Gamma = \frac{1}{2}G_{ab}\dot R^{ab}\,.
\end{align}
Please note here, that the Regulators are defined through
\begin{align}
	\Delta S = \frac12 \Phi_a R^{ab} \Phi_b \,,
\end{align}
which gives an extra minus sign for fermions relative to the inverse propagator.
In particular, the propagators $G_{ba}$ are inverse to $\Gamma^{ab}$ in the sense of
\begin{align}
	G_{ab}\Gamma^{bc} = \gamma^c_{\phantom{c}a}\,.
\end{align}
In the current approach we have to consider general $n$th order $\Phi$-derivatives of functionals $F[\Phi]$, that are themselves scalars, vectors or higher rank tensors in superfield space. Relevant examples for $F$ are the classical and quantum effective action, $F=S[\Phi]$ or $F=\Gamma[\Phi]$ (scalars), the diagrammatic parts of a given flow $F=\textrm{Flow}[\Phi]$, either of $\Gamma$ or its $\Phi$-derivatives (tensors), or the propagator, $F=G[\Phi]$ (second rank tensor). We use the following convention for $n$th order $\Phi$-derivatives of functionals $F[\Phi]$ of the superfield \labelcref{eq:Superfield}, 
\begin{subequations}
\label{eq:NotationDerivatives}
\begin{align}
	F^{a_1a_2\dots a_n}[\Phi] & = \frac{\delta}{\delta\Phi_{a_1}} \frac{\delta}{\delta\Phi_{a_2}} \dots \frac{\delta}{\delta\Phi_{a_n}} F[\Phi] \,. 
\end{align}
We also use the notation 
\begin{align}
F^{(n)}_{\Phi_{a_1}\cdots \Phi_{a_n}}[\Phi] =F^{a_1a_2\dots a_n}[\Phi] \,,
\end{align}
or simply $F^{(n)}[\Phi]$. 
\end{subequations}
All external legs are considered to have incoming momenta, which is equivalent to adopting a positive Fourier convention for all fields, including anti-fermions,
\begin{align}
	\Phi_a(x) = \int \frac{d^d p}{(2\pi)^d} e^{ipx}\Phi_a(p)\,.
\end{align}
Furthermore, we pull the momentum conservation out of the propagator, inverse propagator and regulator to shorten the notation
\begin{align}
	\Gamma^{ab}(q,p) & = (2\pi)^d \delta(p+q)\,\Gamma^{ab}(p) \,,
	\notag                                                        \\[1ex]
	R^{ab}(q,p)      & = (2\pi)^d \delta(p+q)\,R^{ab}(p)\,,
	\notag                                                        \\[1ex]
	G^{ab}(p,q)      & = (2\pi)^d \delta(p+q)\,G^{ab}(p)\,.
\end{align}
%

\subsection{Two-point functions and propagators} 
\label{app:PropagatorsVertices}

Here we collect notational details concerning the propagators of all fields. This Appendix is paired with \Cref{app:regs}, where the regulators are discussed. In the following, the quark index $q$ stands for either $l$ or $s$. We always use the general $q$ when the light and strange sector have the same structure. 

In the present RG-invariant scheme the non-trivial momentum dependence has been absorbed in the definition of the fields and the vertices, see \Cref{sec:RG-invariant expansion}. Then, the RG-invariant two-point functions look like their classical counter-parts and are given by
\begin{align}\nonumber 
	\Gamma^{AA}(p)           & = p^2\,\Pi^\perp(p) + \frac{1}{\xi}\,\Pi^\parallel(p)\,,
	\\[1ex]\nonumber 
	\Gamma^{\bar{c}c}(p)     & = - p^2\,, 	\\[2ex]\nonumber 
		\Gamma^{\bar qq}(p)     & = - {\textrm{i}}\slashed{p} + m_q\,,\\[1ex]\nonumber   
	\Gamma^{\pi\pi}(p)       & = p^2\,+\,\partial_{\rho_l} V_l(\rho_l)\,,	\\[1ex]
	\Gamma^{\sigma\sigma}(p) & = p^2\,+\,\big(\partial_{\rho_l} + 2\rho_l\partial_{\rho_l}^2\big) V_l(\rho_l)\,,
\label{eq:inverseProps}
\end{align}
where we take the Landau gauge limit $\xi \to 0$ after the derivation of equations. The quark and ghost inverse propagators carry a relative minus sign, as we choose the ordering of the indices to be directly compatible with the computer algebra of \textit{QMeS}. The transverse and longitudinal projection operators in \labelcref{eq:inverseProps} are given by,
\begin{align}\nonumber 
	\Pi_{\mu\nu}^\perp(p)     & = \delta_{\mu\nu} - \frac{p_\mu p_\nu}{p^2}\,,\\[1ex]
	\Pi_{\mu\nu}^\parallel(p) & = \frac{p_\mu p_\nu}{p^2}\,. 
	\label{eq:ProjectionOps}
\end{align}
The propagators are given by the inverse of the sum of the two-point functions \labelcref{eq:inverseProps} and the regulators 	\labelcref{eq:RegPhi}, 
\begin{align}
	G_{\bar qq}(p)     & = - \Big({\textrm{i}}\slashed{p}\,\big(1+r_{q,k}(p)\big) - m_q\Big)\, G_{s,q}(m_q^2, p^2)\,,
	\notag                                                                                      \\[1ex]
	G_{AA}(p)           & = \, \Pi^\perp(p)\,G_{s,A}(0, p^2)\,,
	\notag                                                                                      \\[1ex]
	G_{\bar{c}c}(p)     & = -\Pi^\perp(p)\,G_{s,c}(0, p^2)\,,
	\notag                                                                                      \\[1ex]
	G_{\pi\pi}(p)       & = \, G_{s,\pi}(m_\pi^2, p^2)\,,
	\notag                                                                                      \\[1ex]
	G_{\sigma\sigma}(p) & = \, G_{s,\sigma}(m_\sigma^2, p^2)\,.
	\label{eq:DefofProps}
\end{align}
where we have factored out the scalar parts \labelcref{eq:ScalarProp} of the propagators. 
We parametrise them as
\begin{align}
	G_{s,i} (q, m^2) & = \frac{1}{q^2 + m^2 + R_{s,i}(q^2)} \,, \qquad i=A,c,\bar c,q,\bar q\,, 
\label{eq:ScalPProp}
\end{align}
with four-dimensional momentum regulators. With	\labelcref{eq:ScalarProp} and \labelcref{eq:rPhiEqual} we are led to 
\begin{align}\nonumber 
R_{s,i} (q^2)= & k^2 r(x)\,, \qquad i= A,c,\bar c ,\phi\,,\\[1ex]
R_{s,q} (q^2)= & \sqrt{q^2 +k^2 r(x)} -q\,.
\end{align}
At finite temperature and density, Euclidean O(4)-invariance is broken via the rest frame the propagators depend on frequency and spatial momenta separately. This suggests the use of regulators $R_k(q_0,\boldsymbol{q})$ or spatial momentum regulators $R_k(\boldsymbol{q})$. With these regulators, the scalar part of the propagators read 
\begin{align}
	G_{s,i} (q, m^2) & = \frac{1}{q_0^2 + \boldsymbol{q}^2 + m^2 + R_{i}(q_0^2, \boldsymbol{q}^2)} \,, \notag                                   \\[1ex]
	G_{s,i} (q, m^2) & = \frac{1}{\tilde q_0^2 + \boldsymbol{q}^2 + m^2 + R_{i}(\boldsymbol{q}^2)} \,.
\label{eq:ScalPPropT}
\end{align}
Here, the Matsubara frequencies $q_0$ are given by 
\begin{align} 
\tilde q_0 =& 2 \pi T\,\left(n + \frac12 \delta_{iq} \right)+ \textrm{i} \mu_q\delta_{iq}\,,
\label{eq:tildeq0App}
\end{align}
with $n \in \mathbb{Z}$. \Cref{eq:tildeq0App} accommodates all fields: for the quark the Matsubara frequency by is shifted by $\pi T$ (no zero mode). In the presence of the quark chemical potential this shift is augmented by an imaginary one with $\textrm{i}\mu_q$. For $A,c,\bar c,\phi$, four-dimensional regulators are better suited as they do not introduce a further O(4)-breaking on top of the physical one induced by the rest frame. 

Using real four-dimensional regulators for the quark, inevitably breaks the Silver-Blaze property at finite quark chemical potential, $\mu_q\neq0$. The Silver-Blaze property entails that for vanishing temperature, correlation functions only depend on $\tilde q_0$ defined in \labelcref{eq:tildeq0App} and carry no genuine $\mu_q$-dependence below the onset chemical potential. The latter is given by 
In the context of the fRG this is discussed in detail in \cite{Khan:2015puu, Fu:2015naa}. This entails that an expansion the mass of the lowest lying state with non-vanishing quark or baryon number. Evidently, an derivative expansion at $q_0=0$ shifts $\mu_q$-dependences from the Matsubara frequency $\tilde q_0$ to the expansion coefficients. This breaks the Silver-Blaze property, while an expansion about $\tilde q_0=0$ preserves it \cite{Fu:2015naa}. We also note that a Silver-Blaze breaking due to the regulator can be controlled with modified symmetry identities, see \cite{Braun:2020bhy}. This enlarges the set of allowed regulators which may be used for a more rapid convergence of the expansion scheme. Typically, however, one chooses a spatial momentum regulator which guarantees the Silver-Blaze property by definition, if an expansion about $\tilde q_0=0$ is used or the full (complex) frequency dependence on $\tilde q_0$ is captured.

\subsection{Gluonic vertices}
\label{app:AVertices}

In this Appendix we discuss all vertices with gluonic legs used in the present work. These are the ghost-gluon, three-gluon, four-gluon, and quark-gluon vertices. We only take into account the dressings of the classical tensor structures to the gluonic vertices. They are given by
\begin{align}\nonumber
		[\Gamma^{A\bar{c}c}(p,q,r)]^{abc}_\mu  =&\, 
	g_{c\bar{c}A} \mathcal{T}_{A\bar{c}c}^{(1)}(p,q) \,  (2\pi)^d\delta(p+q+r)\,,\\[1ex]\nonumber  
	\Gamma^{AAA}(p,q,r)                   = &\, g_{A^3}\,\mathcal{T}^{(1)}_{AAA}(p,q)\,(2\pi)^d\delta(p+q+r)\,,\\[1ex]\nonumber 
	\Gamma^{AAAA}(p,q,r,s)                 =&\, g_{A^4}^2\,\mathcal{T}^{(1)}_{AAAA}(p,q,r) \\[1ex]\nonumber &\hspace{1cm}\times (2\pi)^d\delta(p+q+r+s)\,,
\\[2ex]
	\Gamma^{A\bar{q}q}(p,q,r)             =&\, g_{q\bar{q}A} \, \mathcal{T}_{A\bar{q}q}^{(1)}(p,q)\, (2\pi)^d\delta(p+q+r) 
	\,,
\end{align}
where the $\mathcal{T}_{\Phi_{a_1}\cdots \Phi_{a_n}}^{(1)}$ are the classical tensor structures of the respective classical vertices $S_{\Phi_{a_1}\cdots \Phi_{a_n}}^{(n)}$. Note, that the $\mathcal{T}^{(1)}$ are but one element of complete bases of tensor structures for the vertices $\Gamma^{(n)}$. We provide the classical tensor structures for the scattering vertices of the fundamental fields for the sake of completeness
\begin{align}\nonumber 	
\left[\mathcal{T}_{A\bar{c}c}^{(1)}\right]^{abc}_\mu(p,q) =&\, 	{\textrm{i}}\,f^{abc} q_\mu \,,\\[1ex] \nonumber 
	\left[\mathcal{T}_{AAA}^{(1)}\right]^{abc}_{\mu\nu\rho}(p,q)           =&\, {\textrm{i}} f^{abc}
	  \\[1ex]\nonumber 
	 &\hspace{-3cm}\times\bigg[(q-p)_\rho\delta_{\mu\nu} + (2 p+q)_\nu\delta_{\mu\rho} - (2 q+p)_\mu\delta_{\nu\rho}\bigg] \,,
	 \\[1ex]\nonumber 
	\left[\mathcal{T}_{AAAA}^{(1)}\right]^{abcd}_{\mu\nu\rho\sigma}(p,q,r)  =& \,
	f^{iab}f^{icd}\left(\delta_{\mu\rho}\delta_{\nu\sigma}-\delta_{\mu\sigma}\delta_{\nu\rho}\right)  \\[1ex]\nonumber 
	&
	+f^{iac}f^{ibd}\left(\delta_{\mu\nu}\delta_{\rho\sigma}-\delta_{\mu\sigma}\delta_{\nu\rho}\right)
	 \\[1ex]\nonumber 
	 &
	+f^{iad}f^{ibc}\left(\delta_{\mu\nu}\delta_{\rho\sigma}-\delta_{\mu\rho}\delta_{\nu\sigma}\right) \,,
	 \\[2ex]
	\left[\mathcal{T}_{A\bar{q}q}^{(1)}\right]^A_\mu(p,q) =& \textrm{i}\,\gamma_\nu\, T_c^A\,, 
\label{eq:classicalTensorsGluon}
\end{align}
where $T^A_c$ are the gauge group generators in the fundamental representation. 
Note that in the Landau gauge, only the transverse projections of the classical tensor structures in \labelcref{eq:classicalTensorsGluon} contribute. 
 
It has been shown in \cite{Gao:2021wun} within the Dyson-Schwinger approach, that the quark-gluon vertex carries three relevant tensor structures out of the complete basis with eight tensors. In our basis, see \Cref{app:quarkGluonBasis}, these three tensor structures are given by the elements (1), (4) and (7). Indeed, if dropping (4,7) in the DSE computation, one has to significantly enhance the dressing of the classical tensor structure (typically a factor greater than two) in order to get the full strength of chiral symmetry breaking. Due to the different resummation schemes in DSE and fRG with emergent composites, part of the dynamics that is carried by the dressings of the elements (4,7) in the DSE, is carried by the four-quark, quark-meson and higher order scatterings in the quark-meson sector. Still, in the present approximation we lack the full dynamics, leading to an infrared enhancement or dehancement factor of a few percent, see \Cref{sec:Ie}.

\subsection{Quark-meson vertices}
\label{app:QMVertices}

We use in the following the shorthand notation
\begin{align}
	\Gamma^{a_1\dots a_n}(p_1,\dots,p_n) = (2\pi)^d\delta\left(\sum_{i=1}^n p_i\right)\,\Gamma^{a_1\dots a_n}(0,\dots)\,,
\end{align}
as we evaluate the following vertices on the zero-momentum configuration.

The dynamics of the (light) quark-meson sector is driven by the quark-meson vertices and the mesonic self-interaction. We list the vertices that feed into the flow of the two- and three-point functions. We need 
\begin{align}
	\Gamma^{\bar ll\sigma}       & = - \left( h_\phi + 2 \rho_l\,  h'_\phi\right)\, \tau_0\,,
	\notag                                                                                                                                   \\[1ex]
	\Gamma^{\bar ll\pi}          & = - h_\phi\,\boldsymbol{\tau}\,,
	\notag                                                                                                                                   \\[1ex]
	\Gamma^{\bar ll\pi\pi}       & = - h'_\phi\,\tau_0\,,
	\notag                                                                                                                                   \\[1ex]
	\Gamma^{\bar ll\pi\sigma}    & = - \sigma_l\, h'_\phi\,\boldsymbol{\tau}\,,
	\notag                                                                                                                                   \\[1ex]
	\Gamma^{\bar{l}l\sigma\sigma} & = - \sigma_l (3 h'_\phi + 2 \rho_l\, h''_\phi)\,\tau_0\,,
	\label{eq:qphin}
\end{align}
where $h'_\phi = \partial_{\rho_l} h_\phi$, $h''_\phi =\partial^2_{\rho_l} h_\phi$. We have evaluated all correlation functions in \labelcref{eq:qphin} on the configuration 
\begin{align}
	\phi_l= \left(\sigma_l, \boldsymbol{\pi}=0 \right)\,.
\label{eq:phiExpand}
\end{align}
The purely mesonic vertices are also evaluated on the configuration \labelcref{eq:phiExpand} and read 
\begin{align}
	\Gamma^{\sigma\sigma\sigma} & =\sigma_l\, ( 3 \, V'' + 2 \rho_l \, V''')  \,,
	\notag                                                                                                              \\[1ex]
	\Gamma^{\pi\pi\sigma}       & = \sigma_l V''\,,
	\notag                                                                                                              \\[1ex]
	\Gamma^{\pi\pi\pi\pi}       & = 3\,V''\,,
	\notag                                                                                                              \\[1ex]
	\Gamma^{\pi\pi\sigma\sigma} & = V'' + 2 \rho_l\,  V'''\,, 
	\label{eq:phin}
\end{align}
with $V''=\partial_{\rho_l}^2 V, V'''=\partial_{\rho_l}^3 V$. This concludes our discussion of the propagators and vertices needed in the flow equations considered in the present work. Due to the full $\rho_l$ dependence of the effective potential $V(\rho)$, the present truncation scheme also takes higher scattering orders of the pions and the $\sigma$-mode into account.

\section{Projections}	
\label{app:Projections}

Here, we provide the full expressions for the projection of the flow equations for the wave functions, the couplings as well as the hadronisation function. 
The full anomalous dimensions of the quark and mesons can be straightforwardly obtained as
\begin{align}\nonumber 
	\eta_l    & = -\frac{1}{4 N_c N_f} {\textrm{tr}}\,\Bigg[
		{\textrm{i}} \frac{\slashed{p}}{p^2}\,\textrm{Flow}^{l\bar l}(p) 
		\Bigg]\,,
	\\[2ex]\nonumber 
	\eta_s    & = -\frac{1}{4 N_c} {\textrm{tr}}\,\Bigg[
	{\textrm{i}} \frac{\slashed{p}}{p^2}\,\textrm{Flow}^{s\bar s}(p) 
	\Bigg]\,,
	\\[2ex]\nonumber 
	\eta_{\phi_l} & = -\frac{1}{N_f^2-1} {\textrm{tr}}\,\Bigg[
		\frac{\textrm{Flow}^{\pi\pi}(p) - \textrm{Flow}^{\pi\pi}(0)}{ p^2}
		\Bigg]\,,
	\\[2ex]
	\Delta\eta_A & = -\frac{1}{3 N_c} {\textrm{tr}}\,\Bigg[
		\frac{\Pi^\perp(p)\textrm{Flow}^{AA,(s)}}{ p^2}
		\Bigg]_{\,p\to0}\,,
	\label{eq:etas}
\end{align}
where $\textrm{Flow}^{AA,(s)}$ is only the diagrammatic contribution of the strange quarks to the gluon two-point flow. The traces $\textrm{tr}$ are full traces in the respective internal spaces. 

The projections on the flow equations of the strong couplings make use of the classical tensor structures they are associated with. These are defined in \labelcref{eq:classicalTensorsGluon}. Furthermore, we use symmetric-point configurations for the momenta of the three- and four-point vertices. In the projection equations they are indicated by the subscript SP.
 For a general $n$-point vertex these configurations are defined by 
\begin{align}
	p^2 = \frac{1}{n}\sum_{i=1}^n p_i^2,\qquad p_i\cdot p_j = - \frac{1}{n-1} p^2\,\,\textrm{for $i\neq j$}\,. 
\label{eq:symmetricPoint}
\end{align}
Here, we use always the vertices of a centred $n$-simplex embedded in $d$ dimensions as the explicit values for the $p_i$. 
In other words, for a three-point function we utilize the vertices of a triangle
\begin{align}\notag 
	p_1 &= |p|\cdot \hat e (\pi/2, \pi/2, 0)\,,          	\\[1ex] \notag
	p_2 &= |p|\cdot \hat e (\pi/2, \pi/2, 2\pi/3)\,,    \\[1ex]
	p_3 &= |p|\cdot \hat e (\pi/2, \pi/2, 4\pi/3)\,, 
\end{align}
where $\hat e(\vartheta_1, \vartheta_2, \varphi)$ is the four-dimensional unit vector in hyperspherical coordinates.

For a four-point function, we use the vertices of a tetrahedron, 
\begin{align}\notag 
	p_1 &= |p|\cdot \hat e (\pi/2, 0, 0)\,,          	\\[1ex] \notag
	p_2 &= |p|\cdot \hat e (\pi/2, \arccos(-1/3), 0)\,,    \\[1ex] \notag
	p_3 &= |p|\cdot \hat e (\pi/2, \arccos(-1/3), 2\pi/3)\,,    \\[1ex]
	p_4 &= |p|\cdot \hat e (\pi/2, \arccos(-1/3), 4\pi/3)\,.
\end{align}

\begin{widetext}
\begin{align}
	\partial_t g_{A^3} =&\,\frac{3}{2}\eta_A\, g_{A^3} - \frac{4}{99 N_c (N_c^2-1) p^2}
	{\textrm{tr}}\,\Big[
	[\mathcal{T}_{AAA}]^{abc}_{\mu\nu\rho}(p_1,p_2,p_3)\, [\textrm{Flow}^{AAA}]^{abc}_{\mu\nu\rho}(p_1,p_2,p_3)
	\Big]_{\textrm{SP}} \,,
	\notag\\[1ex]
	\partial_t g_{c\bar cA} =&\,\left(\frac{\eta_A}{2} + \eta_c \right)\, g_{c\bar cA} -\frac{2}{N_c(N_c^2-1)p^2} 
	  {\textrm{tr}}\,\Big[
	[\mathcal{T}_{A\bar cc}]^{acb}_{\mu}(p_1,p_3,p_2) [\textrm{Flow}^{A\bar cc}]_\mu^{abc}(p_1,p_2,p_3)
	\Big]_{\textrm{SP}}\,,
	\notag\\[1ex]
	\partial_t g_{A^4} =&\,\eta_A\, g_{A^4} + \frac{1}{2 g_{A^4}} \frac{3}{98 N_c^2 (N_c^2-1) p^2}
	  {\textrm{tr}}\,\Big[
	[\mathcal{T}_{AAAA}]^{abcd}_{\mu\nu\rho\sigma}(p_1,p_2,p_3)\, [\textrm{Flow}^{AAAA}]^{abcd}_{\mu\nu\rho\sigma}(p_1,p_2,p_3,p_4)
	\Big]_{\textrm{SP}}\,,
	\notag\\[1ex]
	\partial_t g_{q\bar qA}=&\,\left(\frac{\eta_A}{2} + \eta_q \right)\, g_{q\bar qA} -\frac{1}{6 N_f (N_c^2-1)} 
	{\textrm{tr}}
	\,\Big[
	[\mathcal{T}_{A\bar qq}]^{a}_{\mu}(p_1,p_3,p_2) [\textrm{Flow}^{A\bar qq}]_\mu^a(p_1,p_2,p_3)
	\Big]_{\textrm{SP}}\,.
\end{align}
\end{widetext}
The flow of the Yukawa coupling is given by
\begin{align}
	\partial_t h_\phi = & \left(\frac{\eta_{\phi_l}}{2}+ \eta_q\right)\, h_\phi + \frac{1}{2 N_c}{\textrm{tr}}\Bigg[
		\frac{\tau_0\,\textrm{Flow}^{\bar ll}}{\sigma_l}
		\Bigg] - m_\pi^2 \dot{A}_k\,,
\end{align}
which includes the hadronisation function $\dot{{A}}_k$. It is obtained from the flow of the four-fermi coupling and its projection reads schematically
\begin{align}
	\dot{A}_k = & - \frac{1}{ h_\phi}{\textrm{tr}}\Bigg[
		\frac{\mathcal{P}^{\scriptscriptstyle(\pi-\sigma)}_{\beta\alpha\delta\gamma}\,[\textrm{Flow}^{\bar qq\bar qq}]_{\alpha\beta\gamma\delta}}{\mathcal{P}^{\scriptscriptstyle(\pi-\sigma)}_{\beta\alpha\delta\gamma}\mathcal{O}^{\scriptscriptstyle(\pi-\sigma)}_{\alpha\beta\gamma\delta}}
		\Bigg]\,.
\end{align}
We have defined the scalar-pseudoscalar tensor structure as
\begin{align}
	\mathcal{O}^{\scriptscriptstyle(\pi-\sigma)}_{\alpha\beta\gamma\delta} = [\tau_0]_{\alpha\beta}[\tau_0]_{\gamma\delta} + [\bm\tau]_{\alpha\beta}[\bm\tau]_{\gamma\delta}\,.
\end{align}
The projection operator can be chosen in different ways, one possible choice is
\begin{align}\label{eq:4fermiProjectorOld}
	\mathcal{P}^{\scriptscriptstyle(\pi-\sigma)}_{\beta\alpha\delta\gamma} = \mathcal{O}^{\scriptscriptstyle(\pi-\sigma)}_{\beta\alpha\delta\gamma}\,, 
\end{align}
also used in \cite{Fu:2019hdw}. Note that this choice mixes additional tensor structures into the $\sigma-\pi$-channel, when embedded in a Fierz-complete zero-momentum tensor basis of 10 tensor structures. In our current setup, we define
\begin{align}\label{eq:4fermiProjector}
	\mathcal{P}^{\scriptscriptstyle(a)}_{\beta\alpha\delta\gamma} = \sum_i c_{ai}\  \mathcal{O}^{\scriptscriptstyle(i)}_{\beta\alpha\delta\gamma}\,,
\end{align}
where $i$ iterates over the Fierz-complete basis and the $c_{ai}$ are chosen such that we disentangle the respective channel $a$. Taking four quark-derivatives of the effective action within a Fierz-complete basis leads to 
\begin{align}
	\frac{\delta}{\delta \bar q_\alpha}\frac{\delta}{\delta q_\beta}\frac{\delta}{\delta \bar q_\gamma}\frac{\delta}{\delta q_\delta} \Gamma[\Phi] = -2 \lambda_i \sum_i \Big( \mathcal{O}^{\scriptscriptstyle(i)}_{\alpha\beta\gamma\delta} - \mathcal{O}^{\scriptscriptstyle(i)}_{\alpha\delta\beta\gamma}
	\Big)\,.
\end{align}
Thus, a projection operator onto the $a$-th tensor structure is simply defined by
\begin{align}\label{eq:4fermiProjectorexpl}
	c_{ai} = -\frac12 \left[
	\mathcal{O}^{\scriptscriptstyle(a)}_{\beta\alpha\delta\gamma}\,(
	\mathcal{O}^{\scriptscriptstyle(i)}_{\alpha\beta\gamma\delta} - \mathcal{O}^{\scriptscriptstyle(i)}_{\alpha\delta\beta\gamma})
	\right]^{-1}\,.
\end{align}
We have tested that the difference between disentangling these additional contributions and the simple projection operator choice \labelcref{eq:4fermiProjectorOld} makes only a small difference in the results.

The projection procedure and its intricacies will be discussed in upcoming work \cite{?SW2024}, including an optimisation procedure of the bases.

\section{Regulators}
\label{app:regs}

The regulator $R_k$ is a block diagonal matrix in field space with the matrix elements 
	\begin{align}
R^{AA} =R_A\,, \quad R^{c\bar c} =R_c\,, \quad R^{q\bar q} =R_q\,, \quad R^{\phi\phi} = R_\phi\,,
\end{align}
where we suppressed the momentum dependence as well as the Lorentz, Dirac, flavour and gauge group indices in the notation. All other entries are vanishing or follow by crossing symmetry, leading to 
	\begin{align}
		R_k= \begin{pmatrix}
		R_A & 0 & 0& 0 & 0& 0\\[1ex]
		0 & 0 & -R_c & 0 & 0& 0\\[1ex]
		0 & R_c  &  0  & 0 & 0& 0\\[1ex]
		0 & 0 & 0  & 0 & -R_q & 0\\[1ex]
		0 & 0 & 0 & R_q & 0 & 0\\[1ex]
	0 & 0 & 0 & 0 & 0      & R_\phi
	\end{pmatrix}\,.
\label{eq:QCD-RegulatorMatrix}
\end{align}
Throughout this work we use regulators whose Lorentz, Dirac, flavour and gauge group structure is chosen to be the same as that of the corresponding two-point function \labelcref{eq:inverseProps},
\begin{align}
	R_q(p) =& \,{\textrm{i}}\frac{\slashed{p}}{|p|}\, k\, r_q(x)\,,
	\notag\\[1ex]
	R_A(p) =&\, \left[\Pi_\perp(p) + \frac{1}{\xi} \Pi_\parallel(p)\right]\,k^2 r^{\ }_{A}(x)\,,
	\notag\\[1ex]
	R_{c}(p) = &\,k^2 r_{c}(x)\,,
	\notag\\[1ex]
	R_\phi(p) = &\, k^2 r_{\phi}(x)\,, 
\label{eq:RegPhi}
\end{align}
with 
\begin{align}
	x=\frac{p^2}{k^2}\,.
\label{eq:x}
\end{align}
In	\labelcref{eq:RegPhi} we have introduced the dimensionless shape functions $r_{\Phi_i}(x)$, which only depend on the dimensionless momentum ratio \labelcref{eq:x}. 

There are two commonly used definitions for shape functions. The one used in \labelcref{eq:RegPhi} has a finite infrared limit, typically $r_{\Phi_i}(x\to 0)\to 1$.
Another commonly used definition is $ k^2 r_\phi(x)=  p^2 \tilde r_\phi(x) $ and hence $\tilde r_\phi(x) =r_{\phi}(x)/x$. 
This entails that $\tilde r_\phi(x\to 0) \to 1/x$ and thus the shape function is not regular for $p\to  0$. The appeal of such a split is that the dimensionful prefactor is the full dispersion, i.e.~simply the classical $p^2$ for the mesonic composite, which is convenient for analytic considerations and flows.

We have used the definition in \labelcref{eq:RegPhi} which is better suited for advanced approximations: In these approximations one necessarily uses smooth shape functions and all momentum integrals are done numerically. Splitting integrands in parts that contain products of diverging and vanishing parts is not a numerically convenient choice.

\subsection{Propagators} 
\label{app:Regs+Props}

This Appendix is paired with \Cref{app:PropagatorsVertices}. The independence of observables under variations of shape functions provides us with an important self-consistency check of the results, as well as informing the systematic error estimate, see \Cref{sec:optimisation}.  

As discussed there, momentum locality suggests the use of shape functions that enforce similar physical cutoff scales in all fields. To that end, we use that all propagators can be written as products of a scalar propagator part $G_{s,\Phi_i}$ and a tensorial part, see \Cref{app:PropagatorsVertices}. 

For the following discussion in \Cref{app:CompRegs} we rewrite \labelcref{eq:ScalPProp} 
as 
\begin{align}\nonumber
	G_{s,i} =&\, \frac{1}{p^2\left[1 + \frac{k^2}{p^2} r_i(x)\right] + m^2_{i} }\,,\qquad \textrm{for}\qquad i= A,c,\phi\,,\\[1ex] 
	G_{s,q} = &\, \frac{1}{ p^2\left[1 + \frac{k}{p} r_q(x)\right]^2 + m^2_q }\,, 
\label{eq:ScalarProp}
\end{align}
where $x$ is defined in \labelcref{eq:x}. Similar cutoff scales for all particles can be obtained by choosing identical expressions in the square brackets of all propagators. Consequently, our explicit results are obtained with the following choice of shape functions
\begin{align}\nonumber 
	r_A(x)=&\,r_c(x)=r_\phi(x)= r(x)\,,\\[1ex] 
	r_q(x) = &\,\frac{1}{k} \left( \sqrt{p^2 +k^2 r(x)} -p\right) \,,
\label{eq:rPhiEqual}
\end{align}
with a free overall shape function $r(x)$. Importantly, using the relations in \labelcref{eq:rPhiEqual} for the regulators reduce the part of the error originating in the lack of the full momentum dependence within our approximation.

\subsection{Computationally adapted regulator classes} 
\label{app:CompRegs}

As explained in \Cref{sec:ClassRegulators}, we require that the propagator develops an $n$th order pole at $p=0$ and $m^2 = -k^2$ in order to obtain computationally tractable flows. This entails 
\begin{align}
G_k(p^2\to 0, m^2 \to -k^2) = \textrm{const.} \,\left(\frac{1}{p^2}\right)^n\,.
\label{eq:Propsingsn}
\end{align}
For sufficiently large $n$, \labelcref{eq:Propsingsn} implements the convexity restoring behaviour discussed above such that flows are numerically feasible within double precision. In the present work, we chose $n=8$. 

The general class of shape functions used here is given by
	\begin{align}
	r_n(x) = e^{-f_n(x)}\,, \qquad f_n(x) = \frac{P_n(x)}{Q_{n-1}(x)}\,.
\label{eq:RationalExpShape}
\end{align}
$P_n(x), Q_{n-1}(x)$ are polynomials of order $n$ and $n-1$ respectively, 
\begin{align}
P_n(x) = \sum_{i=1}^{n}p_i x^i\,,\qquad Q_{n-1} = 1 + \sum_{i=1}^{n-1}q_i x^i\,,
\end{align}
where the lowest order coefficients $p_0=0,\, q_0 = 1$ are fixed such that the infrared cutoff in the propagator is given by $k^2$, 
	\begin{align}
	f_n(x\to 0) = O(x)\,\ \Rightarrow \ p^2 + k^2 r(x) = k^2\left(1 + O(x)\right) \,.
\label{eq:LimitPropx} 
\end{align}
The construction of $f_n(x)$ follows from the relations, 
	\begin{itemize}
	\item[(i)]  $r_n(0 ) = 1$\,,
	\item[(ii)]  $\lim\limits_{x\to 0}\partial_x r_n(x) = -1$\,,
	\item[(iii)] $\lim\limits_{x\to 0}\partial_x^i r_n(x)  = 0$ for $i = 2,\dots,n-1$ \,,
\item[(iv)] $\lim\limits_{x\to 0}\partial_x^n r_n(x)  > 0$  \,,
\end{itemize}
where $(i)$ is already guaranteed by \labelcref{eq:LimitPropx}. 
Apart from leading to an $n$th order pole in the propagator, these relations lead to a plateau-like structure around $x=0$, which is comparable to that of the flat regulator. The present class of regulators are smooth versions of the flat or Litim regulator, \cite{Litim:2000ci}, and are located in the neighbourhood of the class of optimised ${\cal C}^\infty$-regulators, see the discussion around \labelcref{eq:funOp} in \Cref{sec:optimisation}. 

The relations (i) - (iv) lead to $n+1$ constraint equations for the $2n - 1 $ parameters in $P_n$ and $Q_{n-1}$. We do not consider the full class of $f_n(x)$ but close the system with a simple Ansatz, 
\begin{align}
Q_{n-1}(x) = 1 + b\, x^{[n / 2]} + q_{n-1} x^{n-1}\,, 
\end{align}
with the floor function $[.]$, $q^{\ }_{[n/2]}=b$ and $b,\,q_{n-1} >0$. The latter condition guarantees the absence of poles in $f_n(x)$ for $x>0$. Moreover, the constraint (iv) enforces directly
	\begin{align} 
	p_n >0\,,
\label{eq:pn>0}
\end{align} 
which also guarantees the required large momentum limit of a vanishing shape function.

We are left with a class of shape functions which is labelled by 
	\begin{align}
	\left\{ n, b =q_{[n/2]}\,,\,c=	\frac{p_n}{q_{n-1}}\right\}\,, \quad \textrm{with}\quad \lim_{x\to \infty} r(x) = e^{-c x}\,, 
\label{eq:regcdef}
\end{align}
with $c>0$, which is satisfied by \labelcref{eq:pn>0}. 
The parameter $c>0$ controls the large momentum tail, while $b$ controls the smoothness of the transition regime about $x\approx 1$ between the flat region behaviour for $x\to 0 $ and the tail with $x\to \infty$.

Implementing the constraints (i)-(iv) above, and using the definitions in \labelcref{eq:regcdef}, we obtain for pole order $n=8$,
\begin{align}
	P_8 &= \frac{(2 b +1) c x^8}{8 (c-1)}+\frac{1}{21} (7 b+3) x^7+\frac{1}{6} (3 b+1) x^6 \notag\\&\hspace{2.14cm} +\left(b+\frac{1}{5}\right) x^5+\frac{x^4}{4}+\frac{x^3}{3}+\frac{x^2}{2}+x\,,
	\notag\\[2ex]
Q_7 &= {\frac{(2 b+1) x^7}{8 (c-1)}+b x^4+1}\,.
\end{align}
For our main results presented in \Cref{sec:Results} we have chosen 
\begin{align}
	n=8\,,\qquad b=0\,,\qquad c=2\,.
\label{eq:RegResults} 
\end{align}
The systematic error estimates are obtained with a variation of the shape-functions at fixed $n=8$. We choose the two extremes
	\begin{align}
(b,c) =(0, 20)\,,\quad (b,c)= (1, 1.5)\,,
\end{align}
where the first is much closer to the flat regulator and thus much steeper, whereas the second one has a smoothened shape around $x=1$ and a long tail. We explicitly show these regulators and compare them to the flat regulator in \Cref{fig:regs}.

\section{Tensor bases}
\label{app:TensorBases}

Here we provide tensor bases of the four-quark interaction and the quark-gluon interaction: this embeds the tensors used in full bases and allows to evaluate the respective projection procedures. Moreover, these full bases will be used in future works. 
Their inclusion, or at least that of the quantitatively most relevant basis elements, is an important next step towards quantitative precision of functional QCD.

\subsection{Four-quark interactions}
\label{app:fourQuarkBasis}

Here, we present a Fierz-complete four-quark interaction basis at vanishing momentum for $N_f = 2$,

\vspace{-4mm}
\begin{alignat}{2}\hspace{5ex}
	 & \mathcal{L}_{\scriptscriptstyle\bar q\bar qqq}^{\scriptscriptstyle(\sigma-\pi)}       &  & = (\bar q q)^2 - (\bar q \gamma_5 \tau_f q )^2
	\,,\notag                                                                                                                                                                             \\[1ex]
	 & \mathcal{L}_{\scriptscriptstyle\bar q\bar qqq}^{\scriptscriptstyle(\eta')}            &  & = (\bar q \tau_f q )^2 - (\bar q \gamma_5 \tau_0 q )^2
	\,,\notag                                                                                                                                                                             \\[1ex]
	 & \mathcal{L}_{\scriptscriptstyle\bar q\bar qqq}^{\scriptscriptstyle(V-A)}              &  & = (\bar{q}\gamma_\mu \tau_0 q)^2 - (\bar{q}i\gamma_\mu\gamma_5\tau_0q)^2
	\,,\notag                                                                                                                                                                             \\[1ex]
	 & \mathcal{L}_{\scriptscriptstyle\bar q\bar qqq}^{\scriptscriptstyle(V+A)}              &  & = (\bar{q}\gamma_\mu \tau_0 q)^2 + (\bar{q}i\gamma_\mu\gamma_5\tau_0q)^2
	\,,\notag                                                                                                                                                                             \\[1ex]
	 & \mathcal{L}_{\scriptscriptstyle\bar q\bar qqq}^{\scriptscriptstyle(V-A)^\text{adj}}   &  & = (\bar{q}\gamma_\mu \tau_0 \tau_f q)^2 - (\bar{q}i\gamma_\mu\gamma_5\tau_0 \tau_f q)^2
	\,,\notag                                                                                                                                                                             \\[1ex]
	 & \mathcal{L}_{\scriptscriptstyle\bar q\bar qqq}^{\scriptscriptstyle(S+P)_-}            &  & = (\bar{q} \tau_0 q)^2 - (\bar{q}\gamma_5 \tau_f q)^2
	\notag                                                                                                                                                                                \\&&&\hspace{4ex} + (\bar q \gamma_5 \tau_0 q )^2 - (\bar q \tau_f q )^2
	\,,\notag                                                                                                                                                                             \\[1ex]
	 & \mathcal{L}_{\scriptscriptstyle\bar q\bar qqq}^{\scriptscriptstyle(S+P)_-^\text{adj}} &  & = (\bar{q} \tau_0 T^a q)^2 - (\bar{q}\gamma_5 \tau_f T^a q)^2
	\notag                                                                                                                                                                                \\&&&\hspace{4ex} + (\bar q \gamma_5 \tau_0 T^a q )^2 - (\bar q \tau_f T^a q )^2
	\,,\notag                                                                                                                                                                             \\[1ex]
	 & \mathcal{L}_{\scriptscriptstyle\bar q\bar qqq}^{\scriptscriptstyle(S+P)_+}            &  & = (\bar{q} \tau_0 q)^2 + (\bar{q}\gamma_5 \tau_f q)^2
	\notag                                                                                                                                                                                \\&&&\hspace{4ex} - (\bar q \gamma_5 \tau_0 q )^2 - (\bar q \tau_f q )^2
	\,,\notag                                                                                                                                                                             \\[1ex]
	 & \mathcal{L}_{\scriptscriptstyle\bar q\bar qqq}^{\scriptscriptstyle(S+P)_+^\text{adj}} &  & = (\bar{q} \tau_0 T^a q)^2 + (\bar{q}\gamma_5 \tau_f T^a q)^2
	\notag                                                                                                                                                                                \\&&&\hspace{4ex} + (\bar q \gamma_5 \tau_0 T^a q )^2 + (\bar q \tau_f T^a q )^2
	\,,\notag                                                                                                                                                                             \\[1ex]
	 & \mathcal{L}_{\scriptscriptstyle\bar q\bar qqq}^{\scriptscriptstyle(S-P)_-}            &  & = (\bar{q} \tau_0 q)^2 + (\bar{q}\gamma_5 \tau_f q)^2
	\notag                                                                                                                                                                                \\&&&\hspace{4ex} - (\bar q \gamma_5 \tau_0 q )^2 - (\bar q \tau_f q )^2
	\,,\notag                                                                                                                                                                             \\[1ex]
	 & \mathcal{L}_{\scriptscriptstyle\bar q\bar qqq}^{\scriptscriptstyle(S-P)_-^\text{adj}} &  & = (\bar{q} \tau_0 T^a q)^2 + (\bar{q}\gamma_5 \tau_f T^a q)^2
	\notag                                                                                                                                                                                \\&&&\hspace{4ex} - (\bar q \gamma_5 \tau_0 T^a q )^2 - (\bar q \tau_f T^a q )^2\,.
\end{alignat}
%

\subsection{Quark-gluon interactions}
\label{app:quarkGluonBasis}

We defined the full transverse quark-gluon vertex with 
\begin{align}\nonumber 
	\Pi^\bot_{\mu\nu}(p) [\Gamma^{A \bar q q}]^A_\nu(p,q,r) = &(2\pi)^d\delta(p+q+r)\Pi^\bot_{\mu\nu}(p) 
	 \\[1ex]  &\hspace{-2cm}\times \sum_{i=1}^8 \lambda^{(i)}_{A\bar q q}(p,q)    	 \left[\mathcal{T}_{A\bar{q}q}^{(i)}\right]^A_\nu(p,q) 
	\label{eq:TransverseAbarqq}
\end{align}
with the transverse projection operator in \labelcref{eq:ProjectionOps}. We have assumed no flavour mixing and $q=l, s$. We can factor out the generator $T_c^A$ in the fundamental representation of the strong gauge group, 
\begin{align}
\left[\mathcal{T}_{A\bar{q}q}^{(i)}\right]^A_\mu(p,q) = T^A \left[\mathcal{T}_{A\bar{q}q}^{(i)}\right]_\mu(p,q)\,.
\label{eq:qbarqAFactor}
\end{align}
see e.g.~\cite{Mitter:2014wpa, Eichmann:2016yit, Cyrol:2017ewj, Gao:2021vsf}. The basis is conveniently written in terms of the mixed transverse projection operator $	\Pi_{\mu\nu}^\perp(p,q) $ and $\sigma_{\mu\nu}$, 
\begin{align}
	\Pi_{\mu\nu}^\perp(p,q)     =&\, \delta_{\mu\nu} - \frac{p_\mu q_\nu}{p\cdot q}\,, \qquad \sigma_{\mu\nu} =&\, \frac{{\textrm{i}}}{2} [\gamma_\mu,\gamma_\nu]\,.
	\label{eq:ProjectionOps2}
\end{align}
With these preparations, the tensor structures of the transverse part of the quark-gluon vertex are then given by the transverse projections with $\Pi^\bot(p)$ of 
\begin{alignat}{2}
	\bar q\slashed D q:\  \left[\mathcal{T}_{A\bar{q}q}^{(1)}\right]_\mu(p,q) =&\,\textrm{i}\,\gamma_\nu
\,,\notag                                                                                              \\[1ex]
	\bar q\slashed D^2 q:\  \left[\mathcal{T}_{A\bar{q}q}^{(2)}\right]_\mu(p,q)  =&\,
	(q-r)_\nu
	\,,\notag                                                                                              \\[1ex]
	\left[\mathcal{T}_{A\bar{q}q}^{(3)}\right]_\mu(p,q) =&\,\textrm{i}\,\sigma_{\nu\rho}(q-r)_\rho
	\,,\notag                                                                                              \\[1ex]
	\left[\mathcal{T}_{A\bar{q}q}^{(4)}\right]_\mu(p,q) =&\,\textrm{i}\,\sigma_{\mu\nu}p_\nu
	\,,\notag                                                                                              \\[1ex]
	\bar q\slashed D^3 q:\  \left[\mathcal{T}_{A\bar{q}q}^{(5)}\right]_\mu(p,q) =&\,\textrm{i}\,p_\mu\,\slashed p
	\,,\notag                                                                                              \\[1ex]
	\left[\mathcal{T}_{A\bar{q}q}^{(6)}\right]_\mu(p,q)  =&\,\Pi^\perp_{\mu\nu}(p,q-r)\,\textrm{i}\,\gamma_\nu\,p\cdot(q-r)
	\notag\\
	&- (p_\sigma (r_\sigma - q_\sigma)) \,\textrm{i}\,\gamma_\nu
	\,,\notag                                                                                              \\[1ex]
	\left[\mathcal{T}_{A\bar{q}q}^{(7)}\right]_\mu(p,q) =&\frac{{\textrm{1}}}{3}\, \Big\{ \sigma_{\alpha\beta}\gamma_\mu + \sigma_{\beta\mu}\gamma_\alpha + \sigma_{\mu\alpha}\gamma_\beta \Big\}\notag\\&\times \,\left({p+q}\right)_\alpha (p-q)_\beta
	\,,\notag                                                                                              \\[1ex]
	\bar q\slashed D^4 q:\  \left[\mathcal{T}_{A\bar{q}q}^{(8)}\right]_\mu(p,q) =&\,\Pi^\perp_{\mu\nu}(p,q-r)\, p\cdot(q-r)\,\textrm{i}\,\sigma_{\nu\rho}p_\rho
	\,,
	\label{eq:Quark-Gluon}
\end{alignat}
where $r = -p - q$ is the quark momentum.

This choice is adapted from \cite{Eichmann:2016yit}, respecting crossing symmetry and ensuring that no kinematic singularities arise in the channels and their projections. In comparison, we modified $\mathcal{T}^{(6)}$ by subtracting its overlap with $\mathcal{T}^{(1)}$. This ensures that the projection of the first tensor structure also does not obtain singularities for the configuration where one quark-leg is zero and the other quark is forward-scattering into a gluon. This configuration and its good resolution is crucial for the glue mass and the quark propagator at low momenta. In our current setup this does not impact our results in any way, but our investigations into fully momentum-dependent systems have shown the importance of great care in the handling of the tensor bases.
This will be discussed in depth in a forthcoming work \cite{?SW2024}.

\section{Pagels-Stokar formula for $f_\pi$}
\label{app:other_fpi}

We compare the 2+1 flavour result obtained in \Cref{sec:resultsfpi} as well as the two-flavour result in \Cref{app:TwoFlavourQCD} with the approximate result obtained via the Pagels-Stokar formula and its extensions. For the sake of convenience we quote the results from  \Cref{sec:resultsfpi} here, 
\begin{subequations}\label{eq:fpiresults2}
	\begin{align}
		\hspace{-1cm}N_f=2+1:\hspace{1.5cm}
		f_\pi &= 97.2^{+4.0}_{-2.2}\,\unit{\MeV}\,,
		\notag \\[2ex]
		f_{\pi,\chi} &= 93.2^{+3.5}_{-3.1}\,\unit{\MeV}\,,
	\end{align}
	\begin{align}
		\hspace{-1.4cm}N_f=2:\hspace{2cm} 	
		f_\pi &= \qty{99.7}{\MeV}\,,
		\notag \\[2ex]
		f_{\pi,\chi} &= \qty{96.7}{\MeV}\,.
	\end{align}
\end{subequations}
We use the Pagels-Stokar formula \cite{Pagels:1979hd} with the renormalisation scale $\mu= 0$,
\begin{align}\label{eq:fpi_PS}
	f_\pi^2 =&\,  \frac{N_c}{4\pi^2} \int_0^\infty  dp^2 \frac{Z_q(\mu)}{Z_q(p)} p^2 \frac{M_q(p)}{(p^2 + M_q(p))^2}\notag                            \\[2ex]
	& \hspace{2cm}\times	\left(M_q(p) - \frac{p^2}{2}\frac{dM_q(p)}{dp^2}\right)\,.
\end{align}
This results in
\begin{subequations}\label{eq:fpi_PSresults}
	\begin{align}
		\hspace{-1cm}N_f=2+1:\hspace{1.5cm} 	
		f_\pi &= 92.7^{+1.7}_{-0.4}\,\unit{\MeV}\,,
		\notag \\[2ex]
		f_{\pi,\chi} &= 89.1^{+1.7}_{-0.3}\,\unit{\MeV}\,,
	\end{align}
	\begin{align}
		\hspace{-1cm}N_f=2:\hspace{2cm} 	
		f_\pi &= \qty{94.6}{\MeV}\,,
		\notag \\[2ex]
		f_{\pi,\chi} &= \qty{91.8}{\MeV}\,.\quad
	\end{align}
\end{subequations}
Furthermore, we also apply an improved formula for $f_\pi$ from \cite{Barducci:1997jh} with $\mu= 0$, given by
\begin{widetext}
	\begin{align}\label{eq:fpi_PS_improved}
		f_\pi^2 =&\,  \frac{N_c}{4\pi^2} \int_0^\infty  dp^2 \frac{Z_q(\mu)}{Z_q(p)}\Bigg[
		p^2 \frac{M_q(p)}{\left[p^2 + M_q(p)\right]^2}	\left(M_q(p) - \frac{p^2}{2}\frac{dM_q(p)}{dp^2}\right)
		\notag\\[1ex]
		&\hspace{4cm} + \frac{p^6 \left(\frac{dM_q(p)}{dp^2}\right)^2 - p^4 M_q^2(p) \left(\frac{dM_q(p)}{dp^2}\right)^2 - p^4 M_q(p) \frac{dM_q(p)}{dp^2}}{2 \,\left[p^2 + M_q^2(p) \right]^2}
		\Bigg]\,.
	\end{align}
\end{widetext}
This leads to 
\begin{subequations}\label{eq:fpi_PS_improved_results}
	\begin{align}
		\hspace{-1cm}N_f=2+1:\hspace{1.5cm} 	
		f_\pi &= 101.4^{+1.7}_{-0.4}\,\unit{\MeV}\,,
		\notag \\[2ex]
		f_{\pi,\chi} &= 97.3^{+1.7}_{-0.3}\,\unit{\MeV}\,,
	\end{align}
	\begin{align}
		\hspace{-1cm}N_f=2:\hspace{2cm} 	
		f_\pi &= \qty{103.9}{\MeV}\,,
		\notag \\[2ex]
		f_{\pi,\chi} &= \qty{100.6}{\MeV}\,.\quad
	\end{align}
\end{subequations}
%

\section{Results for two-flavour QCD}
\label{app:TwoFlavourQCD}

\begin{table}[hb!]\centering
	\begin{tabular}{|>{\centering}m{0.3\linewidth} | >{\centering\arraybackslash}m{0.25\linewidth} |}
		\hline & \\[-1ex]
		Coupling & $N_f=2$ \\[1ex]
		\hline& \\[-1ex]
		$\alpha_{A^3,\Lambda_\textrm{UV}}$  & $0.210$   \\[1.5ex]
		\hline& \\[-1.5ex]
		$\alpha_{c\bar c A,\Lambda_\textrm{UV}}$ &  $0.235$ 
		\\[1.5ex] 
		\hline& \\[-1.5ex]
		$\alpha_{A^4,\Lambda_\textrm{UV}}$  & $0.210$  \\[1.5ex]
		\hline& \\[-1.5ex]
		$\alpha_{l\bar l A,\Lambda_\textrm{UV}}$ & $0.210$
		\\[1.5ex]
		\hline
	\end{tabular}
	\caption{Strong couplings at the initial scale $\Lambda$ for $N_f = 2$.
		\hspace*{\fill} }
	\label{tab:alphasNf2}
\end{table}
\begin{table}[hb!]\centering
	\begin{tabular}{|>{\centering}m{0.24\linewidth} ||>{\centering}m{0.3\linewidth}|| >{\centering\arraybackslash}m{0.4\linewidth} |}
		\hline & & \\[-1ex]
		Observables & Value & Parameter in $\Gamma_{\Lambda_\textrm{UV}}$\\[1ex]
		\hline& & \\[-1ex]
		$m_{\pi,\textrm{pol}}$ [MeV]      & $139(12)$ & $c_{\sigma,l}=\qty{4.76}{\GeV^3}$ \\ [1ex]
		\hline & & \\[-1ex]
		$\alpha_{l\bar l A ,\Lambda_\textrm{UV}}$   &    &  $\alpha_{l\bar l A,\Lambda_\textrm{UV}} = 0.210$ \\[1ex]
		\hline& &\\[-2ex]
		\hline & & \\[-1ex]
		$m_l$ [MeV]      &  $370$  & $a=-0.0220\,\quad b=\qty{2}{\GeV}$
		\\[1ex]  
		\hline& &\\[-2ex]
		\hline & & \\[-1ex]
		$f_\pi$ [MeV]      &  $99.7$ & \noindent\rule{1cm}{0.4pt}		\\[1ex]
		\hline& &\\[-1ex] 
		$m_{\pi,\textrm{cur}}$ [MeV]      &  $138$& \noindent\rule{1cm}{0.4pt}		\\[1ex]
		\hline& &\\[-1ex] 
		$m_\sigma$ [MeV]      &  $393.1$ &	\noindent\rule{1cm}{0.4pt} \\[1ex]  
		\hline& &\\[-1ex] 
		$\sigma_{l,0}$ [MeV]      &  $69.0$ & 	\noindent\rule{1cm}{0.4pt} \\[1ex]  
		\hline
	\end{tabular}
	\caption{IR-Observables and corresponding parameters of the initial UV action for $N_f=2$.
		\hspace*{\fill}}
	\label{tab:resultsNf2} 
\end{table}
\begin{figure*}
	\centering%
	\begin{minipage}[t]{.48\linewidth}
		\includegraphics[width=\textwidth]{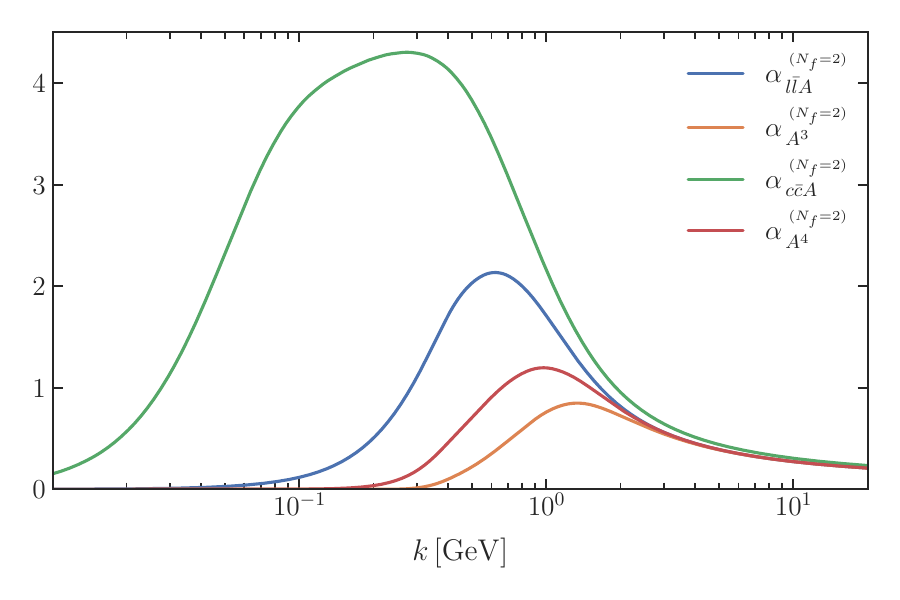}
		\caption{Running of the strong couplings for $N_f=2$.\hspace*{\fill}}
		\label{fig:alphaNf2}
	\end{minipage}%
	\hspace{0.03\linewidth}
	\begin{minipage}[t]{.48\linewidth}
		\includegraphics[width=\textwidth]{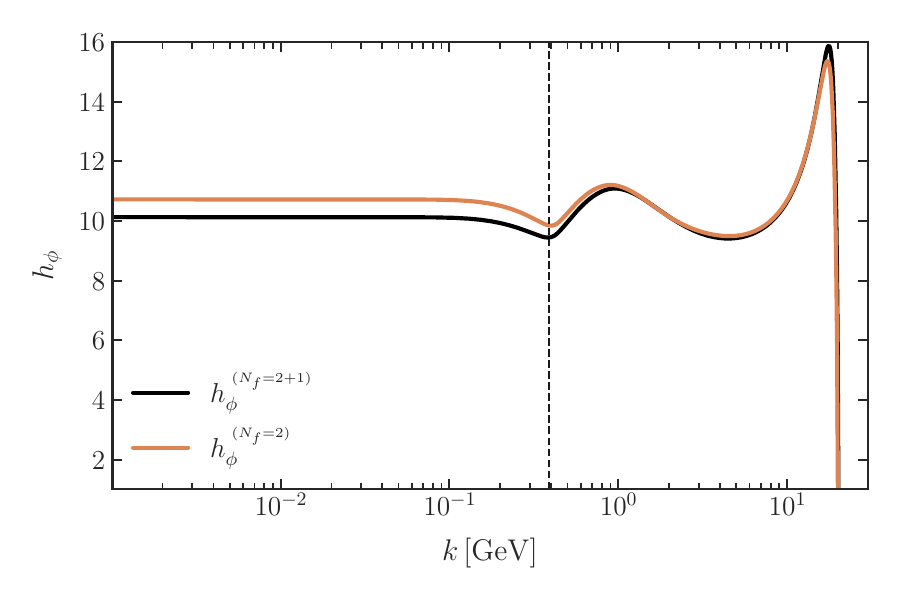}
		\caption{Running of the Yukawa couplings $h_\phi$ in both $N_f=2$ and $N_f=2+1$ for comparison.\hspace*{\fill}}
		\label{fig:hPhiNf2Nf2p1}
	\end{minipage}
	\vspace{10pt}
	\begin{minipage}[t]{.48\linewidth}
		\includegraphics[width=\textwidth]{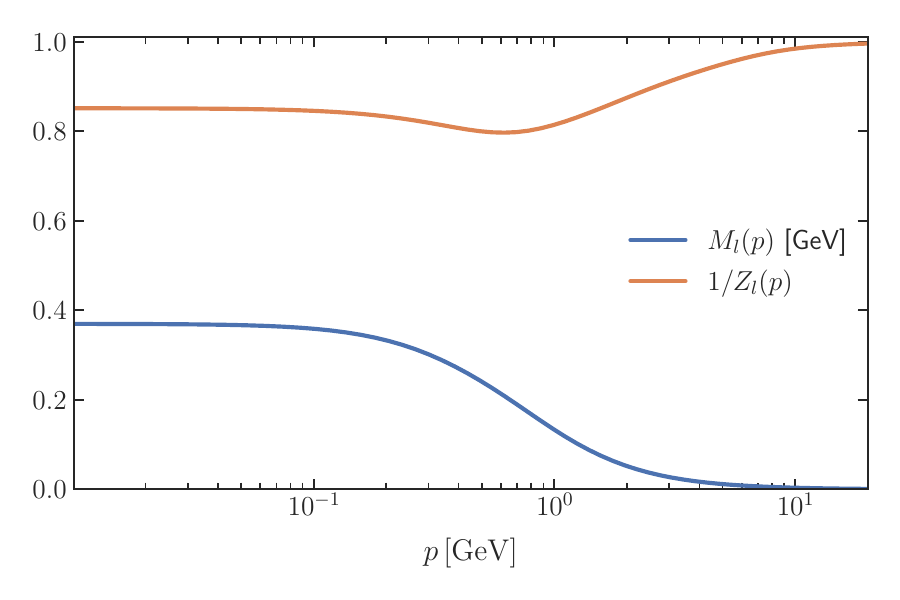}
		\caption{Momentum-dependent quark mass $M_l(p)$ and wave function $1/Z_l(p)$ at $k=0$ for $N_f=2$ flavours.\hspace*{\fill}}
		\label{fig:MqNf2}
	\end{minipage}%
	\hspace{0.03\linewidth}%
	\begin{minipage}[t]{.48\linewidth}
		\includegraphics[width=\textwidth]{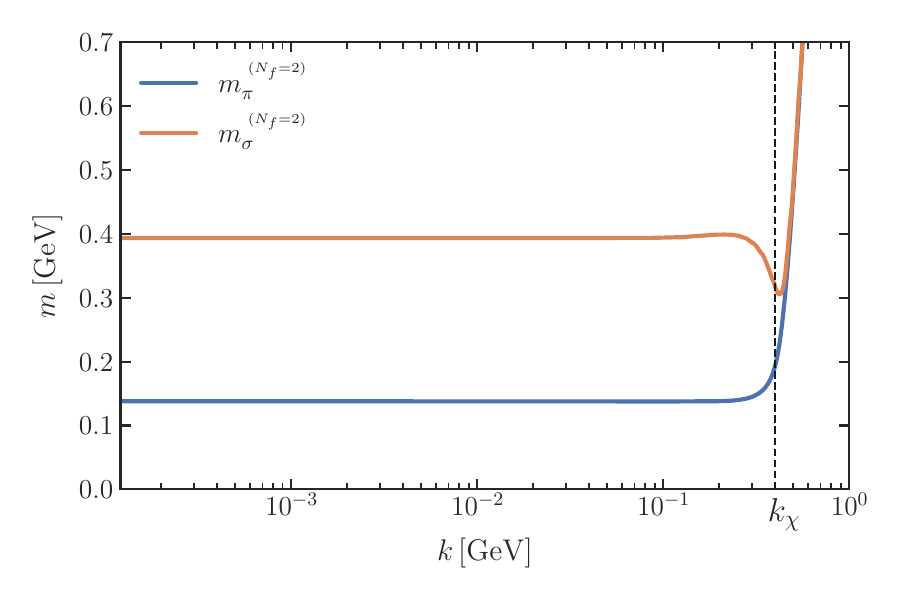}
		\caption{Running of the $N_f=2$-flavour pion and sigma masses.\hspace*{\fill}}
		\label{fig:runPionSigmaNf2}
	\end{minipage}
\end{figure*}%
In this Appendix we gather results and UV parameters for two-flavour QCD. Similar to $N_f=2+1$, we need to choose consistent initial conditions at $\Lambda_{\textrm{UV}}^{(N_f=2)} = \qty{20}{\GeV}$. For the strong couplings they acre collected in \Cref{tab:alphasNf2}. Their evolution is also showcased in \Cref{fig:alphaNf2}.

Note that for $N_f=2$, we have to use a negative modification factor \labelcref{eq:enhancementfunction}, i.e. a 'dehancement' of about 1.5\%. This is compatible with the results of \cite{Fu:2019hdw} where an enhancement of a few percent has been found: In \cite{Fu:2019hdw}, the mixed gluon-meson diagrams for the four-quark coupling have been neglected. If dropped in the current setup, ones is led to a small enhancement factor instead of a dehancement. 

We collect the resulting observables and their corresponding initial conditions in \Cref{tab:resultsNf2}. 
We have not performed a full analysis of the systematic error on the level of the regulator for $N_f=2$ and therefore do not give errors on the results here.
\Cref{tab:resultsChiralNf2} collects the results for $N_f=2$ in the chiral limit. The chiral symmetry breaking scale is slightly higher than in $N_f =2+1$ and is given by
\begin{align}
	k_\chi^{(N_f=2+1)} & = \qty{388}{\MeV}\,,\hspace{1cm}
	k_\chi^{(N_f=2)}    = \qty{397}{\MeV}\,.
	\label{eq:ChiralSymmbScale2}
\end{align}
This scale is also indicated in the plots which show an RG-scale dependence.
\begin{table}[b]\centering
	\begin{tabular}{|>{\centering}m{0.3\linewidth} | >{\centering\arraybackslash}m{0.25\linewidth} |}
		\hline & \\[-1ex]
		Observable & Value \\[0.5ex]
		\hline& \\[-1.5ex]
		$f_{\pi,\chi}$ [MeV] & $96.7$ 	\\[1ex]
		\hline& \\[-1.5ex]
		$m_{l,\chi}$ [MeV]  & $334.1$	\\[1ex]
		\hline& \\[-1.5ex]
		$m_{\sigma,\chi}$ [MeV]  & $247.8$	\\[1ex]
		\hline& \\[-1.5ex]
		$\sigma_{l,0,\chi}$ [MeV] & $67.4$	\\[1ex]
		\hline
	\end{tabular}
	\caption{IR-Observables for $N_f = 2$ in the chiral limit. The corresponding initial parameters are indicated in \Cref{tab:alphasNf2} and \Cref{tab:resultsNf2}. In contrast to the latter table we use $c_\sigma =0$ to evaluate the chiral limit.
		\hspace*{\fill}	\label{tab:resultsChiralNf2} }
\end{table}
\Cref{fig:hPhiNf2Nf2p1} compares the evolution of $h_\phi$ for $N_f=2$ and $N_f=2+1$. The difference in the quark mass almost exclusively originates from the difference in $h_\phi$, since the order parameter $\sigma_0$ is almost identical in both cases. The momentum dependence of the quark mass is shown in \Cref{fig:MqNf2}.
Furthermore, we show the evolution of the pion- and sigma-masses in \Cref{fig:runPionSigmaNf2}.
The pion pole mass can again be extracted from the complex momentum dependence of the propagator. The Padé fit yields
\begin{align}
	m^{(N_f=2)}_{\pi,\textrm{pole}}	= \qty{139(12)}{\MeV}\,,
\end{align}
where we used the average of fit results for the orders $N=2,\dotsc,10$. The error is given by the standard deviation of these fits.

The pion decay constants on the physical point and in the chiral limit in $N_f=2$ are respectively given by
\begin{align} 	
		f_\pi &= \qty{99.7}{\MeV}\,,\hspace{1cm}
		f_{\pi,\chi} = \qty{96.8}{\MeV}\,.
\end{align}
%

\onecolumngrid
\section{Further data}
\label{app:plots}

In this Appendix we collect additional plots and results for $N_f=2+1$, which offer a more detailed understanding of the correlation functions and flows discussed in the present work. 
\vspace{-0.5cm}
\begin{figure}[ht!]
	\centering
	\begin{subfigure}[ht]{.48\linewidth}
		\includegraphics[width=\linewidth]{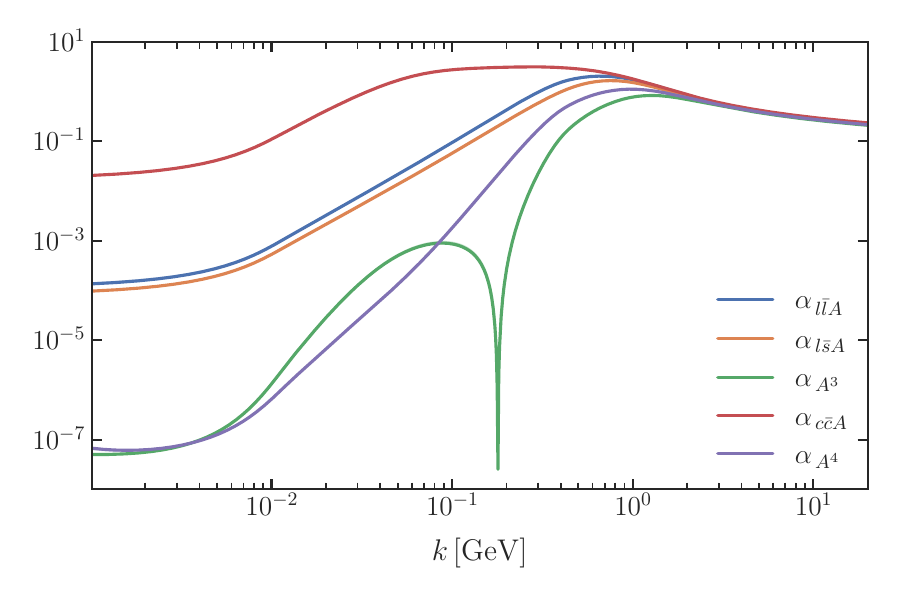}	
		\caption{$k$-dependence of the strong couplings $\alpha_{i,k}$ for $N_f=2+1$ flavours on a double-logarithmic scale. The $\alpha_{A^3}$-coupling becomes negative at $k \approx \qty{0.2}{\GeV}$, for further functional results see e.g.~\cite{Pelaez:2013cpa, Aguilar:2013vaa, Blum:2014gna, Eichmann:2014xya,  Cyrol:2016tym, Cyrol:2017ewj, Athenodorou:2016oyh, Huber:2018ned, Huber:2020keu}.
			\hspace*{\fill}}
		\label{fig:alphasZoomed}
	\end{subfigure}%
	\hspace{0.03\linewidth}%
	\begin{subfigure}[ht]{.48\linewidth}
		\includegraphics[width=\linewidth]{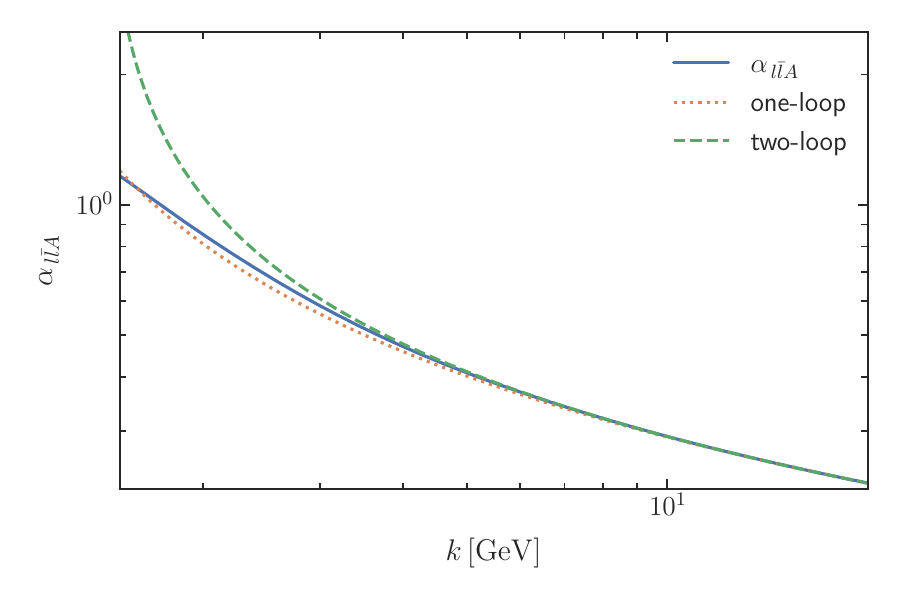}
		\caption{$k$-dependence of the light quark-gluon coupling in comparison to the perturbative one- and two-loop results for $k\geq \qty{1}{GeV}$. The two-loop result matches quantitatively the $k$-dependence of $\alpha_{l\bar l A}$ down to $\Lambda_\textrm{pert} \approx  \qty{5}{\GeV}$, where the perturbative expansion starts to loose its validity, see also~\cite{Gao:2021wun}.\hspace*{\fill}}		\label{fig:pertAlpha}
	\end{subfigure}
	\caption{Cutoff dependence of the avatars of the strong coupling (\Cref{fig:alphasZoomed}), and comparison with perturbation theory for the light quark-gluon coupling (\Cref{fig:pertAlpha}). \hspace*{\fill}}
	\label{fig:CutoffRunningalpha}
	\vspace{-0.5cm}
\end{figure}%
\begin{figure}[ht!]
	\centering%
	\begin{minipage}[t]{.48\linewidth}
		\centering
		\includegraphics[width=\textwidth]{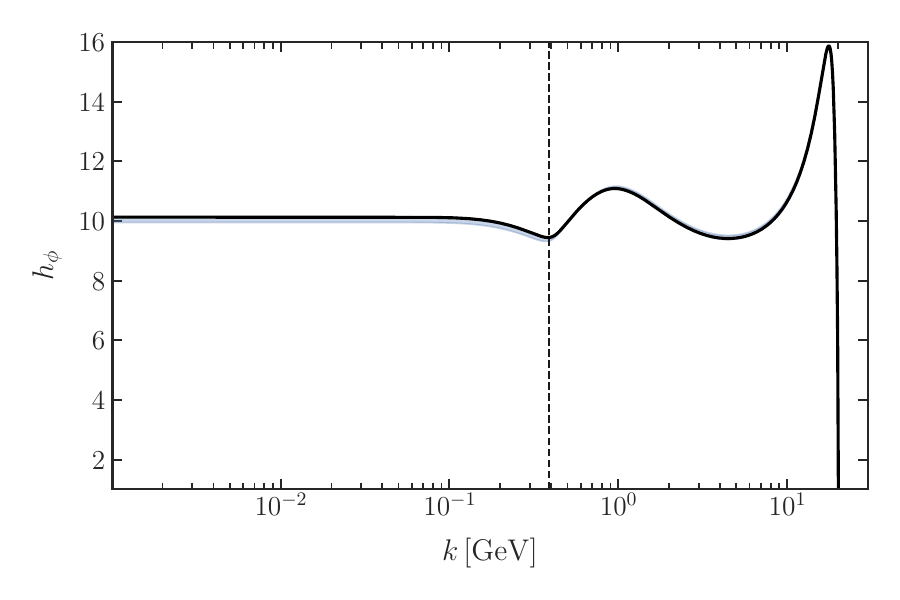}
		\caption{The running of the Yukawa coupling $h_{\phi,k}$ at the physical point. Error bars are given by regulator variation.\hspace*{\fill}}
		\label{fig:hphi}
	\end{minipage}%
	\hspace{0.03\linewidth}%
	\begin{minipage}[t]{.48\linewidth}
		\centering
		\includegraphics[width=\textwidth]{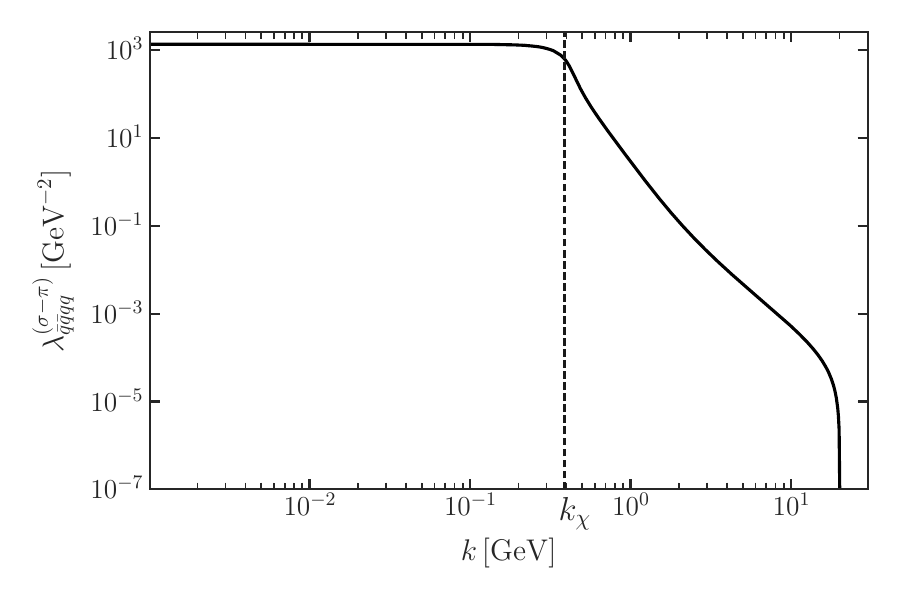}
		\caption{Running of the four-quark vertex at the physical point, which is absorbed into the mesons and $h_{\phi,k}$. We obtain it by transforming back, i.e. $\lambda_{\bar q\bar qqq}^{(\sigma-\pi)} = h_{\phi,k}^2 / 2  m_\pi^2$.\hspace*{\fill}}
		\label{fig:lambda}
	\end{minipage}
\end{figure}%
\vspace{25mm}
\begin{figure}[ht!]
	\centering
	\begin{subfigure}[t]{.48\linewidth}
		\includegraphics[width=\linewidth]{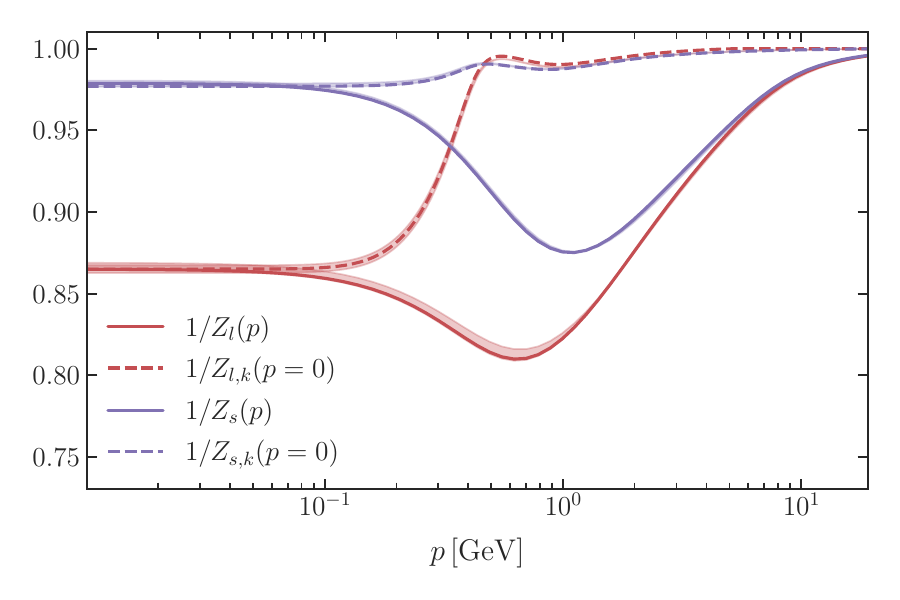}
		\caption{$p$- and $k$-dependent quark wave functions.\hspace*{\fill}}		\label{fig:ZQkvsp}
	\end{subfigure}%
	\hspace{0.03\linewidth}%
	\begin{subfigure}[t]{.48\linewidth}
		\includegraphics[width=\linewidth]{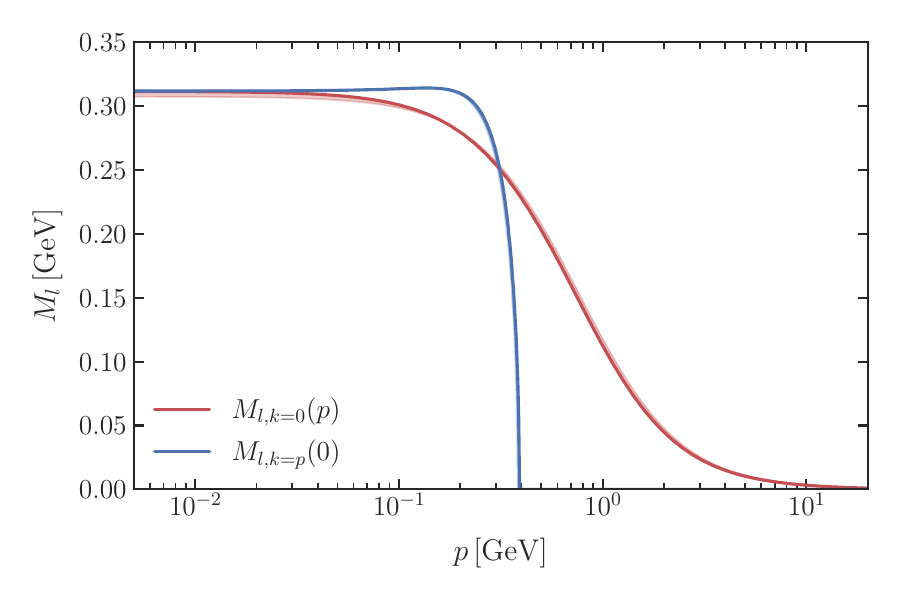}
		\caption{$p$- and $k$-dependent light quark mass function.\hspace*{\fill}}	\label{fig:Mqp-MqkChiralLimit}
	\end{subfigure}
	\caption{Momentum dependent quark wave function and mass function at $k=0$, $Z_q(p)$ and $M_q(p)$ in comparison to the cutoff dependent ones, $Z_{q,k=q}(0)$ and $M_{q=k}(0)$. The errors bars are that obtained by the variation of the regulator. \hspace*{\fill}}
	\label{fig:QuarkPropk-p}
	\vspace{-0.5cm}
\end{figure}%
\begin{figure}[ht!]
	\centering
	\begin{subfigure}[t]{.48\linewidth}
		\includegraphics[width=\textwidth]{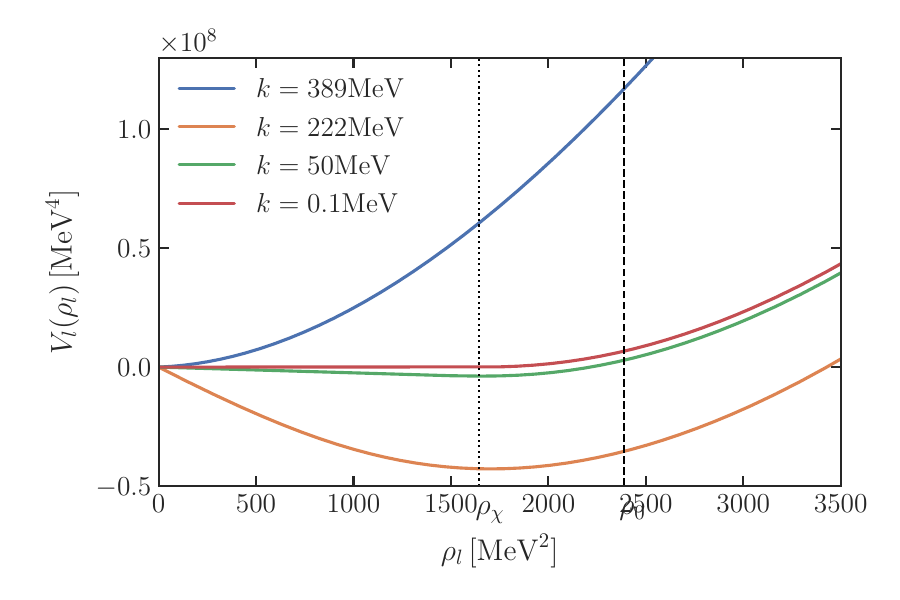}
	\end{subfigure}%
	\hspace{0.03\linewidth}%
	\begin{subfigure}[t]{.48\linewidth}
		\includegraphics[width=\textwidth]{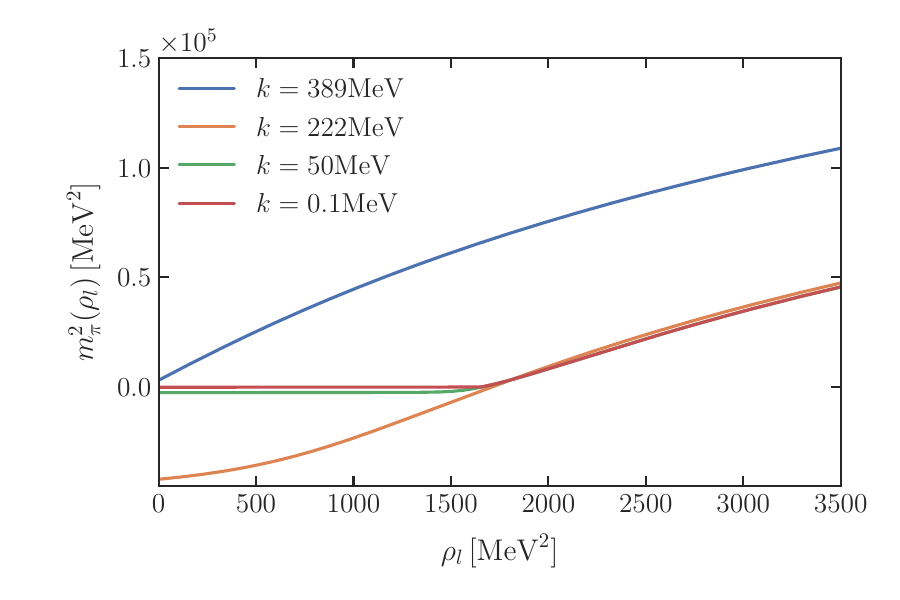}
	\end{subfigure}
	\caption{Time evolution of the effective potential (left) and the pion mass squared (right) at and below the chiral symmetry breaking scale $k_\chi \approx \qty{400}{\MeV}$ in $N_f=2+1$ flavours. \hspace*{\fill}}
	\label{fig:timeevol}
	\vspace{-0.5cm}
\end{figure}%
\begin{figure}[ht!]
	\centering
	\begin{minipage}[t]{.48\linewidth}
		\centering
		\includegraphics[width=\textwidth]{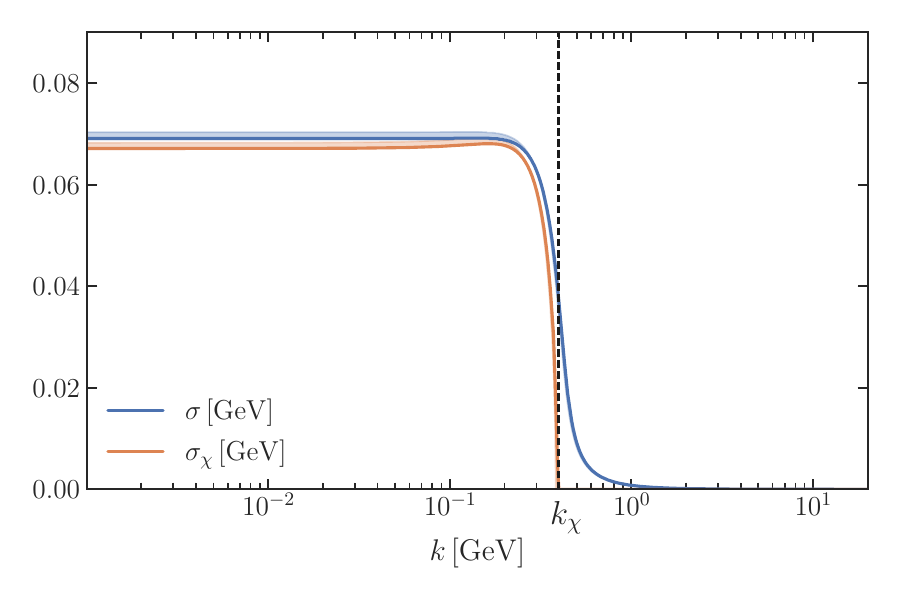}
	\caption{Running of the order parameter $\sigma_l$. The error bars are given by regulator variation.\hspace*{\fill}}
	\label{fig:sigmarunning}

\end{minipage}

\end{figure}%
\twocolumngrid

\endgroup
\clearpage
\bibliographystyle{apsrev4-2}
\bibliography{ref-lib}

\end{document}